\documentclass[twocolumn,twocolappendix]{aastex701}

\usepackage{graphicx}
\usepackage{amsmath}
\usepackage{amssymb}
\usepackage{amsthm}
\usepackage{mathtools}
\usepackage{thmtools}
\usepackage{tcolorbox}
\usepackage{hyperref}
\usepackage{etoolbox}
\usepackage{hyphenat}

\makeatletter
\patchcmd{\ltx@foottext}{%
  .5\textwidth\advance\hsize-18pt}{%
  \linewidth\advance\hsize-1.8em%
}{}{}
\makeatother

\declaretheorem[
    within=section,
    name=Proof
]{myproof}

\renewcommand{\proof}[1][]{\begin{tcolorbox}\begin{myproof}\normalfont}
\renewcommand{\endproof}{\end{myproof}\end{tcolorbox}}

\graphicspath{{./figures/}}

\urldef\nburl\url{https://github.com/straten/epsic/tree/master/notebooks}
\urldef\mwurl\url{https://mathworld.wolfram.com/RodriguesRotationFormula.html}

\newcommand{\mbf}[1]{\mbox{\boldmath $#1$}}
\newcommand{\mbfs}[1]{\mbox{\scriptsize\boldmath $#1$}}

\newcommand{\Eqn}[1]{Equation~(\ref{eqn:#1})}
\newcommand{\Eqns}[3]{Equations~(\ref{eqn:#1}) #2~(\ref{eqn:#3})}
\newcommand{\eqn}[1]{Equation~(\ref{eqn:#1})}
\newcommand{\eqns}[3]{Equations~(\ref{eqn:#1}) #2~(\ref{eqn:#3})}
\newcommand{\eqnp}[2][]{({\ifx\empty#1\empty \else{#1\ }\fi}Eqn.~\ref{eqn:#2})}
\newcommand{\eqnsp}[4][]{({\ifx\empty#1\empty \else{#1\ }\fi}Eqns.~[\ref{eqn:#2}] #3~[\ref{eqn:#4}])}

\newcommand{\Eqnt}[1]{Eqn.~(\ref{eqn:#1})}

\renewcommand{\deg}{\ensuremath{^\circ}}

\newcommand{\peqn}[1]{\text{\Eqnt{#1}}}

\newcommand{\Sec}[1]{Section~\ref{sec:#1}}

\newcommand{\Proof}[1]{Proof~\ref{proof:#1}}
\newcommand{\Proofs}[3]{Proofs~\ref{proof:#1} #2~\ref{proof:#3}} 

\newcommand{\Fig}[1]{Figure~\ref{fig:#1}}
\newcommand{\Figs}[3]{Figures~\ref{fig:#1} #2~\ref{fig:#3}}

\newcommand{\Tab}[1]{Table~\ref{tab:#1}}

\newcommand{\App}[1]{Appendix~\ref{app:#1}}

\newcommand{\Ci}{\ensuremath{i}}

\newcommand{\irow}{\mu} \newcommand{\icol}{\nu}
\newcommand{\jrow}{\kappa} \newcommand{\jcol}{\lambda}

\newcommand{\sech}{\ensuremath{\mathrm{sech} \, }}

\newcommand{\trace}{{\rm Tr}}
\newcommand{\real}{{\rm Re}}
\newcommand{\imag}{{\rm Im}}

\newcommand{\Rotation}{\JM{R}}
\newcommand{\Boost}{\JM{B}}

\newcommand{\tr}[1]{\trace\ensuremath{ \left[ {#1} \right] }}
\newcommand{\re}[1]{\real\ensuremath{ \left[ {#1} \right] }}
\newcommand{\im}[1]{\imag\ensuremath{ \left[ {#1} \right] }}

\newcommand{\sinc}{\mathrm{sinc}}
\newcommand{\bw}{\ensuremath{ \Delta\nu }}

\newcommand{\outerBilinear}[2]{\ensuremath{{#1}\otimes{#2}}}
\newcommand{\stimes}{\ensuremath{\tilde{\otimes}}}
\newcommand{\spinorBilinear}[2]{\ensuremath{{#1}\,\stimes\,{#2}}}

\newcommand{\dcbo}{\mathbin{:}}
\newcommand{\dc}[2]{\ensuremath{{#1}\dcbo{#2}}}

\newcommand{\rotat}{\ensuremath{\vRotation(\phi)}}
\newcommand{\boost}{\ensuremath{\vBoost(\beta)}}

\newcommand{\pauli}[1]{\ensuremath{\cohM{\sigma}_{#1}}}

\newcommand{\pauliI}{
\begingroup
\setlength\arraycolsep{5pt}
\ensuremath{
\begin{pmatrix}
  1 & 0 \\
  0 & 1 
\end{pmatrix}}
\endgroup}

\newcommand{\pauliQ}{
\begingroup
\setlength\arraycolsep{5pt}
\ensuremath{
\begin{pmatrix}
  1 & 0 \\
  0 & -1 
\end{pmatrix}}
\endgroup}

\newcommand{\pauliU}{
\begingroup
\setlength\arraycolsep{5pt}
\ensuremath{
\begin{pmatrix}
  0 & 1 \\
  1 & 0 
\end{pmatrix}}
\endgroup}

\newcommand{\pauliV}{
\begingroup
\setlength\arraycolsep{5pt}
\ensuremath{
\begin{pmatrix}
  0   & -\Ci \\
  \Ci & 0
\end{pmatrix}}
\endgroup}

\newcommand{\Jcol}[1]{\ensuremath{\mbf{#1}}}
\newcommand{\Jrow}[1]{\ensuremath{\mbf{#1}}}
\newcommand{\JM}[1]{{\bf #1}}
\newcommand{\cohM}[1]{\mbf{#1}}
\newcommand{\Pv}[1]{\ensuremath{\underline{\mbf{#1}}}}
\newcommand{\JT}[1]{\ensuremath{\mbf{\mathcal{#1}}}}

\newcommand{\Pvs}[1]{\underline{\mbfs{#1}}}

\newcommand{\Sv}[1]{\underline{\underline{\mbf{#1}}}}

\newcommand{\PM}[1]{\underline{{\bf #1}}}

\newcommand{\MM}[1]{\underline{\underline{{\bf #1}}}}

\newcommand{\vRotation}[1][n]{\ensuremath{\Rotation_{\hat{\Pvs{#1}}}}}
\newcommand{\vBoost}[1][m]{\ensuremath{\Boost_{\hat{\Pvs{#1}}}}}

\shorttitle   {Introduction to Single-Antenna Polarimetry}
\shortauthors {van Straten}
\submitjournal{Publications of the Astronomical Society of the Pacific}

\begin{document}

\title{ An Introduction to Single-Antenna Radio Astronomical Polarimetry }

\author[0000-0003-2519-7375]{Willem van Straten}
\affiliation{Manly Astrophysics, 15/41-42 East Esplanade, Manly, NSW 2095, Australia}
\email{willem.vanstraten@manlyastrophysics.org}

\begin{abstract}

This tutorial reviews the mathematical foundations of single-antenna 
radio polarimetry with the aim of fostering a conceptual understanding
of the relationships between 
%
a physical description of signal propagation
(gain, delay, reflection, down-conversion, etc.),
the corresponding transformations of the electric field vector, and 
the equivalent operations on the Stokes parameters.  
The adopted framework is based on the work of \cite{bri00} and \cite{ham00} and applied to analyze the signal path described by \cite{hbs96} with additional corrections for phase convention and reflection.
Some objective criteria for selecting a model of the instrumental response are introduced and discussed, along with some practical guidelines that facilitate polarimetric calibration.
Further relevant background material and lengthier mathematical proofs are included in the appendix, which introduces the vector, matrix, and tensor notation and concepts of linear algebra used in this work.  The appendix also reviews some of the basics of analog and digital signal processing that are relevant to radio astronomy, and discusses some numerical instabilities that arise when modeling observations.
\end{abstract}


\tableofcontents

\newpage



\setcounter{footnote}{0}

\section{Introduction}
\label{sec:intro}

The polarization of electromagnetic radiation reveals otherwise unattainable details about a wide variety of astrophysical phenomena,
from high-energy cosmic ray air showers \citep{2014PhRvD..89e2002A,2014JCAP...10..014S}
to solar coronal mass ejections \citep{kmo24}
and the relativistic precession of neutron stars \citep{dkl+19}.
On larger scales, radio polarization has been used to study the plasma in the vicinity of a supermassive black hole \citep{2021ApJ...910L..13E},
and the energy density of gravitational waves generated during inflation \citep{pol85,pla20}.

As a signal from an astrophysical source propagates to its point of reception, its state of polarization is altered, which can be used to study the physical properties of the media through which the signal has traveled.
For example, observations of the polarized emission from radio pulsars are used to study the Earth's ionosphere \citep{pmh+23} and map the large-scale structure of the Galactic magnetic field \citep{bbm+96,hmvd18}.
By revealing the strength and orientation of magnetic fields, polarization observations are fundamental to understanding their role in star formation, galaxy evolution, and high-energy phenomena.




Measuring polarization requires purpose-built instrumentation known as a \emph{polarimeter}, and the experimental activities related to measuring polarization are known as \emph{polarimetry}.
A polarimeter typically distorts the polarization state of the signal in a manner that is unintended and cannot be deduced from theory. This unknown component of the instrumental response to an electromagnetic signal must be 
determined experimentally and
calibrated before the polarization intrinsic to the astrophysical source can be interpreted. 

Methods of polarimetric calibration are based on a mathematical model that describes the propagation of a polarized signal along the path between the source and the point of detection.
Accordingly, the primary aim of this paper is to foster a conceptual understanding of the mathematical foundations of polarimetry.
The insights gained through this approach facilitate the development of practical guidelines for use when designing an experiment or analyzing observations.

This review focuses on polarimetric observations of point sources made using a single antenna, such as a single dish or a phased array formed by the phase-coherent addition of signals from multiple elements of an interferometric array.
For a comprehensive introduction to both the theory and practice 
of radio polarimetry using interferometric arrays, including rigorous treatment of arbitrary brightness distributions and direction-dependent effects, the pioneering series by \citet{smi11a,smi11b,smi11c,smi11d}
and the lessons learned by \citet{lab+17} are highly recommended reading.
The formalism employed by Smirnov and the framework presented in this tutorial are both heavily influenced by \cite{ham00}.

Before embarking on this approach,
\Sec{geometry} of this paper reviews the geometric description of polarization introduced 
by \cite{sto52}, beginning with an ideal, monochromatic source of electromagnetic radiation.
In \Sec{statistics}, the Stokes parameters are related to the second-order statistics of 
the electric field as represented by the coherency matrix.
This connection is further explored in \Sec{linear_transformations}, where linear transformations of the electric field are related to Euclidean rotations and Lorentz
boosts of the Stokes parameters.
In \Sec{signal_path}, these transformations are applied to a decomposition of the signal path that extends the \cite{hbs96} model to include the effects of reflection and down-conversion.
Inverting such transformations to recover the original polarization state is discussed in \Sec{calibration}, 
which includes a first-order calibration solution based on the ideal feed assumptions.
\Sec{selection} introduces some criteria for selecting a more complete model of the instrumental response, and \Sec{conclusion} concludes with some pointers to further reading.

Throughout this article, mathematical expressions are greatly simplified by using complex-valued tensors, such as vectors and matrices, and the various operations that are performed on them.  These are described in the appendix with an introduction to
the fundamentals of multivariate linear algebra.
The appendix also reviews the basics of signal processing, such as the analytic signal and down-conversion.
It concludes with some discussion of numerical stability when modeling polarization observations.

\section{Monochromatic Light}
\label{sec:geometry}

At radio frequencies, it is both possible to directly sample the electric field, and sufficient to employ a classical description of electromagnetic radiation \citep{Maxwell1865,bw80}.
At higher frequencies in the electromagnetic spectrum, polarimeters incorporate devices that count photons, such as charge-coupled devices and photomultiplier tubes.
For a comprehensive introduction to the quantum mechanical treatment of the polarization of light, please see
the excellent review by \citet{2021AdOP...13....1G}.

In both classical and quantum approaches to the topic, it is useful to develop some intuition for the geometry of polarization by starting with the nonphysical, idealized case of a purely monochromatic, plane-propagating, transverse electromagnetic wave.
For such a wave, there exist two linearly-independent solutions to Maxwell's equations, representing two oppositely polarized waves.  
Therefore, to fully describe the vector state of any observed signal, radio receiver systems are designed with a pair of receptors that are ideally sensitive to orthogonal senses of polarization.
Typically, the receptors are either circularly polarized (e.g.,  left and right circularly polarized)
or linearly polarized (e.g.,  horizontally and vertically polarized).

The following discussion is based on a Cartesian coordinate system in which the electromagnetic wave propagates in the positive $z$ direction and the electric field is completely described by two orthogonal linearly-polarized components in the $x$-$y$ plane, which is perpendicular to the direction of wave propagation,
\begin{equation}
	\Jcol{\epsilon}(t) =
	\left( 
	\begin{array}{c}
		\epsilon_x(t) \\
		\epsilon_y(t)
	\end{array}
	\right) =
	\left( 
	\begin{array}{c}
		a_x \cos (2\pi \nu t + \phi_x) \\
		a_y \cos (2\pi \nu t + \phi_y)
	\end{array}
	\right). \label{eqn:real_valued_wave}
\end{equation}
Here, $\epsilon_x$ and $\epsilon_y$ are the real-valued
projections of $\Jcol{\epsilon}$ onto the $x$ and $y$ axes.
In the special case of monochromatic light, $\nu$ is the constant frequency of the wave, $a_x$ and $a_y$ are the constant amplitudes of the orthogonal wave components, and $\phi_x$ and $\phi_y$ are the phases of these components at $t=0$.

\begin{figure}
	\centerline{\includegraphics[width=0.7\linewidth]{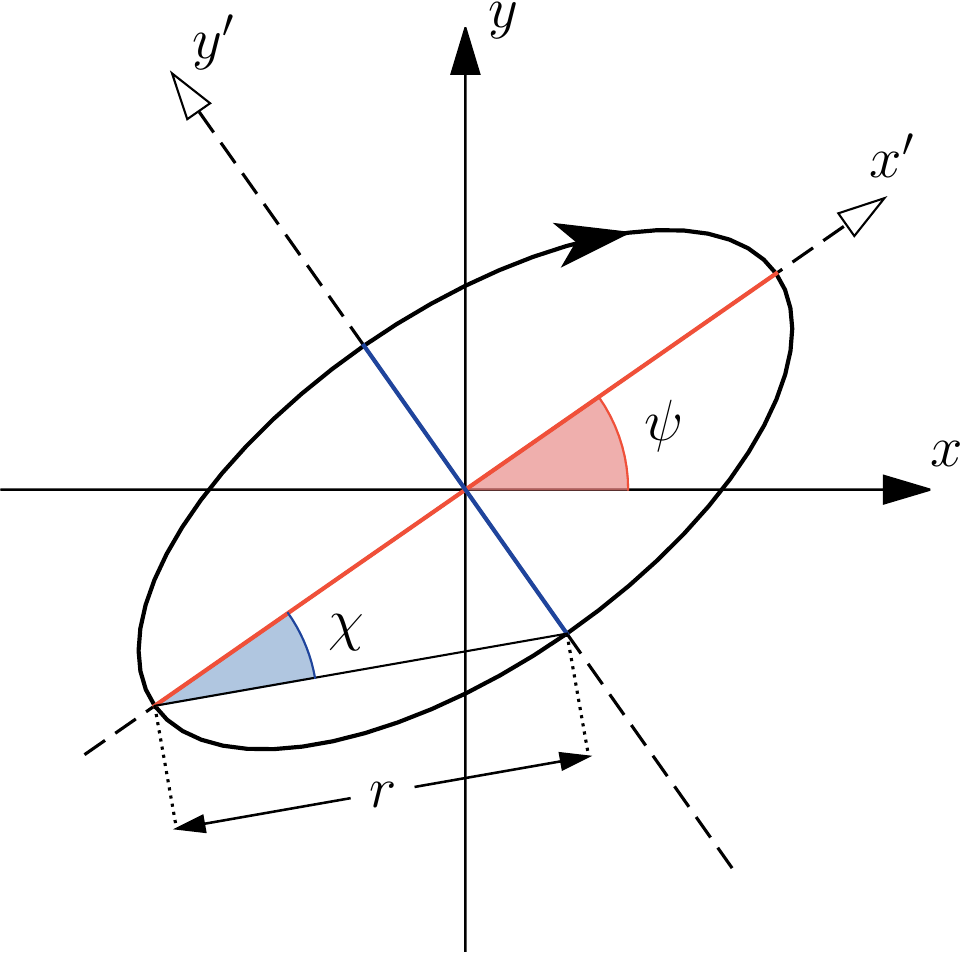}}
\caption{
A left-handed ellipse is traced by the electric field vector on each cycle of a monochromatic wave travelling in the positive $z$ direction, as observed when looking toward the source.
The position angle, $-\pi/2 \le \psi \le \pi/2$, describes the orientation of the semi-major axis of the ellipse with respect to the $x$ axis; it is positive when the $x'$ and $y'$ axes are rotated in a counter-clockwise direction with respect to the $x$ and $y$ axes.
The tangent of the ellipticity angle, $-\pi/4 \le \chi \le \pi/4$, is equal to the ratio between the semi-minor axis and the semi-major axis; it is positive for a left-hand polarized wave and negative for a right-hand polarized wave.
The size of the ellipse is defined by $r$, which is equal to the length of the hypotenuse of the right triangle that contains $\chi$.}
\label{fig:polarization_ellipse}
\end{figure}

On each cycle of the wave, the electric field vector traces an ellipse in the $x$--$y$ plane, as shown in \Fig{polarization_ellipse}.
This figure depicts two angles that describe the geometry of this ellipse: the position angle $\psi$ and the ellipticity angle $\chi$.
\cite{sto52} first demonstrated the utility of expressing the components of the electric field vector in 
terms of these angles, starting with
an intermediate reference frame that is rotated about the 
$z$-axis by the position angle $\psi$, such that
\begin{eqnarray}
	x' &=& x \cos\psi + y\sin\psi \label{eqn:psi_rotation} \\
    y' &=& -x \sin\psi + y\cos\psi.
\end{eqnarray}
In this reference frame, which is depicted with dashed lines
and open arrow heads in \Fig{polarization_ellipse},
the electric field is expressed as in equation~(2) of \cite{sto52}
\begin{equation}
	\Jcol{\epsilon}'(t) =
	\left( 
	\begin{array}{c}
		\epsilon_x'(t) \\
		\epsilon_y'(t)
	\end{array}
	\right) =
	\left( 
	\begin{array}{c}
		r \cos\chi \sin (2\pi \nu t + \phi) \\
		r \sin\chi \cos (2\pi \nu t + \phi)
	\end{array}
	\right),
	\label{eqn:stokes_ellipse}
\end{equation}
where $r$ defines the size of the ellipse and $\phi$ defines the position of the electric field vector at $t=0$.

When $\chi = 0$, $\epsilon_y' = 0$ and the electric field vector oscillates along a line defined by the $x'$ axis, which
is oriented with respect to the $x$ axis by $\psi$.
When $\chi > 0$, 
the electric field travels in a clockwise direction about the ellipse, as seen by an observer looking toward the source, which is defined as a left-hand polarized wave \citep{ieee145}.
%
%
%
When $\chi < 0$, the counter-clockwise traverse of the electric field vector 
is defined as right-hand polarization.
In the special cases of $\chi = \pm \pi/4$, the amplitudes of $\epsilon_x'$ and $\epsilon_y'$ are equal and the electric field traces a circle in the $x$--$y$ plane.
In general, polarization state depends on both the amplitudes of the electric field components and the relative phase between them.  

To move beyond the nonphysical case of perfectly monochromatic waves, for which amplitude and phase are constant as a function of time, it is necessary to characterize random fluctuations of both amplitude and phase using statistical averages.  This is facilitated by adopting the complex-valued analytic representation of a signal, which directly encodes its instantaneous amplitude and phase (see \App{analytic}).
%
%
%
After replacing the real-valued $\epsilon_x(t)$ and $\epsilon_y(t)$ with their associated analytic signals, $e_x(t)$ and $e_y(t)$, the monochromatic wave described by \eqns{real_valued_wave}{through}{stokes_ellipse} is given by the real part of its analytic representation,
\begin{equation}
\label{eqn:analytic_field}
\Jcol{e}(t) =
	\left( 
	\begin{array}{c}
		e_x(t) \\
		e_y(t)
	\end{array}
	\right) = \Jcol{e}_0 \exp(\Ci [2\pi \nu t + \phi - \pi/2]).
\end{equation}
Here, the polarization state of the wave is completely described by the complex-valued two-dimensional Jones vector \eqnp{monochromatic_analytic},
\begin{equation}
\label{eqn:Jones_vector}
\Jcol{e}_0=r\left( \begin{array}{c}
    \cos\psi\cos\chi - \Ci\sin\psi\sin\chi \\
    \sin\psi\cos\chi + \Ci\cos\psi\sin\chi
  \end{array} \right).
\end{equation}
An interactive notebook\footnote{\nburl} demonstrates the relation between $\Jcol{\epsilon}(t)$ and $\Jcol{e}(t)$ by depicting the ellipses inscribed by these functions as $r$, $\psi$ and $\chi$ are varied.



\section{Partial Polarization}
\label{sec:statistics}

No physical signal is ever strictly monochromatic, and 
astrophysical signals are typically an incoherent superposition of waves from a large number of emitting sources.  The polarization of such signals cannot be described as in the previous section.
Furthermore, over most of the spectrum, the electric field fluctuates too rapidly to be directly sampled, and it is generally not possible to determine the geometry of the polarization ellipse directly from measurements of the electric field vector.
Instead, the polarization state must be inferred by other means.  

For example, the position angles of X-ray photons are determined from the trajectories of ejected photoelectrons.
The polarization of optical light is typically estimated by measuring pairs of intensities after the radiation is decomposed into oppositely polarized streams.
(See \App{Stokes_via_intensity_differences} for more detail.)
At radio frequencies, the electric field is directly sampled; however, the astrophysical signal of interest is typically buried in noise and it is necessary to integrate over time and frequency to increase the signal-to-noise ratio.  Such integration can be performed only after squaring the electric field, after which the information about the instantaneous phase of the signal is lost.

%
%

\subsection{Stokes Parameters}
\label{sec:Stokes_parameters}

Although polarization state is not measured directly from the electric field, its interpretation remains founded on the geometry of the polarization ellipse owing to the analysis by \cite{sto52}.
%
%
By considering the conditions under which two independent sources of radiation could be considered equivalent, \cite{sto52} arrived at four measurable quantities that completely describe the state of polarization.
Following refinements by \cite{1942JChPh..10..415P}, the four Stokes parameters are given by
\begin{equation}
\label{eqn:stokes_geometric}
\begin{split}
  S_0&= I                          \\
  S_1&= Ip  \cos2\chi  \cos2\psi   \\
  S_2&= Ip  \cos2\chi  \sin2\psi   \\
  S_3&= Ip  \sin2\chi
\end{split}
\end{equation}
where $I$ is the intensity of the light, $p$ is the degree of polarization (see \Sec{degree_of_polarization}), and
$\chi$ and $\psi$ are the ellipticity and orientation angles that define the polarization ellipse.

\begin{figure}
\centerline{\includegraphics[width=0.7\linewidth]{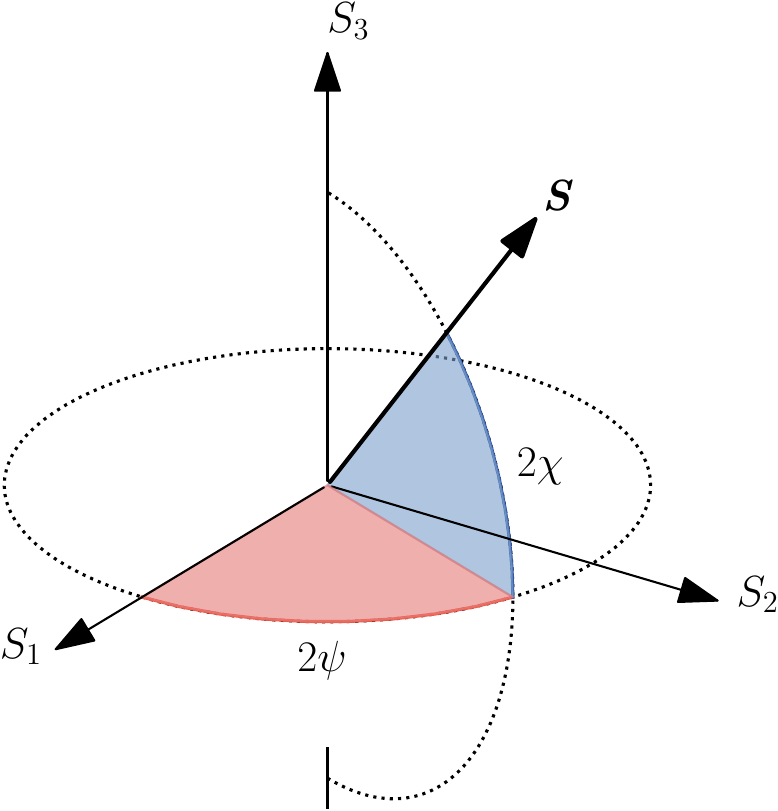}}
\caption{
The spherical coordinates of the polarization vector, $\Pv{S}=(S_1, S_2, S_3)^T=(Q,U,V)^T$, include the vector length $|\Pv{S}|=Ip$, longitude $2\psi$, and latitude $2\chi$.
%
%
%
Points in the $S_1$--$S_2$ plane ($S_3 = 0$) represent linearly 
polarized states, points above this plane ($S_3 > 0$) represent left-hand elliptically 
polarized states and the positive $S_3$ pole represents left-hand circular polarization.
}
\label{fig:polarization_vector}
\end{figure}

The four-dimensional vector of Stokes parameters,
\begin{equation}
	\underline{\underline{S}} =
	\left( 
	\begin{array}{c}
		S_0 \\
		S_1 \\
		S_2 \\
		S_3
	\end{array}
	\right) =
	\left( 
	\begin{array}{c}
		I \\
		Q \\
	 	U \\
		V
	\end{array}
	\right),
\end{equation}
is separable into scalar and vector components, $[S_0,\Pv{S}]$, where
$\Pv{S}=(S_1,S_2,S_3)^T$ is the polarization vector with length $|\Pv{S}|=Ip$ and direction defined by
$2\psi$ and $2\chi$, as depicted in \Fig{polarization_vector}.
Measurement of $\Pv{S}$ yields the orientation and ellipticity angles that define
the geometry of the polarization ellipse,
\begin{eqnarray}
\label{eqn:orientation}
\psi &=& \frac{1}{2} \tan^{-1} \frac{S_2}{S_1} \\
\label{eqn:ellipticity}
\chi &=& \frac{1}{2} \sin^{-1} \frac{S_3}{|\Pv{S}|}.
\end{eqnarray}
%

%
%

\subsection{Degree of Polarization}
\label{sec:degree_of_polarization}

The degree of polarization
\begin{equation}
\label{eqn:degree_of_polarization}
p = \frac{|\Pv{S}|}{S_0}
\end{equation}
is a measure of the fraction of the total intensity that is polarized.
When $p=0$, the light is unpolarized, also defined as \emph{common} light \citep{sto52};
when $p=1$, the light is 100\% polarized, or \emph{purely} polarized;
for partial polarization, $0 < p < 1$.
%

As shown by \eqn{spectral_decomposition_of_coherency_matrix} and the discussion that follows it, unpolarized or partially polarized radiation is equivalent to an incoherent superposition of two orthogonal purely polarized states.
An unpolarized state can be decomposed into any pair of orthogonal purely polarized states of equal amplitude \citep{sto52}.
In contrast, a 100\% polarized wave can be described using a single Jones vector, $\Jcol{e}_0$,
such that
\begin{equation}
\label{eqn:pure_poln_e}
    \Jcol{e}(t) = \Jcol{e}_0 z(t),
\end{equation} 
where $z(t)$ is a complex-valued function that describes stochastic fluctuations of amplitude and phase.
It has a mean of zero and variance of unity; that is, $\langle|z(t)|^2\rangle = 1$, where angular brackets denote an average over time.
Referring to \eqns{analytic_field}{and}{Jones_vector}, monochromatic light is a special case of \eqn{pure_poln_e} in which $z(t)$ is a pure tone with constant frequency; therefore, strictly monochromatic light is 100\% polarized.

%
%

\subsection{Poincar\'{e} Sphere}
\label{sec:Poincare_sphere}

When only the polarization state is of interest, the polarization vector can be normalized by the total intensity, yielding $\Pv{\acute{S}} = \Pv{S}/S_0$, such that $|\Pv{\acute{S}}|=p$.
When $\Pv{\acute{S}}$ is plotted as in \Fig{polarization_vector}, 
every possible polarization state corresponds to a point on or inside a sphere of unity-radius known as the Poincar\'{e} sphere.
In this geometric representation, purely polarized states lie on the surface of the sphere, and the unpolarized state is at its center (the origin).
Partially polarized states are represented by points within the sphere, and their distance from the center is directly proportional to their degree of polarization.

%
%

\subsection{Polarization Ellipse}
\label{sec:polarization_ellipse}

\begin{figure}
\centerline{\includegraphics[width=0.7\linewidth]{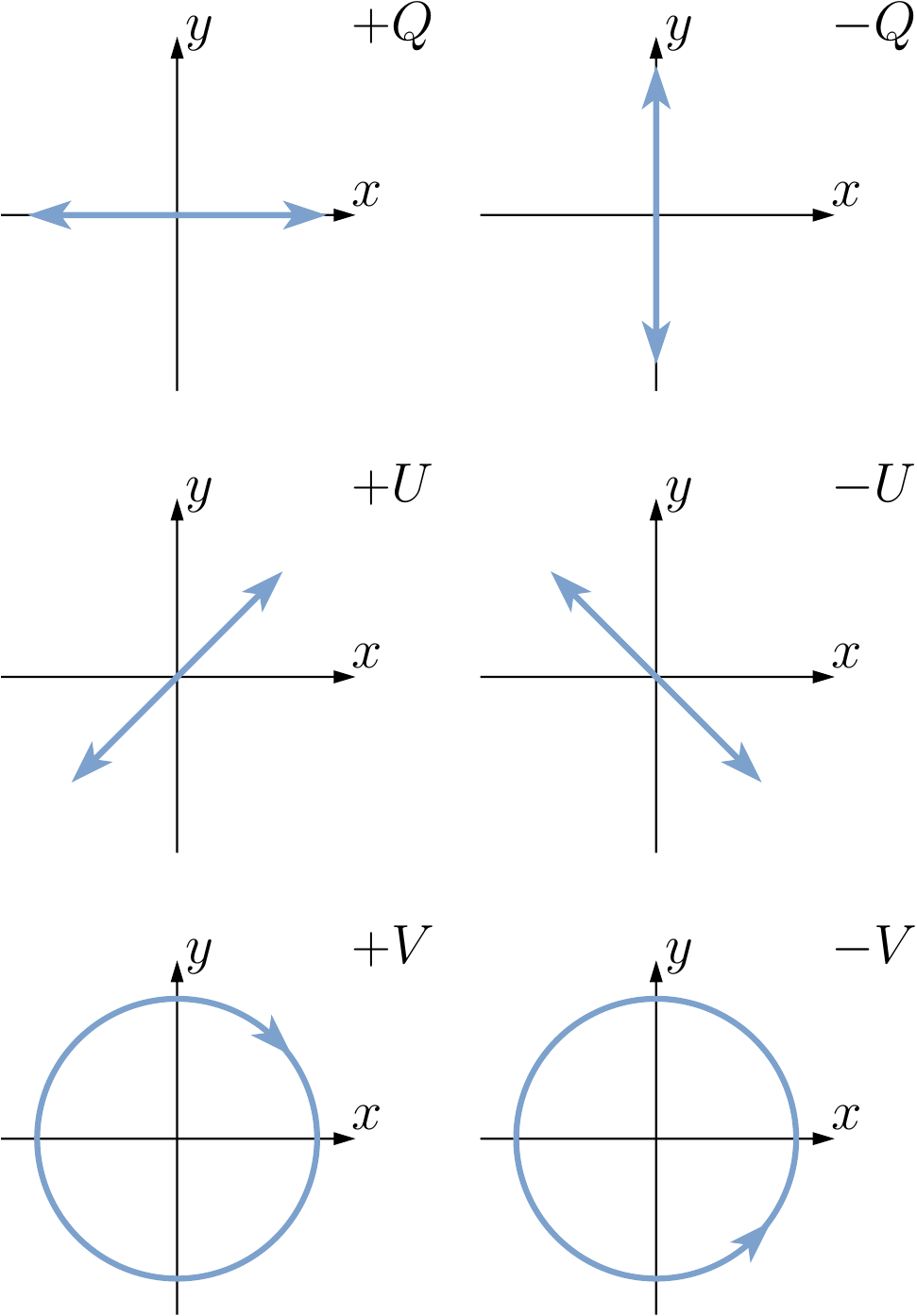}}
\caption{
The polarization ellipses inscribed by the electric field vector for the six special cases of monochromatic light described in \Sec{polarization_ellipse}.
%
}
\label{fig:polarization_cases}
\end{figure}

In this section, the relationship between the measurable Stokes parameters and the inferred geometry of the polarization ellipse described by \eqn{stokes_geometric} is explored using the special case of purely polarized monochromatic light described by \eqn{real_valued_wave}. 
The connection between the \emph{relative} phases\footnote{The Stokes parameters are independent of the absolute phase, $\phi$, that appears in \eqn{stokes_ellipse}. That is, the measured polarization
state does not depend on the choice of time origin.} 
and amplitudes of the components of the electric field vector, the geometry of the ellipse traced by that vector, and the corresponding Stokes parameters
are demonstrated using six special cases of purely polarized radiation: Stokes $\pm$ Q, $\pm$ U, and $\pm$ V. 
These cases are depicted in \Fig{polarization_cases}. \\

%
%

\noindent
{\bf Stokes $\mbf{\pm Q}$}: If $a_y=0$ in \eqn{real_valued_wave}, then the radiation is 100\% linearly polarized with
the electric field vector oscillating along the $x$ axis.
In this case, $\chi=0$, $\psi=0$, Stokes $Q=S_1=S_0$ is positive, and Stokes $S_2=S_3=0$.

If $a_x=0$, then the radiation is 100\% linearly polarized with
electric field vector oscillating along the $y$ axis.
In this case, $\chi=0$, $\psi=\pm\pi/2$, Stokes $Q=S_1=-S_0$ is negative, and Stokes $S_2=S_3=0$.\\

%
%

\noindent
{\bf Stokes $\mbf{\pm U}$}: If $a_x=a_y$ and the relative phase difference, $\Delta\phi = \phi_x - \phi_y = 0$, then $\epsilon_x(t) = \epsilon_y(t)$ 
and the radiation is 100\% linearly polarized with the electric field vector oscillating along 
the line defined by $y=x$ (i.e., offset by $45\deg$ with respect to the $x$ axis).
In this case, $\chi=0$, $\psi=\pi/4$, Stokes $U=S_2=S_0$ is positive, and Stokes $S_1=S_3=0$.

If $a_x=a_y$, and $\Delta\phi = \pm \pi$, then $\epsilon_x(t) = - \epsilon_y(t)$ 
and the radiation is 100\% linearly polarized with the electric field vector oscillating along 
the line defined by $y=-x$  (i.e., offset by $-45\deg$ with respect to the $x$ axis). 
In this case, $\chi=0$, $\psi=-\pi/4$, Stokes $U=S_2=-S_0$ is negative, and Stokes $S_1=S_3=0$. \\

%
%

\noindent 
{\bf Stokes $\mbf{\pm V}$}: If $a_x=a_y$ and $\Delta\phi = -\pi/2$, then the phase of $\epsilon_y$ leads that of $\epsilon_x$ by $90\deg$ and the radiation is 100\% circularly polarized with the electric
field vector tracing a clockwise circle in the x$-$y plane, as seen by an observer looking toward the source;
this is defined as a left-hand circularly polarized (LCP) wave.
In this case, $\chi=\pi/4$, $\psi$ is undefined, Stokes $V=S_3=S_0$ is positive, and Stokes $S_1=S_2=0$.

If $a_x=a_y$ and $\Delta\phi = \pi/2$, then the phase of $\epsilon_x$ leads that of $\epsilon_y$ by $90\deg$ and the radiation is 100\% circularly polarized with the electric
field vector tracing a counter-clockwise circle in the x$-$y plane, as seen by an observer looking toward the source; 
this is defined as a right-hand circularly polarized (RCP) wave.
In this case, $\chi=-\pi/4$, $\psi$ is undefined, Stokes $V=S_3=-S_0$ is negative, and Stokes $S_1=S_2=0$.

%
%

\subsection{Adopted Standard}
\label{sec:adopted_standard}

Apart from the refinements described in \App{refinements}, the conventions and definitions adopted in this article are consistent with both \cite{sto52} and the Institute of Electrical and Electronics Engineers \citep{ieee145}.
In the IEEE standard, the right-hand rule is applied with the thumb pointing in the direction of wave propagation; it defines both the angle measured from a reference axis to the major axis of the polarization ellipse, and the direction in which the electric field rotates for right-hand elliptically polarized states.

The IEEE standard also defines the Poincar\'{e} sphere such that left-hand elliptically polarized states occupy the upper hemisphere, where Stokes V is positive in a right-handed basis defined by Stokes Q, U and V.
As more fully discussed in \citet{hb96} and \citet{vmjr10}, the International Astronomical Union \citep{iau74} defines Stokes $V=-S_3$, which is negative for left-handed circular polarization and opposite to the convention adopted in this article.

%
%

\subsection{Coherency Matrix}
\label{sec:coherency_matrix}

The polarization state of electromagnetic radiation can also be described by the second-order statistics of $\Jcol{e}(t)$, as represented by the complex-valued
$2\times2$ coherency matrix \citep{bw80}
\begin{equation}
  \label{eqn:coherency}
  \cohM{\rho}\equiv\langle\Jcol{e}(t) \otimes \Jcol{e}^\dagger(t) \rangle
=\left( 
    \begin{array}{cc}
    \langle e_0 e_0^* \rangle & \langle e_0 e_1^* \rangle \\
     \langle e_1 e_0^* \rangle & \langle e_1 e_1^* \rangle
    \end{array}
\right).
\end{equation}
Here, the angular brackets denote an average over time, $\otimes$
is the tensor product \eqnp{tensor_product}, and $\Jcol{e}^\dagger(t)$ is the Hermitian transpose
of $\Jcol{e}(t)$ \eqnp{Hermitian_transpose}.  For brevity, explicit dependence on time has been dropped
in the rightmost expression.  Note that each component of $\Jcol{e}(t)$ is multiplied 
by the complex conjugate of either itself or the other component,
a process known as \emph{square law detection}.
Consequently, as for the Stokes parameters,
the components of the coherency matrix are independent of the absolute phase of $\Jcol{e}(t)$.

The coherency matrix is self-adjoint, or Hermitian (i.e., $\cohM{\rho}=\cohM{\rho}^\dagger$), and can be written as a linear combination of four Hermitian basis matrices \citep[e.g.,][]{fan57},
\begin{equation}
  \cohM{\rho} = \frac{1}{2} \sum_{\irow=0}^{3} S_\irow\,\pauli{\irow},    \label{eqn:combination}
\end{equation}
where $S_\irow$ are the four real-valued Stokes parameters
and the basis matrices consist of the $2\times2$ identity matrix 
and the Pauli matrices,
\begin{equation}
\begingroup
\setlength\arraycolsep{5pt}
\begin{array}{ll}
\pauli{0} = \pauliI
&
\pauli{1} = \pauliQ
\\ [5mm]
\pauli{2} = \pauliU
&
\pauli{3} = \pauliV.
\end{array}
\endgroup
\label{eqn:Pauli}
\end{equation}
For brevity in the remainder of this paper, 
the summation symbol is omitted from equations and
Einstein notation is used to imply a sum over repeated indeces.
(See \App{index} for more detail.)

Conversely, the Stokes parameters can be represented as projections of the coherency matrix onto the basis matrices using 
the tensor double contraction \eqnp{double_contraction},
\begin{equation}
S_\irow = \dc{\pauli{\irow}}{\cohM{\rho}}.    \label{eqn:Stokes_contraction}
\end{equation}

\begin{proof}
\begin{align*}
\dc{\pauli{\irow}}{\cohM{\rho}} 
& = \frac{1}{2} S_\jrow \, \dc{\pauli{\irow}}{\pauli{\jrow}}
    & \peqn{combination} \\
&= S_\jrow \delta_{\irow\jrow} & \peqn{basis_orthogonal} \\
&= S_\irow 
\end{align*}
\end{proof}

Substitution of \eqn{coherency} into \eqn{Stokes_contraction} leads to the following expressions for the Stokes parameters.
\begin{eqnarray}
 S_0&=&\langle|e_0(t)|^2\rangle+\langle|e_1(t)|^2\rangle\label{eqn:StokesI} \\
 S_1&=&\langle|e_0(t)|^2\rangle-\langle|e_1(t)|^2\rangle\label{eqn:StokesQ} \\
 S_2&=&2\re{\langle e_0^*(t) e_1(t)\rangle}\label{eqn:StokesU} \\
 S_3&=&2\im{\langle e_0^*(t) e_1(t)\rangle}\label{eqn:StokesV}
\end{eqnarray}

\begin{tcolorbox}
{\bf Example:} Consider $\mu = 3$,
\begin{align*}
& S_3  = \pauli{3} \dcbo \cohM{\rho} & \peqn{Stokes_contraction} \\
& = \pauli{3} \dcbo \langle\Jcol{e} \otimes \Jcol{e}^\dagger \rangle & \peqn{coherency} \\
&= \tr{\pauliV \left( 
    \begin{array}{cc}
    \langle e_0 e_0^* \rangle & \langle e_0 e_1^* \rangle \\
     \langle e_1 e_0^* \rangle & \langle e_1 e_1^* \rangle
    \end{array}
\right)} & \peqn{trace_inner_product} \\
&= -\Ci \langle e_1 e_0^* \rangle + \Ci \langle e_0 e_1^* \rangle \\
    &= 2\, \imag[\langle e_0^*\, e_1 \rangle].
\end{align*}
\end{tcolorbox}

%
%

\subsection{Invariant Interval}
\label{sec:invariant_interval}

As for the Lorentz four-vector in special relativity, the square of the invariant interval of the Stokes parameters is defined by \citep{bar63,bri00}
\begin{equation}
|\Sv{S}|^2 = S_0^2 -|\Pv{S}|^2.
\label{eqn:invariant_interval}
\end{equation}
This interval remains invariant (up to scalar multiples) under linear transformations of the electric field; therefore, $|\Sv{S}|$ is linearly proportional to the scalar amplification of the source.
This property proves useful during numerical analysis and modeling; for example, the invariant can be used to normalize the Stokes parameters and accurately compensate for random scintillation-induced fluctuations of the flux density of the source \citep[as in][]{van04c}.
In high-precision pulsar timing experiments, use of the invariant interval yields arrival time estimates with significantly reduced instrumental artifacts  \citep[e.g.,][]{vbb+01}.

Through its linear relation to the determinant of the coherency matrix via \eqn{determinant_is_invariant},
\begin{equation}
|\cohM{\rho}| = | S_\mu \pauli{\mu} / 2 | 
    = (S_0^2 -|\Pv{S}|^2) / 4 = |\Sv{S}|^2/4,
\label{eqn:invariant}
\end{equation}
the invariant also helps to define the concept of a valid polarization state.  As a direct consequence of the Cauchy-Schwarz inequality for
random variables, a valid polarization state must satisfy
\begin{equation}
|\cohM{\rho}| \ge 0 \quad\text{and}\quad |\Sv{S}| \ge 0.
\label{eqn:valid}
\end{equation}

\begin{proof}
The Cauchy-Schwarz inequality,
\begin{equation}
| \langle e_0 e_1^* \rangle |^2 \le
\langle |e_0|^2 \rangle \langle |e_1|^2 \rangle \nonumber,
\end{equation}
and \eqn{coherency} yield
\begin{equation}
|\cohM{\rho}| = \langle |e_0|^2 \rangle \langle |e_1|^2 \rangle - | \langle e_0 e_1^* \rangle |^2 \ge 0 \nonumber
\end{equation}
\end{proof}
\noindent
Any polarization state that fails to satisfy \eqn{valid} may be called \emph{invalid}, \emph{nonphysical}, and/or \emph{over-polarized}.
Although nonphysical, over-polarized states can arise during numerical analysis.
For example, they can be produced through transformation by an impure Mueller matrix (e.g. \Sec{self_consistency}).
They can also arise when all four Stokes parameters are allowed to vary independently during modeling.
%
%
%

For fully polarized light, $S_0 = |\Pv{S}|$ and
$|\Sv{S}| = |\cohM{\rho}| = 0$.
The invariant interval and determinant are also zero for the \emph{instantaneous} Stokes parameters,
\begin{equation}
\label{eqn:instantaneous_coherency}
  \tilde{s}_\mu \equiv \pauli{\mu}\dcbo\tilde{\cohM{\rho}}
\text{ \quad where \quad} 
    \tilde{\cohM{\rho}} \equiv \Jcol{e} \otimes \Jcol{e}^\dagger,
\end{equation}
because $\tilde{\cohM{\rho}}$ is computed from a single instance of the electric field
and is therefore singular; i.e.,
\begin{equation}
 |\tilde{\cohM{\rho}}| = |\Jcol{e} \otimes \Jcol{e}^\dagger| = e_0 e_0^* \, e_1 e_1^* - e_0 e_1^* \, e_1 e_0^* = 0.
\label{eqn:instantaneous_coherency_determinant}
\end{equation}

%
%

\subsection{Orthogonal Polarizations}
\label{sec:orthogonal_polarization}

%
%
%

\cite{sto52} defined oppositely polarized waves by considering the decomposition of a partially polarized signal into two purely polarized components,
\begin{equation}
 \Jcol{e}_A(t) = \Jcol{e}_a z_a(t) \quad \text{and} \quad \Jcol{e}_B(t) = \Jcol{e}_b z_b(t),
\end{equation}
where $\Jcol{e}_a$ and $\Jcol{e}_b$ are constant Jones vectors that represent the polarizations of the two components,
and $z_i(t) = \Jcol{e}_i^\dagger \cdot \Jcol{e}(t)$ ($i \in \{a,b\}$)
are the projections of $\Jcol{e}(t)$ onto these vectors \eqnsp{Hermitian_transpose}{and}{scalar_product}.
He then considered the superposition of the two components after applying a relative delay  $\tau$ between them,
\begin{equation}
    \Jcol{e}'(t) = \Jcol{e}_A(t) + \Jcol{e}_B(t-\tau),
\end{equation}
and defined $\Jcol{e}_A$ and $\Jcol{e}_B$ as oppositely polarized if and only if the total intensity of $\Jcol{e}'(t)$ is independent of $\tau$.
The total intensity is given by \eqnp[see]{StokesI},
\begin{equation}
    I = \langle \Jcol{e}^\dagger(t) \cdot \Jcol{e}(t) \rangle 
        = \langle |e_0(t)|^2 + |e_1(t)|^2 \rangle;
\end{equation}
therefore, the total intensity of the superposition
\begin{equation} \begin{split}
I' &= \langle \left[\Jcol{e}_A(t) + \Jcol{e}_B(t-\tau)\right]^\dagger \cdot \left[\Jcol{e}_A(t)+ \Jcol{e}_B(t-\tau)\right] \rangle \\
&= \Jcol{e}_a^\dagger \cdot \Jcol{e}_a \langle |z_a(t)|^2 \rangle 
 + \Jcol{e}_b^\dagger \cdot \Jcol{e}_b \langle |z_b(t-\tau)|^2 \rangle \\
& \omit\hfill\ensuremath{ \qquad\qquad\qquad\qquad + 2 \re{\Jcol{e}_a^\dagger \cdot \Jcol{e}_b \, \langle z_a(t) z_b^*(t-\tau) \rangle} }.
\end{split} \end{equation}
Assuming that the signal is stationary, at least in the weak sense, $\langle |z_b(t-\tau)|^2 \rangle = \langle |z_b(t)|^2 \rangle$; therefore, $I'$ is independent of $\tau$, and the two components are orthogonal, if and only if $\Jcol{e}_a^\dagger \cdot \Jcol{e}_b = 0$.

To characterize the Stokes parameters of orthogonally polarized states,
consider the singular coherency matrices of the two purely polarized states,
$$\tilde{\cohM{\rho}}_a \equiv \Jcol{e}_a \otimes \Jcol{e}_a^\dagger \quad \text{and} \quad
\tilde{\cohM{\rho}}_b \equiv \Jcol{e}_b \otimes \Jcol{e}_b^\dagger,$$
and their tensor double contraction,
\begin{equation}
\label{eqn:coherency_projection_field}
\dc{\tilde{\cohM{\rho}}_a}{\tilde{\cohM{\rho}}_b} = | \Jcol{e}_a^\dagger \cdot \Jcol{e}_b |^2.
\end{equation}
\begin{proof}
\begin{align*}
{\tilde{\cohM{\rho}}_a}&:{\tilde{\cohM{\rho}}_b}
= \tr{\tilde{\cohM{\rho}}_a \, \tilde{\cohM{\rho}}_b}
	& \peqn{trace_inner_product} \\
&= \tr{ {\color{blue}\Jcol{e}_a} \Jcol{e}_a^\dagger \, \Jcol{e}_b \Jcol{e}_b^\dagger }
	& \peqn{implicit_tensor_product} \\
&= \tr{ \Jcol{e}_a^\dagger \Jcol{e}_b \Jcol{e}_b^\dagger \, {\color{blue}\Jcol{e}_a} }
	& \peqn{trace_inner_product} \\
&= \tr{ (\Jcol{e}_a^\dagger \Jcol{e}_b)  (\Jcol{e}_a^\dagger \Jcol{e}_b)^* } \\
&= | \Jcol{e}_a^\dagger \Jcol{e}_b |^2
	& \tr{z} = z
\end{align*}
\end{proof}
\noindent
Therefore, the coherency matrices of two purely polarized sources are orthogonal
with respect to the trace inner product \eqnp{trace_inner_product} when the polarizations are orthogonal.
Given the associated Stokes parameters, 
$$\tilde{s}_{a,\mu} = \pauli{\mu} \dcbo \tilde{\cohM{\rho}}_a
\quad \text{and} \quad
\tilde{s}_{b,\mu} = \pauli{\mu} \dcbo \tilde{\cohM{\rho}}_b,$$
the tensor double contraction,
\begin{equation}
\label{eqn:coherency_projection_Stokes}
\dc{\tilde{\cohM{\rho}}_a}{\tilde{\cohM{\rho}}_b} = \frac{1}{2} \left( \tilde{s}_{a,0} \tilde{s}_{b,0} + \tilde{\Pv{s}}_a \cdot \tilde{\Pv{s}}_b \right).
\end{equation}
\begin{proof}
\begin{align*}
{\tilde{\cohM{\rho}}_a}&\dcbo{\tilde{\cohM{\rho}}_b} 
 =  ( \tilde{s}_{a,\mu} \, \pauli{\mu} / 2 ) \dcbo (\tilde{s}_{b,\nu} \, \pauli{\nu} / 2)  \\
&= \frac{1}{4}  \tilde{s}_{a,\mu} \, \tilde{s}_{b,\nu} \, \pauli{\mu}  \dcbo \pauli{\nu}  \\
&= \frac{1}{2} \tilde{s}_{a,\mu} \, \tilde{s}_{b,\nu} \, \delta_{\mu\nu}
   & \peqn{basis_orthogonal}  \\
&= \frac{1}{2} \tilde{s}_{a,\mu} \, \tilde{s}_{b,\mu}
\end{align*}
\end{proof}
\noindent
Equating the right-hand sides of \eqns{coherency_projection_field}{and}{coherency_projection_Stokes},
noting that $\tilde{s}_0 = |\tilde{\Pv{s}}|$, and rearranging yields
\begin{equation}
 \tilde{\Pv{s}}_a \cdot \tilde{\Pv{s}}_b = 2 | \Jcol{e}_a^\dagger \cdot \Jcol{e}_b |^2 - |\tilde{\Pv{s}}_a| |\tilde{\Pv{s}}_b|.
\end{equation}
The angle between $\tilde{\Pv{s}}_a$ and $\tilde{\Pv{s}}_b$ is given by
\begin{equation}
\cos\Theta = \frac{\tilde{\Pv{s}}_a \cdot \tilde{\Pv{s}}_b}{|\tilde{\Pv{s}}_a| |\tilde{\Pv{s}}_b|} = \frac{2 | \Jcol{e}_a^\dagger \cdot \Jcol{e}_b |^2}{|\tilde{\Pv{s}}_a| |\tilde{\Pv{s}}_b|} - 1.
\end{equation}
If $\Jcol{e}_a^\dagger \cdot \Jcol{e}_b = 0$, then $\cos\Theta=-1$ and the angle between $\tilde{\Pv{s}}_a$ and $\tilde{\Pv{s}}_b$ is $180\deg$.  That is, the polarization vectors of orthogonally polarized signals are anti-parallel.

This result is consistent with \citet{sto52}, where it is demonstrated that oppositely polarized waves
trace ellipses with orthogonal major axes ($\left| \psi_A - \psi_B \right| = \pi/2$) and equal and opposite handedness ($\chi_A = - \chi_B$).
That is, on the Poincar\'{e} sphere, oppositely polarized waves occupy antipodal points.



\section{Linear Transformations}
\label{sec:linear_transformations}

Polarimetric studies require modeling and correcting the manner in which the measurement apparatus alter the incident radiation.
In radio astronomy, the instrument includes everything from the antenna (including any reflectors involved in focusing and redirecting the radio waves) to the signal processing system used to compute the Stokes parameters.
It is also necessary to model and correct Faraday rotation, which occurs in the Earth's ionosphere and the interstellar medium.
In this section, the response of a system is described using linear transformations of the electric field and the equivalent linear transformations of the Stokes parameters.


\subsection{Jones Matrices}

In the narrow-band (or quasi-monochromatic) approximation of an electromagnetic wave (see \App{narrow_band_approximation}), the response of a single receptor 
is defined by a complex-valued row vector, $\Jrow{r} = (r_0, r_1)$, such that the voltage induced in the receptor by the incident electric field $\Jcol{e}(t)$ is given by the scalar product, $v(t)=\Jrow{r} \cdot \Jcol{e}(t)$.
The intensity of the response, $I=\langle v^2(t) \rangle$, is maximum when the polarization of the incident wave matches that of the receptor, and reduces to zero when the incident wave is orthogonally polarized.
%

The response of a dual-receptor feed 
is represented by a $2\times2$ complex-valued
Jones matrix with rows equal to the receptor vectors,
\begin{equation}
\label{eqn:feed}
\JM{J}=\left( \begin{array}{c}
	\Jrow{r}_0 \\
    \Jrow{r}_1
\end{array} \right)
=
  \left( \begin{array}{cc}
    r_{00} & r_{01} \\
    r_{10} & r_{11}
  \end{array} \right),
\end{equation}
such that any linear transformation of the electric field vector may be represented by
\begin{equation}
\Jcol{e}'(t)=\JM{J}\Jcol{e}(t).
\label{eqn:field_transformation}
\end{equation}
The receptors in an ideal feed are orthogonal, such the scalar product $\Jrow{r}_0 \cdot \Jrow{r}_1^\dagger=0$.

The response of a system that is composed of a series of elements is represented as a product of Jones matrices.
Matrix multiplication is not commutative, and in general the order in which operations are performed
must correctly reflect the order in which the elements in the signal path are encountered.
For example, consider the system response described by equation~(11) of \citet{hbs96},
\begin{equation}
\Jcol{e}'(t) = \JM{J} \Jcol{e}(t) = {\bf G\, D\, C\, P\, F}\, \Jcol{e}(t).
\label{eqn:chain}
\end{equation}
From the incident electric field $\Jcol{e}(t)$ on the right to the 
observed $\Jcol{e}'(t)$ on the left, astrophysical signals are subjected to

\begin{itemize}
\item Faraday rotation in both the interstellar medium and the ionosphere, \JM{F};
\item the projection between the celestial reference frame and the receptor basis, \JM{P};
\item the nominal antenna and feed configuration, \JM{C};
\item deviations from an ideal feed, \JM{D}; and
\item complex receiver gains, \JM{G}.
\end{itemize}

\begin{figure*}
\centerline{\includegraphics[width=0.6\linewidth]{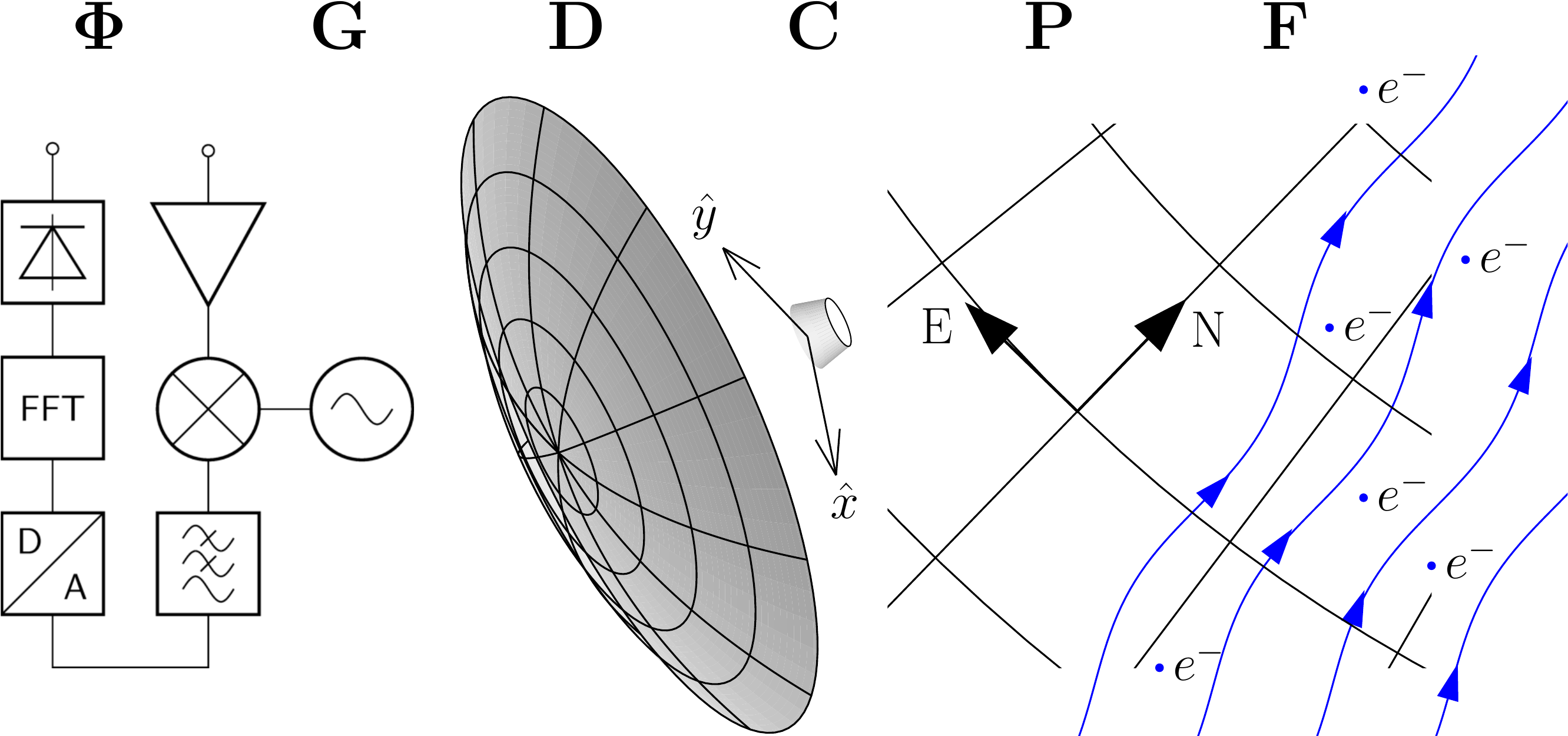}}
\caption{ The radio wave signal path, from free space propagation on the right to detection on the left.  Also from right to left are
Faraday rotation \JM{F} due to free electrons permeated by a magnetic field (represented by blue lines with arrows); the
 coordinate projection \JM{P} between the $x$ and $y$ axes of the receiver and the North and East basis vectors of the celestial reference frame;
 the nominal configuration \JM{C} of an ideal antenna and feed;
unintended deviations \JM{D} from that ideal;
and the gains \JM{G} and phase convention $\MM{\Phi}$ of the down-conversion system and instrumentation used to detect the Stokes parameters.}
\label{fig:signal_path}
\end{figure*}

These transformations are depicted in \Fig{signal_path} and defined and discussed in more detail in \Sec{signal_path}.
Note that \Fig{signal_path} includes an additional phase convention correction $\MM{\Phi}$ that cannot be represented using a Jones matrix and therefore does not appear in \eqn{chain}.  The phase convention correction can be represented by a Mueller matrix and is included in \eqn{Stokes_chain} of \Sec{Mueller}.

\subsection{Congruence Transformations}

Substitution of
$\Jcol{e}'=\JM{J}\Jcol{e}$ into \eqn{coherency} yields the coherency matrix of the transformed electric field, which is described by a congruence transformation,
\begin{equation}
{\cohM{\rho}'}=\JM{J}\cohM{\rho}\JM{J}^\dagger.
\label{eqn:congruence}
\end{equation}

\begin{proof}
\begin{align*}
\cohM{\rho}'
& = \langle\Jcol{e}' \, \otimes \, \Jcol{e}^{\prime\dagger}\rangle & \peqn{coherency} \\
& = \langle\left(\JM{J}\Jcol{e}\right) \otimes \left(\JM{J}\Jcol{e} \right)^{\dagger}\rangle \nonumber &\peqn{field_transformation}\\
& = \langle\JM{J}\Jcol{e} \, \otimes \, \Jcol{e}^{\dagger}\JM{J}^\dagger\rangle & \left(\JM{A}\JM{B} \right)^{\dagger} = \JM{B}^{\dagger}\JM{A}^{\dagger} \\
& = \JM{J} \langle\Jcol{e} \, \otimes \, \Jcol{e}^{\dagger}\rangle \JM{J}^\dagger & \text{\JM{J} is constant} \\
& = \JM{J}\cohM{\rho} \, \JM{J}^\dagger\nonumber & \peqn{coherency}
\end{align*}

\end{proof}

As discussed in more detail in \Sec{rime}, the congruence transformation is a special case of the more general radio interferometer measurement equation for a single point source \citep{ham00,smi11a}.
For a single antenna, the coherency matrix is insensitive to the absolute phase of $\JM{J}$.

\begin{proof}
\label{proof:coherency_matrix_is_detected}
Consider $\JM{J}' = \JM{J}e^{i\phi}$.
\begin{align*}
\cohM{\rho}' 
&= \JM{J}' \cohM{\rho}\JM{J}^{\prime\dagger} 
	& \peqn{congruence} \\
&= \left(\JM{J}e^{i\phi}\right) \cohM{\rho} \left(\JM{J}e^{i\phi} \right)^{\dagger}
	& \\
&= e^{i\phi} \JM{J}\cohM{\rho}\JM{J}^\dagger e^{-i\phi} 
= \JM{J}\cohM{\rho}\JM{J}^\dagger
\end{align*}
\end{proof}

Noting that $(\JM{A}\JM{B})^\dagger = \JM{B}^\dagger \JM{A}^\dagger$,
congruence transformation of the coherency matrix by the 
Jones matrix defined in \eqn{chain} has the following ``onion'' form.
\begin{equation}
{\cohM{\rho}'}=
\JM{G} \left(
\JM{D} \left(
\JM{C} \left(
\JM{P} \left(
\JM{F}
\cohM{\rho} \,
\JM{F}^\dagger \right)
\JM{P}^\dagger \right)
\JM{C}^\dagger \right)
\JM{D}^\dagger \right)
\JM{G}^\dagger
\label{eqn:onion}
\end{equation}
(The matrices in this equation are defined and discussed in more detail in \Sec{signal_path}.)

Polarimetric calibration consists of transforming the observed coherency matrix by the inverse of the system response, 
$\cohM{\rho}={\JM{J}^{-1}}\cohM{\rho}' \JM{J}^{-1\dagger}$, yielding the intrinsic polarization,
\begin{equation} 
\cohM{\rho} =
\cdots
\JM{C}^{-1} \left(
\JM{D}^{-1} \left(
\JM{G}^{-1}
\cohM{\rho} \,
\JM{G}^{-1\dagger} \right)
\JM{D}^{-1\dagger} \right)
\JM{C}^{-1\dagger}
\cdots
\label{eqn:inverse_onion}
\end{equation}
Because $(\JM{A}\JM{B})^{-1} = \JM{B}^{-1} \JM{A}^{-1}$, the onion is turned inside out and operations are inverted in reverse order, beginning with the last element of the signal path in the innermost congruence transformation.

A congruence transformation converts any valid 
polarization state to another valid state.
\begin{proof}
\label{proof:only_valid}
If $|\cohM{\rho}| \ge 0$ \eqnp{valid}, then
\begin{align*}
| \cohM{\rho}' |
&= | \JM{J} \, \cohM{\rho} \, \JM{J}^\dagger |
	& \peqn{congruence} \\
&= | \JM{J} | |\cohM{\rho}| | \JM{J}^\dagger |
	& |\JM{A}\JM{B}| = |\JM{A}||\JM{B}| \\
&= JJ^* |\cohM{\rho}| & J = | \JM{J} | \\
& \ge 0    & |\cohM{\rho}| \ge 0 \text{ and } zz^* \ge 0
\end{align*}
\end{proof}
\noindent

No linear transformation of the electric field can alter the degree of polarization of a purely polarized state.  
\begin{proof}
\label{proof:no_depolarization}
If $\cohM{\rho}$ is purely polarized, then
\begin{align*}
| \cohM{\rho}' |
&= | \JM{J} \, \cohM{\rho} \, \JM{J}^\dagger |
	& \peqn{congruence} \\
&= | \JM{J} | |\cohM{\rho}| | \JM{J}^\dagger |
	& |\JM{A}\JM{B}| = |\JM{A}||\JM{B}| \\
&= 0
	& |{\cohM{\rho}}| = 0
\end{align*}
\end{proof}
\noindent
Conversely, only a singular transformation ($|\JM{J}| = 0$) transforms partial polarization into pure polarization.
%


\subsection{Mueller Matrices}
\label{sec:Mueller}

Using \eqns{combination}{and}{Stokes_contraction}, a congruence
transformation of the coherency matrix can be expressed as an
equivalent linear transformation of the associated Stokes parameters
by a real-valued $4\times4$ Mueller matrix \MM{M}, as defined by
\begin{equation}
 S_\irow' = M_\irow^\icol S_\icol
\label{eqn:linear_to_Mueller}
\end{equation}
where
\begin{equation}
  M_\irow^\icol = \frac{1}{2} \dc{\pauli{\irow}}{\left(\JM{J}\pauli{\icol}\,\JM{J}^\dagger\right)}.
\label{eqn:Mueller_projection}
\end{equation}

\begin{proof}

\begin{align*}
S_\irow' 
&= \dc{\pauli{\irow}}{\cohM{\rho}'} & \peqn{Stokes_contraction} \\
&= \dc{\pauli{\irow}}{\left(\JM{J}\cohM{\rho}\, \JM{J}^\dagger\right)}   & \peqn{congruence} \\
&= \dc{\pauli{\irow}}{\left(\JM{J}\left[S_\icol\,\pauli{\icol}/2\right]\JM{J}^\dagger\right)} & \peqn{combination} \\
&= \frac{1}{2}\dc{\pauli{\irow}}{\left(\JM{J}\pauli{\icol}\, \JM{J}^\dagger\right)}S_\icol & \text{$S_\nu$ is a scalar} \\
&= M_\irow^\icol S_\icol & \peqn{Mueller_projection}
\end{align*}
\end{proof}


Although there is a unique Mueller matrix for every Jones matrix, the converse is not true.
This is most directly understood by noting that a $4\times 4$ real-valued matrix has 16 degrees of freedom (dof), and a $2\times 2$ complex-valued matrix has only 8 dof.
Mueller matrices that do not have an equivalent
Jones matrix are known as ``impure'' or ``depolarizing'' \citep[e.g.,][and \App{impure}]{hbs96,lc96}.
The definition of a pure Mueller matrix is derived in \App{pure_Mueller_matrices}.

The signal path described by \cite{vmjr10} extends the \citet{hbs96} model with an impure correction for the complex phase convention, which is described
in more detail in \Sec{conjugation}.
Therefore, the \cite{vmjr10} calibration model is best summarized
using the equivalent transformation of the Stokes parameters
by a Mueller matrix $\MM{M}$,
\begin{equation}
\Sv{S}' = \MM{M} \, \Sv{S} = \MM{\Phi} \, \MM{G} \, \MM{D} \, \MM{C} \, \MM{P} \, \MM{F}\, \Sv{S}.
\label{eqn:Stokes_chain}
\end{equation}
From right to left, $\Sv{S}$ is the column vector with elements equal to the 4 Stokes parameters intrinsic to the astronomical source;
$\MM{F}$ through $\MM{G}$ are the Mueller matrices derived from the Jones matrices defined in \eqn{chain} and discussed in \Sec{signal_path};
$\MM{\Phi}$ is the phase convention correction
(see \Sec{conjugation});
and $\Sv{S}'$ is the column vector of observed Stokes parameters.
%


\subsection{Polarization Transfer Tensors}
\label{sec:polarization_transfer_tensor}


Linear transformations of polarization state can also be represented 
as a double contraction \eqnp{double_contraction} with a two-dimensional, rank 4 polarization transfer tensor, $\JT{U}$, such that
\begin{equation}
\label{eqn:polarization_transfer_tensor}
\cohM{\rho}' = \JT{U} \dcbo \cohM{\rho}.
\end{equation}
Similar tensors are used in the generalized radio interferometer equation \citep{smi11d} and in descriptions of the propagation of radiation through a magnetized plasma \citep[e.g.,][]{kaw64,zhe68}.  The tensor formalism also simplifies the analysis of the covariances between the Stokes parameters \citep{vt17} and the definition of pure Mueller matrices (\App{pure_Mueller_matrices}).

To relate a congruence transformation of the coherency matrix to the equivalent double contraction with
a polarization transfer tensor, the $\tilde\otimes$ operator is introduced to represent 
a tensor product followed by a transpose over covariant\footnote{ 
Covariant tensor indeces are raised in this work; note that \cite{vt17} incorrectly identified this as a transpose over contravariant tensor indeces.} 
tensor indeces \citep{car91}.
That is, where {\bf A} and {\bf B} are matrices (rank 2 tensors) and
\begin{equation}
\left\{\outerBilinear{\bf A}{\bf B}\right\}_{ik}^{jl}
 \equiv 
A_i^j B_k^l  \label{eqn:indexOuterBilinear}
\end{equation}
is their tensor product,
\begin{equation}
\left\{\spinorBilinear{\bf A}{\bf B}\right\}_{ik}^{jl}
 \equiv 
A_i^l B_k^j. \label{eqn:indexSpinorBilinear}
\end{equation}
Upon double contraction with a matrix {\bf C}, the rank 4 tensors defined by \eqns{indexOuterBilinear}{and}{indexSpinorBilinear} exhibit the following transformation properties (\Proofs{outerBilinear}{and}{spinorBilinear}).
\begin{eqnarray}
\left( \outerBilinear{\bf A}{\bf B} \right) \dcbo {\bf C}
& = &
{\bf A}\left(\dc{\bf B}{\bf C}\right) \label{eqn:outerBilinear} \\
\left( \spinorBilinear{\bf A}{\bf B} \right) \dcbo {\bf C}
& = &
{\bf A}{\bf C}{\bf B} \label{eqn:spinorBilinear}
\end{eqnarray}
Using \eqns{polarization_transfer_tensor}{and}{spinorBilinear}, a congruence transformation by a Jones matrix \JM{J} can be expressed as a double contraction with
\begin{equation}
\label{eqn:Jones_tensor}
    \JT{U}= \spinorBilinear{\bf J}{\bf J}^\dagger.
\end{equation}

\begin{proof}
\begin{align*}
\cohM{\rho}' &= \JT{U} \dcbo \cohM{\rho} & \peqn{polarization_transfer_tensor} \\
&= \left( \spinorBilinear{\bf J}{\bf J}^\dagger \right)\dcbo \cohM{\rho} & \peqn{Jones_tensor} \\
&= \JM{J} \, \cohM{\rho} \, \JM{J}^\dagger. & \peqn{spinorBilinear}
\end{align*}
\end{proof}

The polarization transfer tensor can also replace the congruence transformation in \eqn{Mueller_projection}, thereby yielding the
equivalent Mueller matrix $\MM{M}$,
\begin{equation}
M_\irow^\icol
= \frac{1}{2} \dc{\pauli{\irow}}{\dc{\JT{U}}{\pauli{\icol}}}.
\label{eqn:tensor_to_Mueller}
\end{equation}
The inverse of this mapping is given by
\begin{equation}
\JT{U} = \frac{1}{2} M_\irow^\icol \outerBilinear{\pauli{\irow}}{\pauli{\icol}}.
\label{eqn:Mueller_to_tensor}
\end{equation}

\begin{proof}
\begin{align*}
\frac{1}{2} & \dc{\pauli{\irow}}{\dc{\JT{U}}{\pauli{\icol}}} & \peqn{tensor_to_Mueller} \\
& = \frac{1}{2} \dc{\pauli{\irow}}{\dc{ \left( \frac{1}{2} M_\jrow^\jcol \outerBilinear{\pauli{\jrow}}{\pauli{\jcol}} \right)  }{\pauli{\icol}}} & \peqn{Mueller_to_tensor} \\
& = \frac{1}{4} M_\jrow^\jcol \dc{\pauli{\irow}}{\dc{\left( \outerBilinear{\pauli{\jrow}}{\pauli{\jcol}} \right)}{\pauli{\icol}}} \\
& = \frac{1}{4} M_\jrow^\jcol \left(\dc{\pauli{\irow}}{\pauli{\jrow}}\right)\left(\dc{\pauli{\jcol}}{\pauli{\icol}} \right) & \peqn{outerBilinear} \\
& = M_\jrow^\jcol \delta_{\irow\jrow} \delta_{\icol\jcol} & \peqn{basis_orthogonal}  \\
& = M_\irow^\icol
\end{align*}
\end{proof}

\Eqn{tensor_to_Mueller} expresses the components of $\MM{M}$
as the double projections of \JT{U} onto the Hermitian
basis matrices.
\Eqn{Mueller_to_tensor} represents \JT{U} as a linear combination of
the 16 basis tensors formed by tensor products of the 4
Hermitian basis matrices.

%


\subsection{Comparison with Radio Interferometry}
\label{sec:rime}

As in \cite{smi11a}, the radio interferometer measurement equation for a single point source models the
visibility matrices formed by the outer products of electric field vectors from pairs of elements in an array.  That is, if $\Jcol{e}_A(t)$ and $\Jcol{e}_B(t)$ are the signals from two array elements, the visibility matrix for this pair
\begin{equation}
  \label{eqn:visibility}
  \cohM{V}_{AB} \equiv\langle\Jcol{e}_A(t) \otimes \Jcol{e}_B^\dagger(t) \rangle.
\end{equation}
Furthermore, if $\JM{J}_A$ and $\JM{J}_B$ are the Jones matrices that describe the reception of the incident electric field $\Jcol{e}(t)$ by the two array elements, such that $\Jcol{e}_A(t) = \JM{J}_A \Jcol{e}(t)$ and $\Jcol{e}_B(t) = \JM{J}_B \Jcol{e}(t)$,
then the observed visibility matrix,
\begin{equation}
\cohM{V}_{AB} =
\JM{J}_A \, \cohM{\rho} \, \JM{J}_B^\dagger.
\label{eqn:brightness}
\end{equation}
In the context of interferometry, the coherency matrix $\cohM{\rho}$ is also known as the brightness matrix \citep{smi11a}.
\Eqn{brightness} reduces to \eqn{congruence} when observing a single point source with a single antenna.
In the tensor formalism of the generalized radio interferometer measurement equation \citep{smi11d}, \eqn{brightness} is expressed as
\begin{equation}
\cohM{V}_{AB} =
\dc{\JT{U}_{AB}}{\cohM{\rho}},
\label{eqn:tensor_rime}
\end{equation}
where $\JT{U}_{AB}=\spinorBilinear{\JM{J}_A}{\JM{J}_B^\dagger}$
is the visibility transfer tensor\footnote{A transpose of covariant tensor indeces also appears in equation~(10)
of \cite{smi11d}.} \eqnp[cf.]{Jones_tensor}.

An extended source is described by a brightness matrix that varies with direction, $\cohM{\rho}(\hat{\Pv{z}})$, such that the visibility matrix is given
by the integral
\begin{equation}
\cohM{V}_{AB} = \iint
\JM{J}_A(\hat{\Pv{z}}) \, \cohM{\rho}(\hat{\Pv{z}}) \, \JM{J}_B^\dagger(\hat{\Pv{z}}) \, d\Omega,
\label{eqn:visibility_integral}
\end{equation}
where $\JM{J}_{A,B}(\hat{\Pv{z}})$ are the direction-dependent responses of each
antenna.  For a single-antenna observation of an extended source,
\begin{equation}
\cohM{\rho} = \iint
\JM{J}(\hat{\Pv{z}}) \, \cohM{\rho}(\hat{\Pv{z}}) \, \JM{J}^\dagger(\hat{\Pv{z}}) \, d\Omega.
\label{eqn:coherency_integral}
\end{equation}

There are some key differences between single-antenna and interferometric measurements of polarization.  For a single antenna, the coherency matrix is Hermitian and independent of the absolute phase of $\Jcol{e}(t)$.  For an interferometer, the visibility matrix is not Hermitian and depends on the relative phase between $\Jcol{e}_A(t)$ and $\Jcol{e}_B(t)$.


\subsection{Polar Decomposition}
\label{sec:polar_decomposition}



%
Any invertible Jones matrix $\JM{J}$ can be decomposed into a unique product known as its left polar decomposition,
\begin{equation}
\label{eqn:polar}
\JM{J} = J \, \JM{B} \JM{R},
\end{equation}
where $J=|\JM{J}|^{1/2}$ and $|\JM{J}|$ is the determinant of \JM{J};
\JM{B} is Hermitian 
($\JM{B}^\dagger=\JM{B}$);
\JM{R} is unitary
($\JM{R}^\dagger = \JM{R}^{-1}$);
and both \JM{B} and \JM{R} are unimodular ($|\JM{B}| = |\JM{R}| = 1$).
The requirement for a left polar decomposition\footnote{
A right polar decomposition,
$\JM{J}' = J' \, \JM{R}' \, \JM{B}'$, would be used if receptors were column vectors, such that
$\Jcol{e}'(t)=\JM{J}^{\prime\dagger}\Jcol{e}(t)$ 
and
${\cohM{\rho}'}=\JM{J}^{\prime\dagger}\cohM{\rho}\, \JM{J}'$.  Note that $\JM{J}'=\JM{J}^\dagger = J^* \, \JM{R}^\dagger \JM{B}^\dagger$; therefore,
$J'=J^*$, $\JM{R}'=\JM{R}^{-1}$, and $\JM{B}'=\JM{B}$.
} stems from the definition of a Jones matrix as a pair of receptor row vectors \eqnp{feed}.

Polar decomposition into Hermitian and unitary matrices provides a practical framework for classifying and conceptualizing distinct polarization-altering effects, as discussed in more detail in \Sec{axis-angle}.
Isolating the determinant of the Jones matrix also has practical benefit.
In single-antenna polarimetry, the coherency matrix is insensitive to the phase of $J$ (\Proof{coherency_matrix_is_detected}); therefore, 
without any loss of generality, 
the absolute phase is set to zero
and $J$ is replaced by the real-valued absolute gain, $G=|J|$, thereby eliminating a degenerate dof.

The polar decomposition
also enables unique determination of the Hermitian component of the instrumental response given only an observation of a source that is known to be completely unpolarized.
To demonstrate this, first note that the Hermitian component of any Jones matrix \JM{J}
can be determined via the Gram matrix of its rows,  
$\JM{J}\JM{J}^\dagger$, such that\footnote{
If the Jones matrix is defined as a pair of columnn vectors, then the
Gram matrix of the columns of $\JM{J}'$ would be used to determine its Hermitian component. }
\begin{equation}
\label{eqn:boost_from_polar}
\JM{B}^2 = G^{-2} \JM{J}\JM{J}^\dagger.
\end{equation}
\begin{proof}
\label{proof:self_adjoint_square}
The Gram matrix of the rows of \JM{J},
\begin{align*}
\JM{J}\JM{J}^\dagger
& = J \, \JM{B} \, \JM{R} \left( J \, \JM{B} \, \JM{R} \right)^\dagger \\
& = J \, \JM{B} \, \JM{R} \, \JM{R}^\dagger \JM{B}^\dagger J^* \\
& = |J|^2 \JM{B} \, \JM{R} \, \JM{R}^{-1} \JM{B} \\
& = G^2 \, \JM{B}^2
\end{align*}
\noindent
\end{proof}
\noindent
For an unpolarized source $\cohM{\rho} = I\pauli{0}$; therefore,
\begin{equation}
\cohM{\rho}' = \JM{J} \cohM{\rho} \, \JM{J}^\dagger = I \JM{J}\JM{J}^\dagger = I G^2 \JM{B}^2.
\end{equation}
The unknown factor of $T=IG^2$ can be eliminated by noting that $|\JM{B}|=1$;
therefore, $|\cohM{\rho}'| = T^2$
and $\JM{B}= ({\cohM{\rho}}'/T)^{1/2}.$


\subsection{Axis-Angle Representation}
\label{sec:axis-angle}


As shown in \App{fundamental_transformations}, \JM{B} and
\JM{R} can be expressed using axis-angle representation,
\begin{align}
\label{eqn:Boost}
\boost \equiv e^{\beta \hat{\Pvs{m}} \cdot \Pvs{\sigma}} &= \pauli{0}\cosh\beta + \hat{\Pv{m}} \cdot \Pv{\sigma} \sinh\beta \\
\label{eqn:Rotation}
\rotat \equiv e^{ \Ci \phi \hat{\Pvs{n}} \cdot \Pvs{\sigma} } &= \pauli{0}\cos\phi + \Ci \hat{\Pv{n}} \cdot \Pv{\sigma} \sin\phi,
\end{align}
where $\hat{\Pv{m}}$ and $\hat{\Pv{n}}$ are three-dimensional unit vectors, $\beta$ and $\phi$ are real-valued scalars,
and
$\Pv{\sigma}=\left( \pauli{1}, \pauli{2}, \pauli{3} \right)$ is
a 3-vector whose components are the Pauli spin matrices.
Each axis-angle representation has 3 dof
%
%
in the vector 
that defines it.
Combined with the 2 dof in the real and imaginary parts of the complex-valued scalar $J$, \eqn{polar} has 8 dof,
as expected for a $2\times2$ matrix with independent real and imaginary parts.

The axes and angles in \eqns{Boost}{and}{Rotation} have geometric interpretations in the four-dimensional space of the Stokes parameters.
After a congruence transformation by a Hermitian matrix \JM{B}, the associated Stokes four-vector is transformed by a 
Lorentz boost along the $\hat{\Pv{m}}$ axis by a hyperbolic angle $-2\beta$ (e.g., \App{example_boost}).
A Lorentz transformation of the Stokes four-vector mixes $S_0$ with the polarization vector $\Pv{S}$, thereby altering both the intensity and the degree of polarization.
Consequently, energy is not conserved.
Hermititan matrices are equivalent to the \emph{polconversion} defined by \cite{ham00}; they describe the diattenuation of a system \citep{lc96} due to differential gain and non-orthogonality of the receptors.

In contrast, congruence transformation by a unitary matrix
rotates the Stokes polarization vector $\Pv{S}$ about the $\hat{\Pv{n}}$ axis by
an angle $-2\phi$, using the right-hand rule for rotation (e.g., \App{example_rotation}).
This rotation in three-dimensional space leaves the total intensity and the degree of polarization unchanged. 
Unitary matrices are equivalent to the \emph{polrotation} defined by \cite{ham00}; they describe the retardance of a system \citep{lc96} due to differential phase and they represent any change of basis by projection onto a pair of orthonormal receptors (see \Proof{orthonormal_receptors}).

The axis-angle representations of \boost\ and \rotat\ exhibit a number of properties that prove useful during analysis.
Both transformations are unimodular.
\begin{proof}
\label{proof:unimodular_boost_and_rotation}
%
\begin{align*}
|\boost| 
&= |\pauli{0}\cosh\beta + {\hat{\Pv{m}}\cdot\Pv{\sigma}}\sinh\beta| & \peqn{Boost} \\
&= \cosh^2\beta - |\hat{\Pv{m}}|^2 \sinh^2\beta & \peqn{determinant_is_invariant} \\
	&= 1
\end{align*}
Similarly,
$|\rotat| 
= \cos^2\phi - |\Ci\hat{\Pv{n}}|^2 \sin^2\phi
	= 1$
\end{proof}
\noindent
Therefore, congruence transformation by \boost\ or \rotat\ preserves the determinant of the coherency matrix and the invariant interval of the Stokes parameters. 
\begin{proof}
\label{proof:invariant}
Let $|\JM{U}|=1$ and
\begin{align*}
|{\cohM{\rho}'}| &=|\JM{U}\cohM{\rho}\JM{U}^\dagger| & \peqn{congruence} \\
&=|\JM{U}||\cohM{\rho}||\JM{U}^\dagger| & |\bf{AB}| = |\bf{A}||\bf{B}| \\
&=|\cohM{\rho}| & |\JM{U}| = |\JM{U}|^\dagger = 1
\end{align*}
\end{proof}

Exponentiation of \boost\ or \rotat\ is equivalent 
to multiplying the angle by the power,
\begin{eqnarray}
\label{eqn:boost_power}
\vBoost^y(\beta) = \left( e^{\beta \hat{\Pvs{m}} \cdot \Pvs{\sigma}} \right)^y
= e^{y \beta \hat{\Pvs{m}} \cdot \Pvs{\sigma}} = \vBoost(y\beta) \\
\vRotation^y(\phi) = \left( e^{ \Ci \phi \hat{\Pvs{n}} \cdot \Pvs{\sigma} } \right)^y
= e^{ \Ci y \phi \hat{\Pvs{n}} \cdot \Pvs{\sigma} } = \vRotation(y\phi)
\label{eqn:rotation_power}
\end{eqnarray}
which simplifies the  
derivation of quantities such as the inverse or the Hermitian square root.
Further conceptual benefits of the axis-angle representation are discussed in more detail in \App{symmetry}.

\section{Signal Path}
\label{sec:signal_path}

In this section, each of the elements in the signal path depicted in \Fig{signal_path} are discussed in more detail, beginning with the phase convention $\mbf\Phi$ and ending with Faraday rotation $\bf F$.
%
%
The receiver gains, \JM{G}, and the deviations from an ideal feed, \JM{D}, are analyzed using the polar decomposition and expressed using the axis-angle representation.
Apart from $\mbf\Phi$, the remaining  matrices are purely unitary (rotation) matrices.


\subsection{ Phase Convention }
\label{sec:conjugation}

The phase convention of an instrument is defined by the number of
signal processing steps that result in complex conjugation of the analytic signal, which cannot be represented by a linear transformation of the electric field vector, such as a Jones matrix.

\begin{proof}
If $f(z)=z^*$ and $c$ and $z$ are two complex numbers, $$f(cz) = (cz)^* = c^*z^* \neq cf(z);$$
therefore, complex conjugation is not a linear mapping, which must satisfy
$f(cz)=cf(z)$.
\end{proof}

Referring to \eqns{StokesQ}{through}{StokesV}, complex conjugation of the electric field vector negates the sign of $S_3$, 
a transformation represented by the impure Mueller matrix $\MM{\Phi}$ that appears in \eqn{Stokes_chain}.
Consequently, the measured sign of $S_3$ is impacted by the treatment of the 
complex phase of the electric field, which depends on
the phase convention adopted during the design and implementation 
of instrumentation and signal processing software;
the method used to down-convert the radio signal; and
the Nyquist zone that is digitally sampled.

As noted in Section 2 of \cite{vmjr10}, many signal processing textbooks and software packages adopt the convention that the argument of a complex-valued wave increases linearly with time, such that $z(t) \propto \exp(\Ci\omega t)$.
This convention is adopted in equation~(1) of \citet{sto52} and in the analytic representation of a monochromatic wave presented in \eqn{analytic_field}.
It is opposite to the convention used in many physics text books \citep[e.g.,][]{bw80},
where the solution for a plane-propagating monochromatic electromagnetic wave is 
presented as $\Jcol{e}(x,t) = \Jcol{e}_0\exp(\Ci [kx - \omega t])$.

Choice of phase convention also arises in the implementation of the Discrete Fourier Transform (DFT).
For example, relative to \eqn{Fourier_transform}, which is consistent with the definition adopted by \citet{bra65} and the implementation of commonly used DFT libraries such as FFTW \citep{fj05},
equations~(12.1.7) and~(12.1.9) of Numerical Recipes \citep{pftv86} define the DFT and its inverse using the opposite sign for the argument of the complex exponential.
Application of a DFT that is based on an opposite sign convention is equivalent to negating the frequency, which for a real-valued input signal results in complex conjugation.

The configuration of observatory instrumentation can also negate phase.  
As shown in \App{downconversion}, both complex conjugation and 
negation of frequency occur when either
  lower-sideband down-conversion is used; or,
  during dual-sideband down-conversion, the quadrature component is mixed with a local oscillator that \emph{leads} that of the in-phase component by 90\deg; or
  an even-numbered Nyquist zone is sampled.
%
Complex conjugation at any stage of analysis negates phase in all subsequent processing stages.


\subsection{Receiver Gains}
\label{sec:gain}

A radio receiver converts an electromagnetic wave travelling in free space into two separate signals, each representing an orthogonal component of the electric field vector. These separate signals  propagate along distinct transmission lines and
are independently processed using components such as amplifiers/attenuators, splitters/couplers, and analog-to-digital converters. 
Assuming that there is no further cross-coupling between the two signals, the receiver gains matrix
\begin{equation}
\label{eqn:receiver_gains}
    \JM{G} =
    \left( \begin{array}{cc}
		G_0 & 0 \\
		0   & G_1
	\end{array} \right),
\end{equation}
describes the complex-valued gains, $G_k$, of all components used to process the two separate signals.

Intuition might suggest that it would be fruitful to 
represent the complex-valued gains in polar form, $$G_k = g_k \exp \Ci \phi_k,$$ and decompose \JM{G} into separate amplitude and phase terms,
\begin{equation}
\label{eqn:intuitive_receiver_gains}
    \JM{G} =
    \left( \begin{array}{cc}
		g_0 & 0 \\
		0   & g_1
	\end{array} \right)
    \left( \begin{array}{cc}
    e^{\Ci\phi_0} & 0 \\
    0   & e^{\Ci\phi_1}
\end{array} \right).
\end{equation}
However, when a promising next step is not obvious, 
it pays to employ a systematic approach.  The strategy adopted 
in this work is based on the polar decomposition \eqnp{polar}, the axis-angle representations of the boost and rotation components \eqnsp{Boost}{and}{Rotation},
and determination of the boost component via the Gram matrix \eqnp{boost_from_polar}.

The polar decomposition,
$\JM{G}=G \, \JM{B}_g \JM{R}_g$, includes the absolute gain,
\begin{equation}
\label{eqn:absolute_gain}
G = |\JM{G}|^{1/2} = (G_0 G_1)^{1/2},
\end{equation}
a boost component $\JM{B}_g$, and a rotation component $\JM{R}_g$.
As noted in \Sec{polar_decomposition}, the absolute phase is chosen such that $G$ is real-valued.
The following sub-sections focus on the axis-angle representations of $\JM{B}_g$ and $\JM{R}_g$,
where it is shown that $\JM{B}_g$ characterizes the
ratio of the gain amplitudes, or differential gain, and
$\JM{R}_g$ describes the differential phase.


\subsubsection{Differential Gain}
\label{sec:differential_gain}

Using \eqn{boost_from_polar}, the boost component of \JM{G}
can be derived via the Gram matrix of its rows,
\begin{equation}
\label{eqn:diff_gain_squared}
\JM{B}_g^2 = G^{-2} \, \JM{G}\JM{G}^\dagger
= G^{-2} 
  	\left( \begin{array}{cc}
		g_0^2 & 0 \\
		0   & g_1^2
	\end{array} \right),
\end{equation}
where $g_k = |G_k|$ are the amplitudes of the complex gains, such that $G=(g_0 g_1)^{1/2}$.
Taking the square root of both sides of \eqn{diff_gain_squared} yields
\begin{equation}
\JM{B}_g
 =
  \left( \begin{array}{cc}
    \Gamma & 0 \\
    0   & \Gamma^{-1}
  \end{array} \right),
\label{eqn:boost_decomposed}
\end{equation}
where $\Gamma=(g_0/g_1)^\frac{1}{2}$ is the real-valued amplitude ratio that describes the differential gain.

To arrive at the axis-angle representation of $\JM{B}_g$,
%
%
note that its off-diagonal elements are zero and
%
%
only $\pauli{0}$ and $\pauli{1}$ contribute to the diagonal elements of 
the sum in \eqn{Boost}.
Therefore, $\hat{\Pv{m}}=(1,0,0)^T$ and
\begin{equation}
\JM{B}_g = \pauli{0}\cosh\gamma + \pauli{1}\sinh\gamma = \left( \begin{array}{cc}
		e^\gamma & 0 \\
		0   & e^{-\gamma}
	\end{array} \right).
\label{eqn:diff_gain}
\end{equation}
Equating the right hand sides of \eqns{boost_decomposed}{and}{diff_gain} yields 
\begin{equation}
	\label{eqn:differential_gain}
	\gamma = \ln\Gamma = \frac{1}{2} \ln \frac{g_0}{g_1}.
\end{equation}

Differential gain describes any transformation that subjects opposite polarizations to different levels of amplification, also known as \emph{diattenuation}.
It can vary as a function of both time and frequency for a variety of reasons.  For example, to keep the signal power within the optimal regime of operation, some experiments will set new attenuation levels at the start of each observation.
Some instruments employ active attenuators that introduce differential gain fluctuations on short timescales.  
Furthermore, the mismatched responses of the components in the signal path typically lead to variation of $\beta$ as a function of frequency.


\subsubsection{Differential Phase}
\label{sec:differential_phase}

The axis-angle representation of $\JM{R}_g$ is obtained
by rearranging the polar decomposition of \JM{G} to arrive at
$\JM{R}_g = G^{-1} \JM{B}_g^{-1} \JM{G}$.  Both \JM{G} and $\JM{B}_g$ (and their inverses) are diagonal
and, as a product of diagonal matrices,
$\JM{R}_g$ is also diagonal.
Only $\pauli{0}$ and $\pauli{1}$ contribute to the diagonal elements of 
the sum in \eqn{Rotation}; therefore, $\hat{\Pv{n}}=(1,0,0)^T$ and
\begin{equation}
	\label{eqn:diff_phase}
	\JM{R}_g = \pauli{0}\cos\phi + i \pauli{1}\sin\phi =
	\left( \begin{array}{cc}
		e^{i \phi} & 0 \\
		0   & e^{-i \phi}
	\end{array} \right),
\end{equation}
where $\phi$ describes the differential phase. Subtracting the (degenerate) absolute phase $\phi_\mathrm{abs}=(\phi_0+\phi_1)/2$ from the phases in \eqn{intuitive_receiver_gains} leaves $\phi=(\phi_0-\phi_1)/2.$

Differential phase describes any transformation that subjects opposite polarizations to different delays; it is also known as \emph{retardance} in optics and \emph{cross-hand phase} or delay in interferometry.
It arises when the opposite polarizations propagate along signal paths of different lengths and/or with different speeds.
For a given delay $\Delta t$ between $e_0(t)$ and $e_1(t)$, the corresponding differential phase $\phi = \nu \Delta t$ varies linearly with radio frequency.
Nonlinear spectral variations arise from dispersive effects in the electronics of the receiver, down-conversion system, cables, and other components used to connect observatory instrumentation.
\Sec{Faraday} discusses the differential phase that arises during propagation through the magnetized plasma in the Earth's ionosphere and the interstellar medium.


\subsection{Deviations from an Ideal Feed}

In an ideal feed, the receptors have maximal response to orthogonal senses of either linear or circular polarization.
However, in practice, the receptors are neither perfectly orthogonal nor exactly linearly- or circularly-polarized.
Let non-ideal receptors be represented by unit row vectors, $\hat{\Jrow{r}}_0$ and $\hat{\Jrow{r}}_1$ ($\hat{\Jrow{r}}_i \hat{\Jrow{r}}_i^\dagger = 1$; note that the gains of the receptors are modeled by $G_0$ and $G_1$ in the previous section).
Therefore, the deviations from an ideal feed are given by
\begin{equation}
        \label{eqn:nonorthogonal_Jones}
        \JM{D} =
        \left( \begin{array}{c}
                \hat{\Jrow{r}}_0 \\
                \hat{\Jrow{r}}_1
        \end{array} \right)
        =
	\left( \begin{array}{cc}
		\hat{r}_{00} & \hat{r}_{01} \\
		\hat{r}_{10} & \hat{r}_{11}
	\end{array} \right).
\end{equation}
Deviations from ideal are also known as \emph{cross-talk}, \emph{leakage}, or simply \emph{D-terms}.
The polar decomposition, $\JM{D} = G_d \, \JM{B}_d \JM{R}_d$, includes a scalar gain $G_d$, boost $\JM{B}_d$, and rotation $\JM{R}_d$.
%
%
Rotation matrices describe only orthonormal pairs of receptors (see \Proof{orthonormal_receptors}); therefore,
the orthonormal component of \JM{D} is given by $\JM{R}_d$ and
any non-orthogonality must be described by $\JM{B}_d$,
as detailed in the following sections.


\subsubsection{Non-orthogonal Component}
\label{sec:nonorthogonality}

Using \eqn{boost_from_polar}, the boost component of \JM{D}
can be derived via the Gram matrix of its rows,
\begin{equation}
\label{eqn:nonorthogonal_receptors}
\JM{B}_d^2 = G_d^{-2} \JM{D}\JM{D}^\dagger 
= G_d^{-2} 
  \left( \begin{array}{cc}
1 & \hat{\Jrow{r}}_0\cdot\hat{\Jrow{r}}_1^\dagger \\
\hat{\Jrow{r}}_1 \cdot\hat{\Jrow{r}}_0^\dagger & 1
  \end{array} \right),
\end{equation}
Let $\JM{B}_d^2 = \vBoost(2\beta)$ and note that
%
%
the elements on the diagonal of \eqn{nonorthogonal_receptors} are equal to each other; 
therefore, there can be no contribution from $\pauli{1}$.
Only $\pauli{2}$ and $\pauli{3}$ contribute to the off-diagonal elements of 
the sum in \eqn{Boost}; i.e.,
$\hat{\Pv{m}}=(0,m_2,m_3)^T$, and
\begin{eqnarray}
\vBoost(2\beta) &=& \pauli{0}\cosh2\beta + (m_2\pauli{2}+m_3\pauli{3})\sinh2\beta \nonumber \\
&=& \left( \begin{array}{cc}
\cosh2\beta & z_m^*\sinh2\beta \\
z_m\sinh2\beta & \cosh2\beta
  \end{array} \right),
\label{eqn:nonorthogonal_boost}
\end{eqnarray}
where $z_m = m_2 + \Ci m_3$.
Equating the diagonal elements of \eqns{nonorthogonal_receptors}{and}{nonorthogonal_boost} 
yields $G_d^2 = \sech2\beta$; therefore,
after taking the determinant of both sides of \eqn{boost_from_polar} (noting that $\vBoost(2\beta)$ is unimodular) and rearranging,
\begin{equation}
\label{eqn:det_squared_one}
|\JM{D}\JM{D}^\dagger| = G_d^4 = \sech^2 2\beta = 1 - \tanh^2 2\beta. 
\end{equation}
Similarly, the determinant of \eqn{nonorthogonal_receptors} leads to
\begin{equation}
\label{eqn:det_squared_two}
|\JM{D}\JM{D}^\dagger| = 1 - |\hat{\Jrow{r}}_0\cdot\hat{\Jrow{r}}_1^\dagger|^2.
\end{equation}
Equating the right-hand sides of \eqns{det_squared_one}{and}{det_squared_two} yields
\begin{equation}
\label{eqn:nonorthogonal_beta}
\beta = \frac{1}{2}\tanh^{-1} \left|\hat{\Jrow{r}}_0 \cdot \hat{\Jrow{r}}_1^\dagger \right|.
\end{equation}
When the receptors are orthogonal, $\beta$ is zero
and $\JM{B}_d$ reduces to the identity matrix (no deviations from ideal).

As in Section 4.1 of \cite{van04c}, the 2 dof
in the non-orthogonal component of \JM{D} may be parameterized by 
$\Pv{b} = (0, b_2, b_3)^T = \hat{\Pv{m}} \sinh\beta$, such that
$\sinh^2\beta = \Pv{b}\cdot\Pv{b},$ $\cosh^2\beta=1+\Pv{b}\cdot\Pv{b},$ and
\begin{equation}
\label{eqn:nonorthogonal_component}
\JM{B}_d = \vBoost(\beta) = \pauli{0} (1 + \Pv{b}\cdot\Pv{b})^{1/2} + \Pv{\sigma}\cdot\Pv{b}.
\end{equation}

To first order, this derivation is consistent with Equation (19) of \cite{bri00},
which describes non-orthogonal receptors using a product of two separate boosts along 
the $S_2$ and $S_3$ axes (corresponding to Stokes U and Stokes V, respectively, in a linear basis).
In general, the result of this product includes a rotation about the $S_1$ (Stokes Q) axis;
however, if either of these two boosts is small, then this rotation is negligible.
In contrast, no small-value approximations are made in the derivation of \Eqn{nonorthogonal_component}.
%
%

The instrumental boost due to non-orthogonal receptors can vary as a function of time and frequency.
For example, the parallactic rotation of
the receiver during the transit of the source changes the orientation
of $\hat{\Pv{m}}$ with respect to the celestial coordinate system.
Furthermore, electronic cross-coupling between the two receptors can vary with radio frequency.


\subsubsection{Orthonormal Component}
\label{sec:basis_rotation}

In principle, $\JM{R}_d$ can be obtained
by rearranging the polar decomposition of \JM{D} to arrive at
$\JM{R}_d = G_d^{-1} \JM{B}_d^{-1} \JM{D}$; however, this proves to
be algebraically cumbersome.
Instead, following \cite{bri00}, the transformation is expressed using 
the spherical coordinates $\psi$ and $\chi$ that define the polarization ellipse.

First, in the ideal basis, consider a 100\% polarized signal $\Jcol{e}(t)$ with Stokes polarization vector
\eqnp[see]{stokes_geometric}
\begin{equation}
\Pv{S} = I \left( \cos2\chi  \cos2\psi, \; \cos2\chi  \sin2\psi, \; \sin2\chi \right)^T.
\label{eqn:stokes_vector}
\end{equation}
Next, consider the transformation from the ideal basis to one in which the first receptor responds
maximally to $\Jcol{e}(t)$ and the response in the orthogonal receptor is zero.
In this basis, the measured coherency matrix,
\begin{equation} \begin{split}
{\cohM{\rho}}'
&= \left( 
	\begin{array}{ccc}
	I & \; & 0 \\
	0 & \; & 0
	\end{array}
	\right)
= \left( \pauli{0} + \pauli{1} \right) I / 2
\end{split} \end{equation}
and the polarization vector $\Pv{S}'= I (1,0,0)^T$.

With reference to \Fig{polarization_vector}, $\Pv{S}'$ is obtained by rotating $\Pv{S}$ about the $S_3$ axis by $-2\psi$ in the ideal basis, then rotating the result by $2\chi$ about the $S'_2$ axis in the intermediate basis.  The corresponding
transformation of the electric field, $\Jcol{e}'(t) = \JM{R}_d \, \Jcol{e}(t)$, where
\begin{equation}
\label{eqn:orthonormal_spherical}
 \JM{R}_d  = \vRotation[2](-\chi) \vRotation[3](\psi).
\end{equation}
This is equivalent to equation~(15) of \citet{bri00}, apart
from the negation of $\chi$ that associates
positive values of $\chi$ with $S_3 > 0$.

This is a useful parameterization of the 2 dof in $\JM{R}_d$ because
it reduces to the identity matrix (no deviations from ideal) when $\chi$ and $\psi$ are zero.
Furthermore, because differential phase results in a rotation about the $S_1$ axis, 
it is desirable to describe the deviation from the ideal basis transformation using rotations about the other two axes. Together, the three rotations define the orientation of the basis in a manner similar to the Tait--Bryan angles (yaw, pitch, and roll).


\subsection{Nominal Feed Configuration}
\label{sec:nominal_feed}

As in \cite{vmjr10}, the nominal feed configuration is described by three parameters that are summarized in \Tab{parameters}.
\begin{table*}
\begin{center}
\caption{The effects of phase convention and nominal feed configuration parameters in each basis.}
\label{tab:parameters}
\begin{tabular}{lcc|c}
\hline

Parameter           & Range        & \multicolumn{2}{c}{Effect} \\
                    &              & Linear & Circular          \\

\hline

Phase Convention    & $\pm$ 1      & $\pm$ V     & $\pm$ U   \\

Feed Basis          & LIN or CIRC  & $\Pv{S}=(Q,U,V)^T$ & $\Pv{S}=(V,Q,U)^T$   \\

Feed Hand           & $\pm$ 1      & $\pm$ Q\&V     & $\pm$ U\&V \\

Symmetry Angle      & $-\pi/2<\theta<\pi/2$
& $\vRotation[z](\theta-\pi/4)$ & $\vRotation[z](\theta)$ \\
\hline
\end{tabular}
\end{center}
\end{table*}
The \emph{feed basis} defines the polarizations of the 
receptors (linear or circular);
the \emph{feed hand} defines the handedness of the basis (left or right); and
the \emph{symmetry angle} describes the orientation of the reference frame defined by the receptors.
The corresponding basis transformation
\begin{equation}
    \JM{C} = \JM{R}_h \JM{R}_b \vRotation[z](\Theta)
\end{equation}
is the product of feed hand $\JM{R}_h$ and basis $\JM{R}_b$ transformations, and a rotation about the line of sight $\vRotation[z](\Theta)$ defined by the symmetry angle $\Theta$.  These are detailed in the following sections, starting with the feed basis, which defines both the nominal symmetry angle and the effect of the feed hand transformation.


\subsubsection{Feed Basis}
\label{sec:feed_basis}

Typically, the two receptors in a receiver are either linearly or circularly polarized, and
\Sec{geometry} describes a Cartesian coordinate system defined by linearly polarized receptors.
As shown in \App{circular_basis},
the transformation from this Cartesian basis to one defined by a pair of circularly-polarized receptors is described by
\begin{equation}
\label{eqn:circular}
\JM{R}_b = \frac{1}{\sqrt{2i}}
\left( \begin{array}{cc}
  1 & -i \\
  1 & i
\end{array} \right).
\end{equation}
In the basis defined by $\JM{R}_b$, the Stokes polarization vector $\Pv{S}=(S_1, S_2, S_3)^T=(V,Q,U)^T$ and the effects of all subsequent transformations (from \JM{X} to $\MM{\Phi}$) differ from their impact in the original Cartesian basis.

\begin{figure}
\centerline{\includegraphics[width=0.7\linewidth]{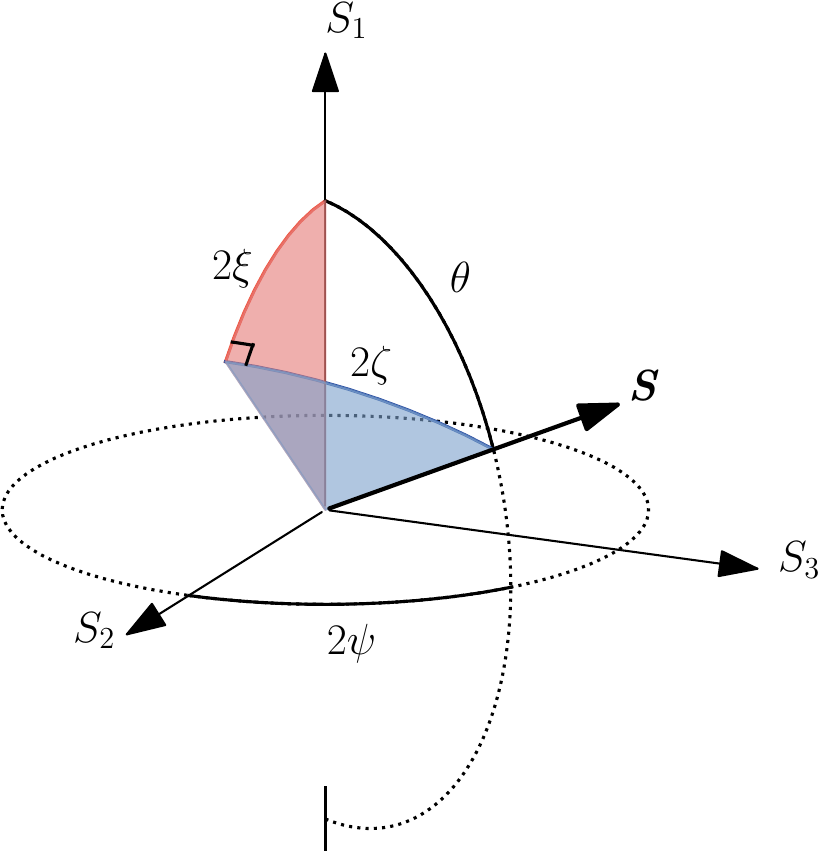}}
\caption{
In the circular basis, the direction of the polarization vector, $\Pv{S}=(S_1, S_2, S_3)^T=(V,Q,U)^T$, is defined by the polar angles, $\xi$ and $\zeta$.
These angles together define a right triangle on the surface of the Poincar\'{e} sphere with hypotenuse equal to the colatitude, $\theta=\pi/2 - 2\chi$, measured with respect to the pole defined by positive $S_1$ (Stokes V). 
Also depicted is the longitude $2\psi$, measured with respect to positive $S_2$ (Stokes Q) toward positive $S_3$ (Stokes U).
%
%
}
\label{fig:polarization_vector_circular}
\end{figure}

For example, in the circular basis, the rotations around the $S_2$ and $S_3$ axes that define both the polarization ellipse \eqnp{stokes_ellipse} and the orthonormal component of deviation \eqnp{orthonormal_spherical} are better described by pair of polar angles, $\xi$ and $\zeta$, respectively.
These angles, depicted in \Fig{polarization_vector_circular}, are related to the orientation and ellipticity angles by
\begin{eqnarray}
    \tan2\xi &=& \cot2\chi \cos2\psi \\
    \sin2\zeta &=& \cos2\chi \sin2\psi.
\end{eqnarray}

In the circular basis, the differential phase $\phi$ describes rotation about the Stokes V axis, which is equivalent to a physical rotation about the line of sight (\Proof{Faraday_rotation}).
Later sections describe the impacts of the circular basis on the feed hand $\JM{R}_h$ (\Sec{feed_hand}), axis of symmetry $\vRotation[z](\Theta)$ (\Sec{symmetry}), and phase convention $\MM{\Phi}$ (\Sec{calibration_phase_convention}) transformations.


\subsubsection{Feed Hand / Basis Reflection}
\label{sec:feed_hand}

The feed hand is determined by the design and integration of the receiver, such as the sign convention for the voltages
output by an orthomode transducer, which determines the directions of the $x$ and $y$ axes in \Fig{polarization_ellipse}; or any cable swaps that might occur between the receiver and the backend, which swap the $x$ and $y$ axes.
Feed hand is also impacted by the total number of reflections in the antenna structure, which depends upon the placement of the feed at the primary or secondary focus, and the number of reflections in any beam waveguide structure \citep[e.g.,][]{gaa+04,pnw+15} or corrective mirrors designed to optimize sensitivity \citep[e.g.,][]{kbh94,gj97}. 



%
%
%
%

Normal reflection by a conducting surface, such
that the angles of incidence and reflection equal zero 
and the incident and reflected rays are anti-parallel, 
can be conceptualized in two different ways that arrive at the same result.
One way to model the observation of reflected rays is by turning over the feed horn, which originally points up at the sky, to point down at the reflector.  
This approach is detailed in \App{yaw_rotation}.

Alternatively, the change in wave direction is modeled by negating the $z$-axis, thereby producing a left-handed coordinate system.
The handedness of the reference frame is also negated by exchanging the components of the electric field.
With reference to \eqns{StokesQ}{through}{StokesV},
swapping $e_x$ and $e_y$ negates Stokes Q and V,
which is equivalent to a $\pm 180\deg$ rotation about the Stokes~U axis.
In a basis defined by circularly-polarized receptors, 
swapping $e_L$ and $e_R$
reverses the signs of both Stokes U and V,
which is equivalent to a $\pm 180\deg$ rotation about the Stokes~Q axis.
In both cases, the rotation is given by $\JM{R}=\pauli{2}$; it negates both the ellipticity 
$\chi$ \eqnp{ellipticity} and the position angle $\psi$ \eqnp{orientation}.

Although the number of reflections in an antenna can be easily counted, it is generally less feasible to account for voltage negation or cable swapping.
Therefore, the feed hand is typically determined experimentally with reference to previously published polarization data.
For example, as both $\chi$ and $\psi$ are negated upon reflection, the feed hand can be unambiguously determined by observing a source with known ellipticity or Faraday rotation measure.
Given the total number of hand reversals $N$, the feed hand matrix is
defined by 
\begin{equation}
\JM{R}_h = \pauli{2}^N.
\end{equation}

The negation of the position angle
after normal reflection 
from a metal surface 
is noted in the context of a more richly detailed treatment of the physics of reflection and refraction by \citet{bw80}.
However, it appears to be unrecognized in some important works on radio polarimetry, which mention only the negation of Stokes V \cite[e.g.,][]{hbs96,rh21}.
%
%
Appendix A of \cite{cla10} provides a carefully detailed derivation of
both the linear and circular polarization of radiation that is observed after reflection,
including a historical account of some confusion that has persisted on this topic.
%

%
%
%
%
%
%


\subsubsection{Axis of Symmetry}
\label{sec:symmetry}

The axis of symmetry in the $x$-$y$ plane depicted in \Fig{polarization_ellipse} is defined as the
position angle of a linearly-polarized wave that produces equal responses
in each ideal receptor.
In a basis defined by linearly-polarized receptors, a linearly-polarized wave that oscillates along 
$x = y$ (positive Stokes $U$) will produce
equal and in-phase responses in each receptor; therefore, the symmetry angle has a nominal value of 45\deg.
In a basis defined by circularly-polarized receptors, a linearly-polarized wave that oscillates along the $x$ axis will produce equal and in-phase responses in each receptor, and the symmetry angle has a nominal value of 0\deg.

The symmetry axis of a receiver on steerable mount describes its rotation about the line of sight with respect to a reference point on the antenna structure.  For a fixed dipole on the ground, the symmetry axis describes its rotation about the zenith.  Once established, the symmetry angle is typically treated as a constant.  Any unintended rotation (either geometric or apparent; e.g.,  owing to cross-coupling of the receptors) must be determined through calibration of the orthonormal component of deviation, which includes an orientation angle $\psi$ \eqnp{orthonormal_spherical}. Other known rotations are
included in the projection transformation described in \Sec{projection}.

%


\subsection{Projection onto the Celestial Sphere}
\label{sec:projection}

For radio astronomical observations, the incident electric field measured in the reference frame of the antenna must be transformed to the celestial reference frame adopted by the IAU.
In this coordinate system, the electric field vector is described by its projection onto the plane of the sky, with basis vectors $\hat{\mbf x}$ and $\hat{\mbf y}$ pointing North and East, respectively.
When observing with a fully-steerable antenna, this transformation is described by a rotation about the line of sight through the parallactic angle, which is well determined by geometry.
In the {\sc psrchive} software \citep{hvm04}, the transformation to celestial coordinates can be computed for a wide variety of antenna mounts,
or it can optionally be replaced by a user-supplied table of transformations that describe the direction-dependent response of the antenna,
as in the case of a dipole array \citep[e.g.,][]{2016MNRAS.462.4482A,2017PASA...34...62S}.

Without a detailed electromagnetic model of a fixed dipole, the direction-dependent response
can be described to first order as a geometric projection of the receptors onto the sky.
Such a transformation is a product of 
    a rotation about the Stokes V axis that describes parallactic rotation about the line of sight, and 
    a boost along an axis in the Stokes $Q$ -- $U$ plane that describes the combined effects of differential gain and apparent nonorthogonality due to foreshortening of the projected receptors.
%
The combination of these effects causes the symmetry axis (see \App{symmetry}) of the projection transformation to vary with the direction to the source.

Therefore, for a fixed dipole array,
there is no fundamental degeneracy as described in \App{degeneracy}, which is a consequence of a single constant axis of symmetry.
That is, in principle, it is possible to determine 6 dof of the response (all but the absolute gain) by observing a polarized point source with a phased array of fixed dipoles over a wide range of hour angles.
However, as described in \App{eigenvectors_of_projection}, near multicollinearity may arise when the rapidity of the boost along the Stokes V axis is allowed to vary, causing numerical instability and inflated uncertainty of the rapidity estimate.

The Lorentz boost caused by projection of fixed dipoles onto the celestial sphere converts significant levels of unpolarized flux into linearly-polarized flux.  
In wide-field interferometric imaging experiments, this can be exploited to fully constrain the instrumental response using the unpolarized sky \citep[e.g.,][]{bmh+22,kvd+25}.
A similar approach is generally not possible in single-antenna observations owing to a combination of stray radiation \citep[e.g.][]{2002ASPC..278..397L}, radio frequency interference \citep[e.g.][]{2010MNRAS.405..155O}, and the unknown variation of the instrumental response over the beam \eqnp{coherency_integral}.


\subsection{Faraday Rotation}
\label{sec:Faraday}

During propagation
through the interstellar medium or the Earth's ionosphere, an astrophysical signal experiences Faraday rotation.
This phenomenon, also known as circular birefringence, arises because the natural modes of propagation in a magnetized (non-relativsitic and collisionless) plasma are circularly polarized, such that LCP and RCP components of the electric field propagate with different speeds.
At a given frequency, the resulting differential delay between LCP and RCP components causes the electric field to rotate about the line of sight.
For example, if 
the signal propagates in the direction of the magnetic field vector, then LCP is delayed with respect to RCP \citep{man72,2021MNRAS.507.4968F}; the resulting transformation
is equivalent to rotating the electric field by $\Delta\Psi$ around the line of sight (counter-clockwise as seen by the observer).
\begin{proof}
\label{proof:Faraday_rotation}
In the circular basis \eqnp{circular}, apply differential phase \eqnp{diff_phase}, and return to the original Cartesian basis.  
\begin{align*}
\JM{F}&(\Delta\Psi)
 = \JM{R}_b^{-1} \JM{R}_g \JM{R}_b \\
&= \frac{1}{2}
    \left( \begin{array}{cc}
      1 & 1 \\
      i & -i
    \end{array} \right) 
    \left( \begin{array}{cc}
      e^{-\Ci \Delta\Psi} & 0 \\
      0 & e^{\Ci \Delta\Psi}
    \end{array} \right) 
    \left( \begin{array}{cc}
      1 & -i \\
      1 & i
    \end{array} \right) \\
&= \left( \begin{array}{cc}
      \cos\Delta\Psi & -\sin\Delta\Psi \\
      \sin\Delta\Psi & \cos\Delta\Psi
    \end{array} \right) \\
&= \vRotation[z](-\Delta\Psi)
\end{align*}
\end{proof}
The change in the position angle of the radiation,
\begin{equation}
    \label{eqn:Faraday_rotation}
    \Delta\Psi(\nu) = \mathrm{RM} \, c^2 \nu^{-2},
\end{equation}
where RM is the rotation measure, $c$ is the speed of light, and $\nu$ is the frequency of the radiation.
Note that birefringence is dispersive and the phase shift / rotation is proportional to $\nu^{-2}$; this differs from the differential phase
due to a path length difference, which is proportional to $\nu$.
The RM is proportional to the path integral of the density of free electrons, $n_e$, times the strength of the magnetic field, $\Pv{B}$, parallel to the direction of propagation $d\Pv{z}$ \citep[e.g.,][]{hml+06}
\begin{equation}
\mathrm{RM} = C \int_0^D n_e \, \Pv{B} \cdot  d\Pv{z},
\end{equation}
where $D$ is the distance to the astrophysical source and
%
\begin{equation}
C = 0.81 \frac{\mathrm{rad\, m}^{-2}}{ \mathrm{cm}^{-3} \, \mu\mathrm{G \, pc} }.
\end{equation}

If the magnetic or ionic properties of the ionosphere fluctuate rapidly, either on timescales
shorter than the interval over which the Stokes parameters are integrated or on spatial scales smaller than
the volume sampled by multipath propagation \cite[e.g.,][]{css16,sc19}, then the signal
will be depolarized by stochastic Faraday rotation \citep{mm98}, resulting in an impure
Mueller matrix with no equivalent Jones matrix (see \App{impure}).


\section{ Calibration }
\label{sec:calibration}




The response of each element of the signal path may include an unknown component that must be determined experimentally.  
This section briefly reviews methods of calibrating
each component in \Eqn{Stokes_chain} and \Fig{signal_path},
focusing primarily on the response of a single antenna (single dish or phased array).  
Where relevant, additional notes on interferometric calibration are included.


\subsection{ Phase Convention }
\label{sec:calibration_phase_convention}

For linearly-polarized receptors, $S_3$ equals Stokes~V, and complex conjugation
negates the ellipticity angle.
%
For circularly-polarized receptors, $S_3$ equals Stokes U, and the position 
angle is negated by conjugation.
%
%
Complex conjugation of the electric field vector
also negates frequency in the spectral domain.
Therefore, the phase convention used to represent the electric field can be determined by incorporating information from other sources.
For example, owing to dispersion, frequency negation is immediately apparent during radio pulsar signal analysis,
and the phase convention must be known to perform phase-coherent dispersion removal in radio pulsar observations \cite[e.g.,][]{vb11}.
Frequency negation can also be detected in continuum observations over regions of the radio spectrum that include strong, well-defined spectral lines.
In an interferometer, complex conjugation of the visibility matrix computed for each antenna pair results in a 180$\deg$ rotation of the image of the brightness distribution on the plane of the sky.


\subsection{ Receiver Gains }
\label{sec:calibration_gains}

The receiver gains matrix, \JM{G}, is often determined using observations of an artificial reference source with a well-defined polarization state.
For example, many receivers have a built-in source of broadband noise
(e.g.,  a noise diode coupled to the receptors) that can be used to estimate gains as a function of radio frequency.
Determination of \JM{G} using only an artificial noise source is based on the \textit{ideal feed assumptions}, in which the receptors are orthogonal and there is no cross-coupling between them (i.e., \JM{D} is the identity matrix);
and the artificial reference source illuminates the receptors equally (i.e., the receptors have identical responses to the reference source, in both amplitude and phase).
Although these assumptions are common, it is more accurate to experimentally determine both \JM{D} and the Stokes parameters intrinsic to the artificial reference source \citep[e.g.,][]{van04c}, as described in \Sec{calibration_deviation}.

Even when fully modeling the deviations from an ideal feed and polarization of the artificial reference source, the ideal feed assumptions can provide a useful, and sometimes necessary, first guess for \JM{G}.
Therefore, they are an important part of polarimetric calibration for both single antennas \citep[e.g.,][]{hx97,nms+97,gl98,hdvl09} and interferometric arrays \citep[e.g.,][]{jkmg08,sjk+21}.
As shown in \Sec{gain}, the polar decomposition of the complex gains,
\begin{equation}
    \JM{G} = G \, \JM{B}_g(\gamma) \JM{R}_g(\phi),
\end{equation}
has 3 dof: the absolute gain $G$, the differential gain $\gamma$, and the differential phase $\phi$.
To experimentally determine $G$ requires a \textit{standard candle}, i.e., a source with known flux density at the radio frequency of the observations.
Absolute gain calibration (also known as flux calibration) is described in more detail in Section 7.2 of \cite{vdo12}.
In an interferometer, unique values of the complex gains must be determined for each element in the array.

Differential gain mixes only $S_0$ and $S_1$, and differential phase mixes only $S_2$ and $S_3$.
Therefore, given observations of an ideal artificial reference source with known polarization, 
the two unknown values of $\gamma$ and $\phi$ can be solved independently. 
%
The following two subsections derive solutions for 
$\gamma$ and $\phi$ based on the assumption that
a pure linearly polarized reference source
induces identical signals in the orthogonal receptors
of an ideal feed.
In the linear basis, the Stokes parameters intrinsic to such a source,
\begin{equation}
    \label{eqn:ideal_noise_diode}
    \Sv{S}_\mathrm{ref} = I_\mathrm{ref} [1, 0, 1, 0]^T,
\end{equation}
where $I_\mathrm{ref}$ is the intensity of the reference source.

\subsubsection{ Differential Phase }
\label{sec:ideal_differential_phase}

As a rotation about the $S_1$ axis, differential phase mixes only $S_2$
and $S_3$, and to determine $\phi$ requires observations of a polarized reference source with known intrinsic values of
$S_2$ and $S_3$.  In the linear basis, $S_2$ and $S_3$ correspond to Stokes U and Stokes~V,
respectively; in the circular basis, $S_2$ and $S_3$ correspond to Stokes Q and Stokes U.
Given the measured Stokes parameters of the reference source, with observed values of $S_2'$ and $S_3'$, \eqns{StokesU}{and}{StokesV} can be solved for the intrinsic and observed values of $\langle e_0^*(t)e_1(t)\rangle$; e.g.
\begin{equation}
\langle e_0^*(t)e_1(t)\rangle = (S_2 + \Ci S_3) / 2.
\end{equation}

To relate the intrinsic values to their observed values, 
consider the transformation of the intrinsic electric field vector \Jcol{e}(t)
into the observed electric field vector as described by \eqn{diff_phase},
\begin{equation}
\Jcol{e}'(t) =
	\left( \begin{array}{cc}
		e^{i \phi} & 0 \\
		0   & e^{-i \phi}
	\end{array} \right) \Jcol{e}(t).
\end{equation}
The observed value of 
\begin{equation}
\langle e_0^{\prime*}(t)e'_1(t) \rangle = e^{-2\Ci\phi} \langle e_0^*(t)e_1(t)\rangle
\end{equation}
can then be used to solve for $\phi$; i.e.,
\begin{equation}
\phi = -\frac{1}{2} \arg\left(
\frac{\langle e_0^{\prime*}(t)e'_1(t) \rangle}{\langle e_0^*(t)e_1(t)\rangle}\right),
\end{equation}
where $\arg(z)$ is the argument of complex number $z$.

If the reference source produces an in-phase signal in each receptor, 
such that the differential phase between $e_0(t)$ and $e_1(t)$ equals zero,
then $\langle e_0^*(t)e_1(t)\rangle$ is real-valued and
\begin{equation}
\phi = -\frac{1}{2} \arg\left(\langle e_0^{\prime*}(t)e'_1(t) \rangle\right)
= -\frac{1}{2} \tan^{-1}(S_3'/S_2').
\end{equation}

\subsubsection{ Differential Gain }
\label{sec:ideal_differential_gain}

As a boost along the $S_1$ axis, differential gain mixes only $S_0$ and $S_1$;
therefore, determination of $\gamma$ requires observations of a polarized reference source with
known intrinsic values of $S_0$ and $S_1$, where $S_1$
corresponds to Stokes Q in the linear basis and to Stokes~V in the circular basis.
Given the measured Stokes parameters of the reference source, with observed values of $S_0'$ and $S_1'$, \eqns{StokesI}{and}{StokesQ} can be solved
for the intrinsic and observed values of $\langle|e_0(t)|^2\rangle$ and $\langle|e_1(t)|^2\rangle$; e.g.
\begin{eqnarray}
    \langle|e_0(t)|^2\rangle &=& (S_0 + S_1)/2 \\
    \langle|e_1(t)|^2\rangle &=& (S_0 - S_1)/2.
\end{eqnarray}
To relate the intrinsic values to their observed values, 
consider the transformation of the intrinsic electric field vector \Jcol{e}(t)
into the observed electric field vector as described by \eqn{boost_decomposed},
\begin{equation}
\Jcol{e}'(t)
 = G
  \left( \begin{array}{cc}
    \Gamma & 0 \\
    0   & \Gamma^{-1}
  \end{array} \right) \Jcol{e}(t).
\end{equation}
The observed values of 
\begin{equation}
\begin{split}
\langle|e'_0(t)|^2\rangle =& G^2 \Gamma^2 \langle|e_0(t)|^2\rangle \\ \langle|e'_1(t)|^2\rangle =& G^2 \Gamma^{-2} \langle|e_1(t)|^2\rangle
\end{split}
\end{equation}
can be used to solve for $\Gamma$; i.e.,
\begin{equation}
\Gamma = \left(
\frac{\langle|e'_0(t)|^2\rangle}{\langle|e'_1(t)|^2\rangle} 
\frac{\langle|e_1(t)|^2\rangle}{\langle|e_0(t)|^2\rangle} \right)^{1/4}
\end{equation}
If the reference source produces a signal with equal power in each receptor, 
then $\langle|e_1(t)|^2\rangle=\langle|e_0(t)|^2\rangle$ and
\begin{equation}
\Gamma = \left(
\frac{\langle|e'_0(t)|^2\rangle}{\langle|e'_1(t)|^2\rangle} \right)^{1/4}
=\left(
\frac{S_0' + S_1'}{S_0' - S_1'} \right)^{1/4}.
\end{equation}


\subsection{ Deviations from Ideal }
\label{sec:calibration_deviation}

For a single antenna, the deviations from an ideal feed, \JM{D}, can be 
determined by modeling variations in the observed polarization of an astrophysical source as a function of parallactic angle \cite[e.g.,][]{scr+84,xil91,mck92,hpn+01,joh02,van04c}.
In some treatments, \JM{D} is estimated after the complex receiver gains, \JM{G}, have been calibrated; in others, \JM{D} and \JM{G} are jointly determined.
Some of the models used to describe $\JM{D}$ are discussed and compared in \Sec{selection}.
In an interferometer, unique values of $\JM{J}_i = \JM{G}_i \JM{D}_i$ must be determined
for each element in the array, indexed by $i$.  When used as a phased-array, only the Jones matrix of the sum, $\JM{J} = \sum_i \JM{J}_i,$
must be determined.

More broadly, deviations from the ideal feed assumptions include the 
unintended polarization intrinsic to the artificial reference source, $\mbf{\rho}_\mathrm{ref}$.
If observations of the reference source are available, then $\mbf{\rho}_\mathrm{ref}$ can be included in the model and jointly determined along with \JM{G} and \JM{D}.
When doing so, the reference source can be modeled
as a free-space signal that is transmitted into the feed, or
as a guided signal on a transmission line that is coupled after reception.

A free-space signal may be transmitted into the feed using a dipole with known orientation.
%
Typically, the polarization of such an artificial reference signal is distorted by near-field effects, mutual coupling, resonances, and standing waves between the transmitting dipole, feed horn and antenna structure. 
%
%
When modeled as a free-space signal, the artificial reference source
is transformed by the product, \JM{GD}, as in \cite{van04c}.

Alternatively, the artificial reference source may be coupled to the astronomical signal after reception, at which point the two polarizations propagate as a pair of guided signals on separate transmission lines.
This requires splitting the artificial reference source signal and coupling it identically to the two orthogonally-polarized signals, ideally before any amplification.
Typically, the polarization of such a reference signal is distorted by mismatched impedance, insertion losses, reflections, and standing waves.
When modeling an artificial reference source that is guided and coupled,
$\mbf{\rho}_\mathrm{ref}$ is transformed by only
\JM{G}, as in \cite{bja+20}.

In either case (free-space or guided transmission), the instrumental response to an artificial noise source typically differs from its response to an astrophysical source.
However, in many experiments, the absolute polarization of the artificial reference source is irrelevant.
For example, techniques used to update the differential gain and phase given known values of $\mbf{\rho}_\mathrm{ref}$ and \JM{D} \citep[e.g.,][]{ovhb04} rely only on the temporal stability of these quantities.
Therefore, any differences in the instrumental response to astrophysical and artificial sources can be absorbed by redefinition of $\mbf{\rho}_\mathrm{ref}$.

This distinction is particularly important when using an astrophysical source as a polarized reference source \citep[e.g.,][]{van13}.  In this case, $\JM{D}$ may be corrupted by unmodeled ionospheric Faraday rotation \cite[e.g.,][]{rvg+24}.
To avoid also corrupting $\mbf{\rho}_\mathrm{ref}$, it is necessary to model the artificial reference source as though it is coupled after $\JM{D}$, regardless of the actual receiver design.


\subsection{ Nominal Feed Configuration }
\label{sec:calibration_configuration}

Even the nominal feed configuration may require some reasoning, informed by observations, about the handedness and symmetry axis of the receiver.
For example, any linear transformation of the electric field that negates either the ellipticity or the position angle must negate both.
Therefore, regardless of the polarization of the receptors, if the sign of either the ellipticity or position angle of an observed source is known, then the handedness of the receptor basis can be unambiguously determined.
Even without an estimate of the absolute value of the position angle, its sign can be inferred from either the RM or the slope of the canonical S-shaped sweep of position angle described by the Rotating Vector Model \citep[RVM;][]{rc69a} in longitude-resolved observations of the polarized emission from radio pulsars.


\subsection{ Celestial Sphere Projection }
\label{sec:calibration_projection}

In principle, the projection between the celestial reference frame and the nominal receptor basis may be completely determined by known geometry; however, it may also require sophisticated modeling of direction-dependent effects.  
For example, gravitational deformation of an antenna (both the reflecting surface and the support structure) can cause the polarimetric response to vary with pointing direction
\citep{rh21,2024SPIE13094E..3GI}.
Furthermore, the response of a dipole array varies with direction to the source in a manner that is not well-described by the first-order geometric effects of projection.
Therefore, the accuracy with which the instrumental response can be determined will be limited by the fidelity of the electromagnetic model used to describe direction-dependent effects specific to the antenna design \citep[e.g.,][]{wbv11,mwb+12,2015RaSc...50...52S}.


On some radio telescopes, either the entire reflector can be mechanically rotated about the line of sight \cite[e.g.,][]{2021PASA...38....9H} or the receiver can be rotated about the line of sight with respect to the reflector \cite[e.g.,][]{swb+96}.
This feature is typically used to compensate for the parallactic rotation of the observatory with respect to the sky; however, it can also be used to simulate observation over a range of parallactic angles \citep[e.g.,][]{gcv+23}.
For such systems, the mechanical rotation can be included as a component of the projection matrix.
For example, if the feed horn is rotated with respect to the reflector, the projection matrix may be decomposed as
\begin{equation}
\JM{P} = \vRotation[z](\theta_\mathrm{feed}) 
\JM{J}(l,m) \vRotation[z](\theta_\mathrm{para}),
\end{equation}
where $\theta_\mathrm{feed}$ is the feed horn rotation,
$\JM{J}(l,m)$ models direction-dependent effects of the antenna (e.g., as a function of the direction cosines $l$ and $m$ that describe angular offsets from the primary axis), and $\theta_\mathrm{para}$ is the parallactic angle \citep[e.g.,][]{gvc+25}.


\subsection{ Faraday Rotation }
\label{sec:calibration_Faraday}

To calibrate Faraday rotation, it is necessary to first estimate the Faraday Rotation Measure (RM).  A wide variety of techniques have been developed for RM estimation, including 
directly modeling the variation in observed position angle $\Delta\Psi$ as a function of radio frequency \citep[e.g.,][]{njkk08};
iteratively computing the weighted
mean position-angle difference between
two radio frequency bands
\citep{hml+06,cvk+19};
and searching for the peak in linearly-polarized flux after integrating over radio frequency as a function of trial RM \citep[e.g.,][]{hbo05,2005A&A...441.1217B}.

In principle, Faraday rotation can be corrected by rotating the Stokes polarization vector $\Pv{S}(\nu)$ observed at each frequency by $-\Delta\Psi(\nu)$ about the Stokes V axis.
This effectively yields the position angle as though observed at infinite radio frequency.
However, such a correction would later have to be inverted when refining the RM estimate, which requires a set of observations  made at finite radio frequencies.
Therefore, the Faraday rotation at each frequency $\nu$ is corrected with respect to the linear polarization state observed at some fiducial frequency $\nu_0$ (typically the centre frequency of the band); i.e.,
\begin{equation}
\label{eqn:constant_RM}
    \Delta\Psi'(\nu) = \mathrm{RM} \, c^2 \left( \nu^{-2} - \nu_0^{-2} \right).
\end{equation} 

The RM may also vary as a function of time over the duration of the integration.
For example, the ionosphere can cause the RM to vary by several rad m$^{-2}$ on timescales of hours.  These variations can be predicted using a model of the geomagnetic field and a map of ionospheric electron content \citep[e.g.,][]{pnt+19}.
Temporal variations in RM also cause the position angle observed at the fiducial frequency to vary with time; therefore, Faraday rotation at time $t$ is corrected by
\begin{equation}
\label{eqn:variable_RM}
\Delta\Psi(\nu,t) 
  = \mathrm{RM}_0 \, c^2 \left( \nu^{-2} - \nu_0^{-2} \right)
  + \Delta\mathrm{RM}(t) \, c^2 \nu^{-2}.
\end{equation} 
where 
$\mathrm{RM}_0$ is constant 
and
$\Delta\mathrm{RM}(t)=\mathrm{RM}(t) - \mathrm{RM}_0$.


\section{ Polarimeter Model Selection }
\label{sec:selection}


Accurate polarimetric calibration requires a 
model of the instrumental response that includes deviations from an ideal feed.
Several different approaches to decomposing the single-antenna
instrumental response have been proposed in the published literature.
Broadly, these models can be divided into those based on Jones matrices and
those based on Mueller matrices; e.g., 
the Jones matrix equation~(11) of \citet[][hereafter HBS96]{hbs96};
the Jones matrix equation~(19) of \citet[][hereafter B2000]{bri00}; and
the Mueller matrix equation~(22) of \citet[][hereafter H2001]{hpn+01}.
%
%
In principle, a Mueller matrix has an additional 9 dof and can describe any linear transformation of the Stokes parameters, including non-linear transformations
of the electric field such as complex conjugation.  However, these additional dof are not employed in H2001, where the Mueller matrices are derived from Jones matrices.

Regardless of the objects used to represent polarization state and transformations,
when applied to experimental data analysis, a mathematical model of the 
instrumental response should be
\begin{enumerate}
        \item physically motivated, such that model components describe elements of the signal path;
	\item surjective, such that the parameter space spans all possible transformations of interest;
        \item injective, such that each transformation is described by a unique set of parameters;
	\item self-consistent, such that the assumed properties of the system are preserved; and
	\item numerically stable, at least in the vicinity of the anticipated solution.
\end{enumerate}

In the following subsections, these criteria are discussed in the context of
single-antenna observations of a point source, primarily using the models proposed by 
HBS96, B2000, and H2001 as examples.
When referring to the model parameters used in these works, 
the original mathematical symbols are retained;
these may conflict with the symbols employed in the previous
sections of this paper.

\subsection{Physical Motivation}

\citet{ham00} proposed a purely algebraic decomposition of the 
instrumental response using a single polar decomposition, and this 
approach is one of the two implemented by \citet{van04c}.
Although simple and useful, the polar decomposition is not directly amenable to physically meaningful interpretation.  It does not
model the order in which elements in the signal chain are encountered,
and it does not permit separation of backend and frontend transformations.

In practice, it proves useful to decompose the instrumental Jones matrix
into a product of backend and frontend transformations, such as \JM{G} and \JM{D}, respectively, of HBS96.
The frontend describes the non-ideal cross-coupling between the receptors, and is typically assumed to remain stable over timescales of the order of days.
Therefore, the frontend can be calibrated less frequently than the backend component, which describes the complex gains applied to the receptors.  
These gains are typically adjusted at the start of each observation,
especially if the system equivalent flux density varies strongly with position on the sky.
HBS96, B2000, and H2001 describe models that reflect the order in 
which transformations occur, and that can be decomposed into frontend 
and backend components.

\subsection{Surjectivity}

When considering only linear transformations of the electric field by
a single antenna, as represented using Jones matrices, a surjective
model must have 7 independent dof.
If the absolute gain is treated as a scalar multiplier, then the
matrix component of the transformation must be described by six parameters.
In the framework developed by B2000, these 6 dof describe
three Lorentz boosts that mix the total intensity with Stokes Q, U, and V
and three Euclidean rotations about the Stokes Q, U, and V axes.

The model developed by H2001 includes only 5 of these 6 dof:
the differential gain, $\Delta G$ (eq.~20) describes the mixing between Stokes I and Q;
the symmetric cross-coupling amplitude and phase, $\epsilon$ and $\phi$ (eq.~18) describe the mixing of Stokes I with U and V;
the differential phase, $\psi$ (eq.~20) describes a rotation about the Stokes Q axis, which mixes Stokes U and V; and
the ellipticity angle, $\alpha$ (eq.~15) describes a rotation about the Stokes U axis, which mixes Stokes Q and V.
Missing from this list is the rotation about the Stokes~V axis
that corresponds to physically rotating the receiver about the line 
of sight, which mixes Stokes Q and U.
As noted in H2001, this rotation can be constrained only through
observation of a source with an accurately known position angle.
The same conclusion is reached in Appendix B of \cite{van04c}.

In HBS96, the product of \JM{G} and \JM{D} has 8 dof (four complex-valued matrix elements: $g_p$, $g_q$, $d_p$, and $d_q$) because the relative phase between each pair of array elements is important when computing the visibility matrix \eqnp{visibility}.  The absolute phase can be eliminated by replacing \JM{G} with its polar decomposition.

\subsection{Injectivity}
\label{sec:injectivity}

A model that does not injectively map the vector space of its
parameters to that of the transformations they describe
leads to ambiguous solutions.  
For example, for circularly-polarized receptors, both the differential
phase and the orientation of the receiver about the line of sight correspond to rotations about the Stokes~V axis.  Therefore, it is
necessary to ensure that the rotation axes in a model of the receiver
are defined with respect to the ($S_1$, $S_2$, $S_3$) basis and not
the ($Q$, $U$, $V$) basis.

Section 5.1 of H2001 describes ambiguity in the 
ellipticity angle, $\alpha$, and differential phase, $\psi$, that arises in the case of linearly-polarized receptors.
They assert that two possible solutions, 
$(\alpha_1, \psi_1) = (0, \psi_0)$ and 
$(\alpha_2, \psi_2) = (\pi/2, \psi_0 + \pi)$,
%
%
differ by only a 180$\deg$ rotation about the Stokes~V axis, which negates the unknown values of Stokes Q and U of the calibrator source.
Referring to \App{analytic_polarization_ellipse} and Section~4 of B2000, this ambiguity can be eliminated by restricting the ellipticity angle to the 
interval $-\pi/4 \le \alpha \le \pi/4$.

\subsection{Self-consistency}
\label{sec:self_consistency}

Many treatments begin with a description of linear transformations of
the electric field, as represented by Jones matrices.  In this case,
the corresponding Mueller matrices should be pure.  However, to simplify the
manual expansion of products of these Mueller matrices, some authors
introduce small-value approximations.  These typically result in an 
impure Mueller matrix that is inconsistent with the assumed linear 
response to the electric field.

For example, after assuming that the differential gain $\Delta G$ is small,
H2001 arrive at equation~(20),
\begin{equation}
\label{eqn:hpn+01_eqn20}
{\MM{M}}_A=\left( \begin{array}{cccc}
1 & \Delta G/2 & 0 & 0 \\
\Delta G/2 & 1 & 0 & 0 \\
0 & 0 & \cos\psi & -\sin\psi \\
0 & 0 & \sin\psi & \cos\psi
\end{array} \right).
\end{equation}
In principle, this Mueller matrix can be shown to be impure
by computing its associated target coherency matrix and applying
the test defined by \eqn{pure_target_coherency}.
In this particular case, it is more instructive and equally valid to show that this Mueller matrix cannot be pure because it can transform a purely polarized state into a nonphysical state that fails to satisfy \eqn{valid}.

Consider the observation of an ideal reference source \eqnp{ideal_noise_diode} with Stokes parameters,
$\Sv{S}=I[1,0,1,0]^T$.
The transformed Stokes parameters,
\begin{equation}
\Sv{S}'={\MM{M}}_A \Sv{S} = I[1,\Delta G/2,\cos\psi,\sin\psi]^T,
\end{equation}
have a degree of polarization,
\begin{equation}
p' = \sqrt{1+\frac{\Delta G^2}{4}} > 1.
\end{equation}
As no linear transformation of the electric field vector can convert a valid polarization state into an over-polarized state (\Proof{only_valid}), ${\MM{M}}_A$ cannot have an equivalent Jones matrix and therefore must be impure.
Not all impure Muller matrices result in over-polarization; some impure transformations depolarize a purely polarized state (e.g.\ see \App{impure}), and no linear transformation of the electric field is able to do so (\Proof{no_depolarization}).
This specific example demonstrates the potential pitfalls of using Mueller matrices and small-value approximations during numerical analysis.

\subsection{Numerical Stability}

Numerical stability of a mathematical model is critical when fitting the model to experimental data.
The fundamental problems caused by numerical instability impact on any method of model fitting; however, for the purpose of concrete illustration, this section and the analysis in \App{instability} focus on  
techniques that optimize a merit function by inverting a Hessian matrix that encodes its local curvature.
In this case, a model is unstable when the Hessian matrix becomes ill-conditioned
(i.e., prone to large numerical errors during inversion) or singular
(i.e., non-invertible).
The Hessian is ill-conditioned when two or more model parameters are highly
collinear; it is singular when the merit function is effectively independent of one or more
of its parameters, such that the partial derivatives of the model 
with respect to its degenerate parameters are zero.


Three types of numerical instability --
fundamental, extrinsic, and intrinsic --
are distinguished as follows.
Fundamental instability arises when the available data
do not constrain all of the dof in a surjective model; 
without any further assumptions or constraints, 
the degenerate dof will cause numerical methods to fail.
An extrinsically unstable model provides no means
of isolating degenerate dof, such that necessary constraining assumptions can be introduced when fundamental instability is encountered.
A model is intrinsically unstable if it causes numerical methods to fail even when there are sufficient constraints on all dof.

\subsubsection{Fundamental Instability}

Appendix B of \citet{van04c} describes the fundamental instability
caused by 2 degenerate dof that arise when the only available
experimental constraints are observations of unknown sources made 
at multiple parallactic angles.
In H2001, these degeneracies are avoided by introducing two assumptions. 
The rotation about the line of sight is assumed to be zero; 
and the calibrator source is assumed to have zero circular polarization.
The latter assumption is typically invalid when the calibrator source
is a pulsar; therefore, additional observations of other sources with
either known or assumed circular polarization must be incorporated to
constrain the instrumental mixing between I and V \cite[e.g.,][]{lck+16}.

The analysis of fundamental instability is extended in \App{instability}, which identifies additional degenerate dof in impure Mueller matrices, owing to unknown components of the instrumental response that commute with the matrix argument. It also presents an example of parameter collinearity that arises when an unknown component of the Stokes parameters intrinsic to the observed sources is an eigenvector of the matrix argument.
Owing to the degenerate sign of Stokes~V identified in \App{degeneracy}, the phase convention of the backend cannot be determined using only unknown sources observed at multiple parallactic angles, regardless of the basis defined by the polarization of the receptors.

Section 5.2 of H2001 identifies this degeneracy only
for the circular basis, and incorrectly concludes that two solutions, one with $\alpha_1 = -\pi/4$ and the other with $\alpha_2 = \pi/4$, differ only by negation of Stokes Q.
Stokes~V is also negated by the 180$\deg$ rotation about the Stokes U axis that describes the transformation from $\alpha_1$ to $\alpha_2$.
H2001 also incorrectly asserts that negating Stokes Q  
rotates the position angle by $90\deg$. Negating Stokes Q also negates the position angle, such that $\mathrm{P.A.}_1 = \pi/2 - \mathrm{P.A.}_2$.

\subsubsection{Extrinsic Instability}

Extrinsic instability arises when the degenerate dof in a model cannot be isolated from other dof.
For example, the 4 dof of the frontend transformation must effectively model two Euclidean rotations and two Lorentz boosts.
When the receptors are linearly polarized, these correspond to a rotation about and a boost along the Stokes U axis, and 2 potentially degenerate dof including a rotation about and a boost along the Stokes~V axis.
These rotations and boosts are clearly separated in equation~(19) of B2000.
In H2001, the Stokes~V boost is effectively differentiated from the Stokes U boost by the cross-coupling phase, and the Stokes~V rotation 
is assumed to be zero.
In the deviations from an ideal feed \JM{D} employed by HBS96, the 4 dof are inseparably combined in the two complex-valued elements, $d_p$, and $d_q$.
Consequently, the HBS96 model can be applied to 
single-antenna observations of point sources only if sufficient observational constraints eliminate the known degeneracies.

\subsubsection{Intrinsic Instability}

Appendix II of \cite{ck69} describes an intrinsically unstable model of undesirable cross coupling between the two receptors of an imperfect feed.
This model is adapted in equation~(A2) of \cite{scr+84} 
and equation~(16) of H2001,
\begin{equation}
\Jcol{e}'=\left(
 \begin{array}{cc}
1 & \epsilon_1 e^{i\phi_1} \\
\epsilon_2 e^{-i\phi_2} & 1
\end{array}
 \right) \Jcol{e}.\label{eqn:hpn+01_eqn16}
\end{equation}
The equation is intrinsically unstable in the vicinity of the 
ideal solution owing to the polar coordinate singularity at the origin,
where the phase angle $\phi_k$ becomes 
undetermined as the cross-coupling amplitude $\epsilon_k$ 
approaches zero.  
This problem persists in equation~(18) of H2001, where
$\phi$ is poorly constrained when $\epsilon$ is small. 
\cite{joh02} addresses this problem by directly modeling the real 
and imaginary components of all complex-valued quantities,
instead of their amplitude and phase.
%
%
In contrast, equation~(15) of B2000 is
unstable only at the poles, where $\chi = \pm\pi/4$, 
and the orientation $\theta$ becomes degenerate
with the differential phase.  When the basis is correctly defined, these poles are well away from the region 
in which the anticipated solution lies.


\section{Conclusion}
\label{sec:conclusion}

The geometry of the polarization ellipse 
that defines the Stokes parameters is elegantly connected
to that of the coherency matrix via the Pauli matrices.
Geometry also relates linear transformations of the electric field 
to their impact on the Stokes parameters, leading to a physically meaningful
and broadly applicable classification of polarimetric transformations.
%
%
%
The properties of these transformations simplify the analysis of the elements of the signal path and yield both theoretical insights and practical guidelines that can be applied during calibration.

These guidelines include criteria for evaluating and selecting a model of the instrumental response that performs well during numerical analysis of experimental data.
Among those considered,  the model introduced by \citet{bri00} satisfies all of the selection criteria for single-antenna observations of a point source.
In this physically-motivated and self-consistent model, there is a one-to-one mapping between parameters and linear transformations, which are described
using equations that remain numerically stable during modeling.

This introduction focuses primarily on the mathematical and conceptual foundations of polarimetry, with emphasis on the numerical analysis of radio astronomical observations.  For a broader review of both the physical processes that produce polarized radiation and the astrophysical phenomena that can be studied through polarimetry, the review by \cite{tri14} is exceptionally thorough and insightful.

Radio polarimetry is a vibrant field of research that continues to yield transformative insights and discoveries in astrophysics, driven by both technological advances and innovative techniques.  For example, the relatively recent discovery of polarization closure traces \citep{bp20,snt22} enabled the first images of magnetic field structure near the M87 black hole using very long baseline interferometry \citep{2021ApJ...910L..13E}.
Novel techniques of high-fidelity polarimetry led to the discovery of a surprisingly high degree of linear polarization in low-frequency radio images of solar bursts \citep{dkom25}.
At even higher time resolution, measurements of the polarized radiation from pulsars, magnetars and fast radio bursts continue to reveal new insights into their nature and extreme environments \citep[e.g.,][]{crh+06,2013Natur.501..391E,2018Natur.553..182M,ptv+22}.

Current and future scientific goals, such as detecting the neutral Hydrogen emission from the epoch of reionization \citep[e.g.,][]{2018ApJ...867...15T,2020MNRAS.493.1662M} and understanding the origin and evolution of interstellar and intergalactic magnetic fields \citep[e.g.,][]{2017ARA&A..55..111H}, motivate the development of next-generation facilities and methods of polarimetric analysis, such as Faraday tomography \citep[e.g.,][]{2000JGR...10516063G,2017A&A...597A..98V}.
Furthermore, the unprecedented sensitivity, resolution, and fields of view of new observatories necessitate further research and development of novel methods of direction-dependent interferometric calibration \citep[e.g.,][]{2015MNRAS.449.2668S} and correction of ionospheric Faraday rotation \citep[e.g.,][]{2019A&A...622A...5D}.
Improved polarimetric fidelity also has the potential to increase the sensitivity of pulsar timing array experiments to the low-frequency stochastic gravitational wave background \citep[e.g.,][]{gcv+23,rvg+24,gvc+25}.

Given the breadth and potential impact of scientific knowledge that remains to be discovered through polarimetry, I hope that this introduction might help others to further pursue the subject.

\begin{acknowledgments}

I'm grateful to the mentors who sparked my interest in polarimetry and the colleagues who share their enthusiasm for the subject, especially \nohyphens{Matthew Britton, Paul Demorest, Russell Edwards, Marisa Geyer, Lucas Guillemot, Johan Hamaker, JinLin Han, Simon Johnston, Aris Karastergiou, Vivek Venkatraman Krishnan, Dick Manchester, Maaijke Mevius, Aris Noutsos, Steve Ord, Stefan Os{\l}owski, Aditya Parthasarathy, John Reynolds, Maciej Serylak, Ben Stappers, and Caterina Tiburzi}.  Marisa's visit in 2018 prompted the development of this tutorial, and Caterina and Maaijke provided insightful feedback and brilliant ideas that significantly enhanced its elaboration.
The comprehensive and constructive criticism of the anonymous reviewer led to further additions and improvements.

\end{acknowledgments}

\newpage

\begin{appendix}

\renewcommand{\theHequation}{\thesection.\arabic{equation}}
\renewcommand{\theHmyproof}{\thesection.\arabic{myproof}}



\setcounter{equation}{0}

\section{Signal Processing Fundamentals}
\label{app:foundations}

This section reviews the theoretical basis for using complex numbers to represent signals and Jones matrices to represent linear transformations of the electric field.

\subsection{The Analytic Signal}
\label{app:analytic}

The voltage signal from each receptor is a real-valued function of
time, or process, that may be represented by its associated analytic
signal.
The analytic signal, also known as Gabor's complex signal, is a
complex-valued representation of a real-valued process that provides
its instantaneous amplitude and phase.  
The analytic signal associated with a process, $x(t)$, is defined by
\begin{equation}
  z(t) = x(t) + \Ci\hat x(t),
\end{equation}
where $\hat x(t)$ is the Hilbert transform of $x(t)$ \citep{pap65}.
Following a Fourier transform, defined by
\begin{equation}
  X(\nu) = \int_\infty^\infty x(t) \exp(- \Ci 2\pi \nu t) dt,
  \label{eqn:Fourier_transform}
\end{equation}
the Hilbert transform is equivalent to
\begin{equation}
\hat{X}(\nu)=H(\nu)X(\nu),    
\end{equation}
where
\begin{equation} H(\nu) = \left\{ \begin{array}{cc}
                -\Ci  & \nu > 0 \\
                \Ci & \nu < 0
                \end{array} \right.
\end{equation}
is known as the quadrature filter \citep{pap65}.  
The quadrature filter can be understood as a
90\degr\ phase shifter by noting that 
the Hilbert transform of $\cos(\nu_0t)$ is equal to
$\sin(\nu_0t)$ (see \Tab{fourier}).  
Using the quadrature filter, it can also be
shown that the Fourier transform of the analytic signal, $Z(\nu)$, is
equal to zero for $\nu < 0$.
\begin{equation}
\begin{split}
 Z(\nu) & =  X(\nu) + \Ci\hat X(\nu) \\
	& =  X(\nu) + \Ci H(\nu)X(\nu) \\
	& =  \left \{ \begin{array}{cc} 
	2X(\nu)  & \nu > 0 \\
	0 & \nu < 0 
	\end{array} \right.
\end{split}
\end{equation}
Conversely, the analytic signal associated with $x(t)$ is
produced
by suppressing the negative frequencies in $X(\nu)$.

As it is derived from the real-valued process, the analytic signal
does not contain any additional information.  However, the analytic
signals associated with two orthogonal senses of polarization, $e_0(t)$ and
$e_1(t)$, facilitate calculation of the coherency matrix.  

\begin{table}
\caption
{Useful Fourier Transform pairs. The left column lists functions of
time.  In the right column, the corresponding Fourier transform is
given as a function of oscillation frequency, $\nu$. 
}
\label{tab:fourier}
\begin{center}
\begin{tabular}{c|c}
\hline
\hline
$x(t)$ & $X(\nu)$ \\
\hline \\ [-3mm]
$\cos(2\pi\nu_0t)$ & $(\delta(\nu+\nu_0)+\delta(\nu-\nu_0)) / 2$ \\ [3mm]
$\sin(2\pi\nu_0t)$ & $\Ci (\delta(\nu+\nu_0)-\delta(\nu-\nu_0)) / 2$ \\ [2mm]
\hline
\multicolumn{2}{c}{Quadrature Filter} \\
\hline \\ [-3mm]
$h(t)=(\pi t)^{-1}$ & 
$H(\nu) =\left\{ \begin{array}{cc}
            -\Ci & \nu > 0 \\
            \Ci  & \nu < 0 \end{array} \right.$ \\ [4mm]
\hline
\multicolumn{2}{c}{Lowpass Filter} \\
\hline \\ [-3mm]
$\pi(t)=\sinc(\pi\bw t)$ & 
$\Pi(\nu/\bw) = \left\{ \begin{array}{cc}
            0  & |\nu/\bw| > 1/2 \\
            1/2& |\nu/\bw| = 1/2 \\
            1  & |\nu/\bw| < 1/2 \end{array} \right.$  \\ [6mm]
\hline
\multicolumn{2}{c}{Shift Theorem} \\
\hline \\ [-3mm]
$x'(t)=x(t + \tau)$ & 
$X'(\nu) = X(\nu) \exp(\Ci \pi \nu \tau)$  \\ [2mm]
\hline
\end{tabular}
\end{center}
\end{table}



\subsection{Narrow-band Approximation}
\label{app:narrow_band_approximation}

Presented with a scalar input signal, $e(t)$, the output of a one-dimensional linear time-invariant system is given by the convolution,
\begin{equation}
e'(t)=(j*e)(t) \equiv \int_{-\infty}^{\infty}{x(\tau)j(t-\tau)d\tau}.
\label{eqn:convolution}
\end{equation}
where $j(t)$ is the one-dimensional impulse response function.  In a two-dimensional
system, each output signal is given by a linear combination of the input
signals,
\begin{equation}
\begin{split}
e_1'(t)=(j_{00}*e_0)(t) + (j_{01}*e_1)(t), \\
e_2'(t)=(j_{10}*e_0)(t) + (j_{11}*e_1)(t).
\end{split}
\label{eqn:convolution2}
\end{equation}

Let $e_0(t)$ and $e_1(t)$ be the components of the electric field vector $\Jcol{e}(t)$ and $j_{mn}(t)$ be the elements of the $2\times2$ impulse response matrix, $\JM{j}(t)$.
By the convolution theorem, \eqn{convolution2} is equivalent to
\begin{equation}
\Jcol{E}'(\nu)=\JM{J}(\nu)\Jcol{E}(\nu),
\label{eqn:convolution_nu}
\end{equation}
where $\JM{J}(\nu)$ and $\Jcol{E}(\nu)$ are the Fourier transforms of $\JM{j}(t)$ and $\Jcol{e}(t)$. Under the assumption that $\JM{J}(\nu)$ is
approximately constant over a sufficiently narrow range of frequencies, convolution reduces to multiplication by a Jones matrix. 



\setcounter{equation}{0}
\section{Vector, Matrix, and Tensor Notation}
\label{app:index}

The following typographical conventions are used to indicate the dimensions of mathematical symbols.  Scalar quantities such as the real value $x$ or the complex value $z$ are typeset in italics.
All vectors are typeset with a bold italic font.
Two-dimensional vectors, such as the electric field $\Jcol{e}$, are not underlined.
Three-dimensional vectors, such as the Stokes polarization vector $\Pv{S}$, have
a single underline.
Four-dimensional vectors, such as the Stokes four-vector $\Sv{S}$,
have a double underline.

All matrices have a bold roman font; 
$2\times2$ matrices like the Jones matrix $\JM{J}$ are not underlined;
$3\times3$ matrices like the depolarizer matrix $\PM{M}_\Delta$ are underlined once; and
$4\times4$ matrices like the Mueller matrix $\MM{M}$ are underlined twice.
Rank 4 tensors, typeset using a bold calligraphic font, 
include the two-dimensional polarization transfer tensor $\JT{U}$ and the
four-dimensional $\MM{\JT{T}}$ in \App{pure_Mueller_matrices}.

Unit vectors, such as a receptor with unity gain
$\hat{\Jrow{r}}$ or a rotation axis $\hat{\Pv{n}}$, are decorated with a hat.
A tilde is used to denote both singular matrices and null four-vectors, such as the instantaneous coherency matrix $\tilde{\cohM{\rho}}$ and  Stokes parameters $\tilde{\Sv{s}}$ computed from a single instance of the electric field.

The elements of a column vector are labeled with a sub-script, as in
the two components of the electric field, 
%
\begin{equation}
	\Jcol{e} =
	\left( 
	\begin{array}{c}
		e_x \\
		e_y
	\end{array}
	\right).
\end{equation}
and the four components of the Stokes parameters, $S_\mu$.
To distinguish between Jones vectors that represent the electric field and those that represent receptors, the latter are described using row vectors with a super-scripted index; e.g.
\begin{equation}
	\Jrow{r} = \left( r^x, \, r^y \right).
\end{equation}
The Hermitian transpose converts a column vector to a row vector (and vice versa) and converts each component to its complex conjugate; e.g.
\begin{equation}
	\Jcol{e}^\dagger = \left( e_x^*, \, e_y^* \right).
	\label{eqn:Hermitian_transpose}
\end{equation}

The scalar product (or dot product) between
a row vector and a column vector is given by
\begin{equation}
	\Jrow{r} \cdot \Jcol{e} = r^x e_x + r^y e_y = r^i e_i ,
	\label{eqn:scalar_product}
\end{equation}
where the last equality uses the Einstein convention that implies summation over the repeated index $i$.
A scalar product is implied when a row vector on the left
is multiplied by a column vector on the right; i.e.,
\begin{equation}
	\Jrow{r} \Jcol{e} = \Jrow{r} \cdot \Jcol{e} = r^i e_i .
	\label{eqn:implicit_scalar_product}
\end{equation}
%

Using the same index notation, the element in the $\irow^\mathrm{th}$ row and
$\icol^\mathrm{th}$ column of a matrix \JM{A} is
\begin{equation}
	A_\irow^\icol
\end{equation}
and the $\irow^\mathrm{th}$ row and
$\icol^\mathrm{th}$ column of a matrix product is given by the sum
\begin{equation}
\left\{ \JM{A} \cdot \JM{B} \right\}_\irow^\icol 
 = A_\irow^\gamma B_\gamma^\icol.
	\label{eqn:matrix_product}
\end{equation}
Similarly, the $\irow^\mathrm{th}$ row of a matrix times a column vector 
is given by
\begin{equation}
\left\{ \JM{A} \Jcol{x} \right\}_\irow
 = A_\irow^\gamma x_\gamma.
	\label{eqn:matrix_times_vector}
\end{equation}

Vectors and matrices can be seen as rank 1 and rank 2 tensors, respectively, where the rank of a tensor is equal to the number of indeces required to define its elements.  
%
%
The elements of a rank 4 tensor are described using four indeces; e.g.
\begin{equation}
	A_{\irow\jrow}^{\icol\jcol}.
\end{equation}

Given two tensors, a rank $N$ tensor $A$ and a rank $M$ tensor $B$, the tensor product $A \otimes B$ yields a rank $N+M$ tensor.  
For example, the tensor product of two matrices (each a rank 2 tensor) yields a rank 4 tensor with elements given by
\begin{equation}
\label{eqn:matrix_tensor_product}
	\left\{ \JM{A} \otimes \JM{B} \right\}_{\irow\jrow}^{\icol\jcol} = A_\irow^\icol B_\jrow^\jcol.
\end{equation}
The tensor product of a column vector and a row vector (each a rank 1 tensor) yields a matrix (rank 2 tensor)
with elements 
\begin{equation}
	\left\{ \Jcol{e} \otimes \Jrow{r}\right\}_i^j = e_i r^j.
	\label{eqn:tensor_product}
\end{equation}
A tensor product is implied when a column vector on the left
is multiplied by a row vector on the right; e.g.
\begin{equation}
	\Jcol{e} \Jrow{r} = \Jcol{e} \otimes \Jrow{r}.
	\label{eqn:implicit_tensor_product}
\end{equation}
%
%
A contraction of two tensors, $A\cdot B$, is defined as a tensor product followed by summation over a pair of matching indeces, thereby reducing the rank of the result of the tensor product by two.
For example, the scalar product of a row vector and column vector is equivalent to a contraction.  Each is a rank 1 tensor and their tensor product yields a rank 2 tensor; summation over their indeces, as in \eqn{scalar_product}, yields a scalar (rank 0 tensor).
Similarly, the product of two matrices is equivalent to a tensor contraction.  Each is a rank 2 tensor and their tensor product yields a rank 4 tensor; summation over a pair of indeces, as in \eqn{matrix_product}, yields a rank 2 tensor (another matrix).

A double contraction of two tensors, $A \dcbo B$, is defined as a tensor product followed by summation over two pairs of matching indeces, thereby reducing the rank of the result of the tensor product by four.
For example, the double contraction of a rank 4 tensor $\JT{U}$ with a (rank 2) matrix \JM{C} yields another (rank 2) matrix with elements given by
\begin{equation}
	\left\{ \mathcal{U} \dcbo \JM{C} \right\}_\irow^\icol = \mathcal{U}_{\irow\jrow}^{\icol\jcol} C_\jcol^\jrow.
	\label{eqn:double_contraction}
\end{equation}

Using the Einstein summation convention, the trace of a matrix \JM{A} is
\begin{equation}
	\tr{\JM{A}} = A^\beta_\beta
	\label{eqn:trace}
\end{equation}
and the trace of a matrix product
\begin{equation}
\tr{ \JM{A} \JM{B} } = \tr{ \JM{B} \JM{A} } = A_\beta^\gamma B_\gamma^\beta = \JM{A} \dcbo \JM{B} = \JM{B} \dcbo \JM{A}.
\label{eqn:trace_inner_product}
\end{equation}
That is, the trace of a matrix product is equivalent to the double contraction of the matrices, also known as the projection with respect to the trace inner product.
Like the scalar product of two vectors, this projection is symmetric
and matrices \JM{A} and \JM{B} are orthogonal with respect to the trace inner product if $\JM{A} \dcbo \JM{B} = 0$. 
%

\Eqns{outerBilinear}{and}{spinorBilinear} introduce transformation properties exhibited by
two tensor double contractions.  \Eqn{outerBilinear}
follows from the definition of the tensor product \eqnsp{indexOuterBilinear}{and}{matrix_tensor_product}.

\begin{proof}
\label{proof:outerBilinear}
\begin{align*}
\left\{ \left( \outerBilinear{\bf A}{\bf B} \right) \dcbo \JM{C} \right\}_\irow^\icol 
& = \left\{ \outerBilinear{\bf A}{\bf B} \right\}_{\irow\jrow}^{\icol\jcol} C_\jcol^\jrow
& \peqn{double_contraction} \\
& = A_\irow^\icol B_\jrow^\jcol C_\jcol^\jrow & \peqn{matrix_tensor_product} \\
& = A_\irow^\icol (\JM{B} \dcbo \JM{C}) & \peqn{trace_inner_product} \\
& = \left\{ {\bf A}\left(\dc{\bf B}{\bf C}\right) \right\}_\irow^\icol 
\end{align*}
\end{proof}

Likewise, \Eqn{spinorBilinear}
follows from the definition of the $\tilde\otimes$ operator \eqnp{indexSpinorBilinear}.
\begin{proof}
\label{proof:spinorBilinear}
\begin{align*}
\left\{ \left( \spinorBilinear{\bf A}{\bf B} \right) \dcbo \JM{C} \right\}_\irow^\icol 
& = \left\{ \spinorBilinear{\bf A}{\bf B} \right\}_{\irow\jrow}^{\icol\jcol} C_\jcol^\jrow
& \peqn{double_contraction} \\
& = A_\irow^\jcol B_\jrow^\icol C_\jcol^\jrow & \peqn{indexSpinorBilinear} \\
& = A_\irow^\jcol \left\{ \JM{C} \cdot \JM{B} \right\}_\jcol^\icol  & \peqn{matrix_product} \\
& = \left\{ {\bf A} \JM{C} \JM{B} \right\}_\irow^\icol & \peqn{matrix_product}
\end{align*}
\end{proof}



\setcounter{equation}{0}
\section{Linear Algebra Fundamentals}
\label{app:linear_algebra}

This section reviews the elements of linear algebra used
in this work, beginning with the Hermitian basis matrices.
The spectral decomposition is used
to derive the axis-angle representations of rotations and boosts,
the symmetry properties of which are used to identify
matrices that commute and the eigenmatrices of a congruence transformation.
Finally, the symmetry between \eqns{indexOuterBilinear}{and}{indexSpinorBilinear} is used to define a pure Mueller matrix and its associated Jones matrix, and an example of a depolarizing impure Mueller matrix is presented.


\subsection{Hermitian Basis Matrices}
\label{app:basis}

The Hermitian basis matrices \eqnp{Pauli} have a number of useful algebraic properties that
enable meaningful geometric interpretations of the equations used in this work.
When the basis matrices are indexed with a Greek character, the index is understood
to span all dimensions including the identity; e.g.,  $\mu \in \{0,1,2,3\}$.  When Latin characters are used,
the index is understood to span only the Pauli matrces; e.g.,  $m \in \{1,2,3\}$.
The following identities prove useful in this work.
\begin{eqnarray}
    \tr{\pauli{i}}&=&0 \label{eqn:Pauli_trace} \\
    |\pauli{i}|&=&-1 \label{eqn:Pauli_determinant} \\
    |\pauli{\mu} + \pauli{\nu} | &=& |\pauli{\mu}| + |\pauli{\nu}|; \ensuremath{\quad \mu \ne \nu} \label{eqn:basis_sum_determinant} \\
    \pauli{\mu} \dcbo \pauli{\nu} &=& 2 \delta_{\mu\nu} \label{eqn:basis_orthogonal} \\
    \pauli{i}\,\pauli{j} &=& \delta_{ij}\,\pauli{0} + i\epsilon_{ijk}\,\pauli{k} \label{eqn:Pauli_product}
\end{eqnarray}
In the last identity, $\epsilon_{ijk}$ is the permutation symbol and summation over the index $k$ is implied.
\Eqn{Pauli_trace} expresses the traceless property of the Pauli matrices.
\Eqns{Pauli_determinant}{and}{basis_sum_determinant} can be used to show that
\begin{equation}
| x_\mu \pauli{\mu} | = x_\mu^2 |\pauli{\mu}| = x_0^2 -x_1^2 -x_2^2 - x_3^2.
\label{eqn:determinant_is_invariant}
\end{equation}
That is, the determinant of a linear combination of the Hermitian
basis matrices is equal to the invariant interval of the 
four-vector of the scalar coefficients, $x_\mu$.
%
%
\Eqn{basis_orthogonal} states that the Hermitian basis matrices
are mutually orthogonal with respect to the trace inner product.
The first term on the right-hand side of \eqn{Pauli_product} identifies
the Pauli matrices as the square roots of $\pauli{0}$.  
The second term describes how the Pauli matrices
behave like the basis vectors of a right-handed coordinate system, 
where $\hat{\Pv{v}}_i \times \hat{\Pv{v}}_j = \epsilon_{ijk}\,\hat{\Pv{v}}_k$.
Using this identity, the product of two matrices,
\begin{equation}
\begin{split}
\JM{A}\JM{B} &=
 (a\,\pauli{0}+\Pv{a}\cdot\Pv{\sigma})(b\,\pauli{0} + \Pv{b}\cdot\Pv{\sigma}) \\
        &=
 (ab + \Pv{a}\cdot\Pv{b})\pauli{0} + (a\Pv{b} + b\Pv{a}
    + \Ci \Pv{a}\times\Pv{b}) \cdot \Pv{\sigma},
    \label{eqn:multiplication_rule} 
\end{split}
\end{equation}
where $\Pv{\sigma}=\left( \pauli{1},  \pauli{2}, \pauli{3} \right)$ is
the Pauli vector that appears in \eqns{Boost}{and}{Rotation}.

\subsection{Basis Transformations}
\label{app:basis_transformations}

A pair of orthonormal receptors, $\hat{\Jrow{r}}_0$ and $\hat{\Jrow{r}}_1$, have unit length and are orthogonal; that is, they satisfy
\begin{equation}
\hat{\Jrow{r}}_i \cdot \hat{\Jrow{r}}_j^\dagger = \delta_{i,j}.
\label{eqn:orthonormality}
\end{equation}
The projection of the electric field vector onto a pair of orthonormal receptors defines a basis transformation that is equivalent to a unitary Jones matrix.
\begin{proof}
\label{proof:orthonormal_receptors}
Let 
$$\JM{R} = \left( \begin{array}{c}
		\hat{\Jrow{r}}_0 \\
		\hat{\Jrow{r}}_1
	\end{array} \right)$$
such that 
\begin{align*}
\left\{ \JM{R} \JM{R}^\dagger \right\}_i^j 
& = \left\{ \JM{R} \right\}_i^k \left\{ \JM{R}^\dagger \right\}_k^j
  & \peqn{matrix_product} \\
& = \left\{ \hat{\Jrow{r}}_i \right\}^k \left\{ \hat{\Jrow{r}}_j^\dagger \right\}_k  \\
& = \hat{\Jrow{r}}_i \cdot \hat{\Jrow{r}}_j^\dagger 
  & \peqn{scalar_product} \\
& = \delta_{i,j} 
  & \peqn{orthonormality}
\end{align*}
That is, $\JM{R} \JM{R}^\dagger = \pauli{0}$; 
therefore, $\JM{R}^\dagger=\JM{R}^{-1}$.
\end{proof}
In general, a unitary matrix has a determinant equal to $\exp(\Ci\phi)$, and
multiplication by $\exp(-\Ci\phi/2)$ yields a unimodular matrix
(with $|\JM{R}|=1$).
Basis transformations defined by unitary matrices are used extensively throughout both \Sec{signal_path} and the following sections of the appendix.  

\subsection{Spectral Decomposition}
\label{app:spectral_decomposition}

%
The eigenvalues $\lambda$ and eigenvectors $\Jcol{v}$ of
a matrix \JM{A} satisfy the following (logically-equivalent) equations,
\begin{equation}
    \label{eqn:eigen}
    \JM{A}\Jcol{v} = \lambda\Jcol{v} \quad \Leftrightarrow \quad
    \left(\JM{A} - \lambda\JM{I} \right) \Jcol{v} = 0,
\end{equation}
where \JM{I} is the identity matrix with the dimensions of \JM{A}.
\Eqn{eigen} can be expressed simultaneously for all eigenvalue/eigenvector pairs by $\JM{A} \JM{R} = \JM{R} \cohM{\Lambda}$, where the $k^\mathrm{th}$
column of {\bf{R}} is the unit eigenvector $\hat{\Jcol{v}}_k$, and the diagonal matrix $\cohM{\Lambda}$ is defined by $\Lambda_k^j = \delta_{j,k} \lambda_k$.
This can be rearranged to yield the eigendecomposition of \JM{A},
\begin{equation}
\JM{A} = \JM{R} \cohM{\Lambda} \JM{R}^{-1} \label{eqn:eigendecomposition}.
\end{equation}

If \JM{A} is normal (i.e., $\JM{A}^\dagger\JM{A}=\JM{A}\JM{A}^\dagger$),
and $\Jcol{v}$ is an eigenvector of \JM{A} with associated eigenvalue $\lambda$, 
then $\Jcol{v}$ is also an eigenvector of $\JM{A}^\dagger$ with associated eigenvalue $\lambda^*$.

\begin{proof}
\label{proof:normal_eigenvalue}
Define $\JM{L}=\JM{A} - \lambda\JM{I}$ and consider
\begin{align*}
|\JM{L}^\dagger \Jcol{v}|^2
  &= \left( \JM{L}^\dagger \Jcol{v} \right)^\dagger \left( \JM{L}^\dagger \Jcol{v} \right) \\
  &= \Jcol{v}^\dagger \JM{L} \JM{L}^\dagger \Jcol{v} \\
  &= \Jcol{v}^\dagger \JM{L}^\dagger \JM{L} \Jcol{v} & \text{\JM{L} is normal} \\
  &= \left( \JM{L}\Jcol{v} \right)^\dagger \JM{L} \Jcol{v} \\
  &= |\JM{L} \Jcol{v}|^2 \\
  &= |\left(\JM{A} - \lambda\JM{I} \right) \Jcol{v}|^2 = 0 & \peqn{eigen}
\end{align*}
Therefore $\JM{L}^\dagger \Jcol{v} = \left(\JM{A}^\dagger - \lambda^*\JM{I} \right) \Jcol{v} = 0.$
\end{proof}
\noindent
Furthermore, if \JM{A} is normal, then \eqn{eigendecomposition} is equivalent to a
congruence transformation by a unitary matrix known as its spectral decomposition,
\begin{equation}
\label{eqn:congruence_eigendecomposition}
\JM{A} = \JM{R} \cohM{\Lambda} \JM{R}^\dagger.
\end{equation}

\begin{proof}
\label{proof:spectral_decomposition}
If \JM{A} is normal, then
\begin{align*}
( \JM{A}^\dagger \hat{\Jcol{v}}_i )^\dagger \hat{\Jcol{v}}_j
&= \hat{\Jcol{v}}_i^\dagger \JM{A} \hat{\Jcol{v}}_j  \\
( \lambda_i^* \hat{\Jcol{v}}_i)^\dagger \cdot \hat{\Jcol{v}}_j
&= \hat{\Jcol{v}}_i^\dagger \cdot ( \lambda_j \hat{\Jcol{v}}_j)
 & \text{\Proof{normal_eigenvalue}} \\
\lambda_i \hat{\Jcol{v}}_i^\dagger \cdot \hat{\Jcol{v}}_j
&= \lambda_j \hat{\Jcol{v}}_i^\dagger \cdot \hat{\Jcol{v}}_j \\
(\lambda_i-\lambda_j) \hat{\Jcol{v}}_i^\dagger \cdot \hat{\Jcol{v}}_j &= 0
\end{align*}
If $\lambda_i \ne \lambda_j$, then
$\hat{\Jcol{v}}_i^\dagger \cdot \hat{\Jcol{v}}_j = 0$ and
\begin{align*}
\left\{ \JM{R}^\dagger \JM{R} \right\}_i^j 
& = \left\{ \JM{R}^\dagger \right\}_i^k \left\{ \JM{R} \right\}_k^j \\
& = \left\{ \hat{\Jcol{v}}_i^\dagger \right\}^k \left\{ \hat{\Jcol{v}}_j \right\}_k 
  = \hat{\Jcol{v}}_i^\dagger \cdot \hat{\Jcol{v}}_j = \delta_{i,j}
\end{align*}
That is, $\JM{R}^\dagger \JM{R} = \JM{I}$; therefore, $\JM{R}^\dagger=\JM{R}^{-1}.$
\end{proof}

In the natural basis defined by ${\bf{R}}^\dagger$, the matrix $\JM{A}$ becomes the diagonal matrix,
\begin{equation}
\label{eqn:diagonal_decomposition}
\cohM{\Lambda} = \JM{R}^\dagger \JM{A} \JM{R}.
\end{equation}
Both Hermitian and unitary matrices are normal, and each takes
diagonal form in the natural basis defined by its eigenvectors.  Consequently, any Hermitian matrix can be diagonalized such that it is equivalent to a differential gain transformation (see \Sec{differential_gain}) in its natural basis.
Similarly, any unitary matrix is equivalent to differential phase (see \Sec{differential_phase}) in its natural basis.
%
This is applied in \Proof{Faraday_rotation}, where rotation about the line of sight is shown to be equivalent to differential phase between the circularly-polarized natural modes of a magnetized cold plasma.

The spectral decomposition can also be written as a linear combination of outer products,
\begin{equation}
\JM{A} = \lambda_k\, \hat{\Jcol{v}}_k \otimes \hat{\Jcol{v}}_k^\dagger.
\label{eqn:spectral_decomposition}
\end{equation}

\begin{proof}

\begin{align*}
A_i^j
&= \left\{ {\bf{R}} \cohM{\Lambda} {\bf{R}}^\dagger \right\}_i^j 
    & \peqn{congruence_eigendecomposition} \\
&=  \left\{ {\bf{R}} \right\}_i^k
    \left\{ \cohM{\Lambda} \right\}_k^l
    \left\{ {\bf{R}}^\dagger \right\}_l^j 
    & \peqn{matrix_product} \\
&= \left\{ \hat{\Jcol{e}}_k \right\}_i \; \delta_{k,l} \; \lambda_k \left\{ \hat{\Jcol{e}}_l^\dagger \right\}^j \\
&= \lambda_k \;\left\{ \hat{\Jcol{e}}_k \right\}_i \left\{ \hat{\Jcol{e}}_k^\dagger \right\}^j \\
&= \lambda_k \; \left\{ \hat{\Jcol{e}}_k \otimes \hat{\Jcol{e}}_k^\dagger \right\}_i^j
    & \peqn{implicit_tensor_product}
\label{eqn:eigen_sum}
\end{align*}
\end{proof}
\noindent
If the eigenvalues are distinct, then
\begin{equation}
\label{eqn:projection}
\JM{P}_k=\hat{\Jcol{v}}_k \otimes \hat{\Jcol{v}}_k^\dagger
\end{equation}
form a complete set of mutually orthogonal projection matrices; i.e.,
\begin{equation}
\label{eqn:idempotent_orthogonal}
\JM{P}_i \JM{P}_j = \delta_{i,j} \JM{P}_j
\end{equation}
and
\begin{equation}
\sum_i \JM{P}_i = \JM{I}
\label{eqn:projection_complete}
\end{equation}
\begin{proof}
\label{proof:orthogonal_idempotent}
$\JM{P}_k$ are orthogonal and idempotent.
\begin{align*}
\JM{P}_i \JM{P}_j &= 
(\hat{\Jcol{v}}_i  \hat{\Jcol{v}}_i^\dagger)
(\hat{\Jcol{v}}_j  \hat{\Jcol{v}}_j^\dagger) 
    & \peqn{implicit_tensor_product} \\
&= \hat{\Jcol{v}}_i ( \hat{\Jcol{v}}_i^\dagger
\hat{\Jcol{v}}_j ) \hat{\Jcol{v}}_j^\dagger
    & \text{Associativity} \\
&= \delta_{i,j} \hat{\Jcol{v}}_i \hat{\Jcol{v}}_j^\dagger
   & \hat{\Jcol{v}}_i^\dagger \hat{\Jcol{v}}_j = \delta_{i,j} \\
&= \delta_{i,j} \JM{P}_j & \peqn{projection}
\end{align*}
\end{proof}
\begin{proof}
$\JM{P}_k$ form a complete set.

Let $\Jcol{w} = c_j \hat{\Jcol{v}}_j$ and consider
\begin{align*}
\sum_i \JM{P}_i \Jcol{w} 
&= \left( \hat{\Jcol{v}}_i \hat{\Jcol{v}}_i^\dagger \right)
\left( c_j \hat{\Jcol{v}}_j \right) 
  & \peqn{projection} \\
&= c_j \hat{\Jcol{v}}_i \hat{\Jcol{v}}_i^\dagger \hat{\Jcol{v}}_j 
  & \text{Commutativity} \\
&= c_j \hat{\Jcol{v}}_i \delta_{i,j} 
  & \hat{\Jcol{v}}_i^\dagger \hat{\Jcol{v}}_j = \delta_{i,j} \\
& = c_j \hat{\Jcol{v}}_j \\
& = \Jcol{w}
\end{align*}
\end{proof}

The $2 \times 2$ Hermitian coherency matrix has two eigenvalues, $\lambda_k = (S_0 \pm |\Pv{S}|)/2$, and
can be expressed as a spectral decomposition,
\begin{equation}
\begin{split}
\label{eqn:spectral_decomposition_of_coherency_matrix}
\cohM{\rho} 
&=  \lambda_0\, \hat{\Jcol{e}}_0 \otimes \hat{\Jcol{e}}_0^\dagger
    + \lambda_1\, \hat{\Jcol{e}}_1 \otimes \hat{\Jcol{e}}_1^\dagger \\
&=  \lambda_0\,\tilde{\cohM{\rho}}_0 + \lambda_1\,\tilde{\cohM{\rho}}_1,
\end{split}
\end{equation}
where $\tilde{\cohM{\rho}}_k = \hat{\Jcol{e}}_k \otimes \hat{\Jcol{e}}_k^\dagger$ correspond to purely polarized states,
as in \eqn{instantaneous_coherency}.
That is, any partially polarized state can be represented as an incoherent superposition of a pair of purely polarized states
that are orthogonally polarized ($\hat{\Jcol{e}}_0^\dagger\hat{\Jcol{e}}_1 = 0$ and $\tilde{\cohM{\rho}}_0 \tilde{\cohM{\rho}}_1 = 0$).
If the signal is unpolarized, then the eigenvalues are equal and
any non-zero vector is an eigenvector.
If $\cohM{\rho}$ is purely polarized, then one of the eigenvalues equals zero, and the polarization state is completely described by the Jones vector associated with the non-zero eigenvalue.

In the natural basis defined by ${\bf{R}}^\dagger$, the eigenvalues
$\lambda_m$ are equal to the variances of two uncorrelated signals
received by orthogonally polarized receptors described by
the Hermitian transposes of the eigenvectors.  In this basis,
the total intensity, $S_0=\lambda_0+\lambda_1$; the
polarized intensity, $S_1=|\Pv{S}|=\lambda_0-\lambda_1$; and
$S_2=S_3=0$.
That is, ${\bf{R}}^\dagger$ rotates the basis such that the
polarization vector points along $S_1$.  In this basis, it is clear that the degree of polarization
\begin{equation}
\label{eqn:degree_of_polarization_eigen}
p = \frac{|\Pv{S}|}{S_0} = \frac{|\lambda_0-\lambda_1|}{\lambda_0+\lambda_1},
\end{equation}
is equal to zero when the intensities of the orthogonal modes are equal ($\lambda_0 = \lambda_1$);
in contrast, $p=1$ when the signal consists of a single purely polarized mode (one of the eigenvalues is zero).

\subsection{Axis-angle Representation}
\label{app:fundamental_transformations}

\Eqns{Boost}{and}{Rotation} are defined using the matrix exponential, which is given by the power series,
\begin{equation}
    \exp(\JM{A}) \equiv \sum_{k=1}^\infty \frac{\JM{A}^k}{k!}.
    \label{eqn:matrix_exponential}
\end{equation}
Here, $\JM{A}^k$ is the $k^\mathrm{th}$ power of $\JM{A}$; 
$\JM{A}^0 = \JM{I}$, where \JM{I} is the identity matrix with the dimensions of \JM{A};
and $k!$ is the factorial of $k$.
If \JM{A} is normal, then its spectral decomposition
\eqnsp{spectral_decomposition}{and}{projection},
\begin{equation}
    \JM{A} = \lambda_k \JM{P}_k,
\end{equation}
such that
\begin{equation}
    \JM{A}^n = (\lambda_k)^n \JM{P}_k.
    \label{eqn:matrix_power}
\end{equation}

\begin{proof}
Using induction, start with the base case and apply \eqn{projection_complete} to yield
\begin{align*}
\JM{A}^0 = (\lambda_k)^0 \JM{P}_k = \sum_k \JM{P}_k = \JM{I}
\end{align*}
If $\JM{A}^n = (\lambda_k)^n \JM{P}_k$, then
\begin{align*}
\JM{A}^{n+1} &= \JM{A}\JM{A}^n \\
&= \left( \lambda_j \JM{P}_j \right) \left( (\lambda_k)^n \JM{P}_k \right)  \\
&= \lambda_j (\lambda_k)^n \JM{P}_j \JM{P}_k   \\
&= \lambda_j (\lambda_k)^n \delta_{jk} \JM{P}_k & \peqn{idempotent_orthogonal} \\
&= (\lambda_k)^{n+1} \JM{P}_k. & \hfill \square
\end{align*}
\end{proof}

\noindent
Substitute \eqn{matrix_power} into \eqn{matrix_exponential} to yield
\begin{align}
\exp(\JM{A}) 
& = \left( \sum_{k=1}^\infty \frac{(\lambda_j)^n}{k!} \right) \JM{P}_j 
  = \exp(\lambda_j) \JM{P}_j.
\end{align}

To arrive at \eqns{Boost}{and}{Rotation}, consider the eigenvalues of 
$\JM{A} = \Pv{a}\cdot\Pv{\sigma}$, which must satisfy
%
\begin{equation}
    |\JM{A} - \lambda \JM{I}| = \left|\Pv{a}\cdot\Pv{\sigma} - \lambda \pauli{0} \right| = \lambda^2 - |\Pv{a}|^2 = 0.
\end{equation}
That is, there are two eigenvalues given by $\lambda = \pm |\Pv{a}|$.
Let $a=|\Pv{a}|$ and $\hat{\Pv{a}}=\Pv{a}/a$, and use the spectral decomposition,
\begin{equation}
\JM{A} = a\JM{P}_0 -a\JM{P}_1,
\end{equation}
to show that
\begin{equation}
\JM{P}_0 + \JM{P}_1 = \pauli{0} \quad \text{and} \quad
\JM{P}_0 - \JM{P}_1 = \hat{\Pv{a}}\cdot\Pv{\sigma};
\end{equation}
therefore
\begin{align}
e^{\Pvs{a}\cdot\Pvs{\sigma}}
&= \exp(a) \JM{P}_0 + \exp(-a) \JM{P}_1 \nonumber \\
&= \cosh(a) (\JM{P}_0 + \JM{P}_1) 
+ \sinh(a)(\JM{P}_0 - \JM{P}_1) \label{eqn:Pauli_exponential} \\
&= \cosh(a) \pauli{0}
+ \sinh(a) \hat{\Pv{a}}\cdot\Pv{\sigma}. \nonumber
\end{align}
Setting $\Pv{a} = \beta \hat{\Pv{m}}$ in \eqn{Pauli_exponential} 
yields the axis-angle representation of Hermitian matrices \eqnp{Boost};
likewise, setting $\Pv{a} = \Ci \phi \hat{\Pv{n}}$ yields the axis-angle representation of unitary matrices \eqnp{Rotation}.


\subsection{Congruence Eigenmatrices}
\label{app:symmetry}

The unit vector in the axis-angle representation of a matrix defines an axis of symmetry in the three-dimensional space of the Stokes polarization vector.
This symmetry axis can be exploited to rearrange and simplify matrix equations and identify degenerate systems of equations \citep[as in Appendix B of][and \App{degeneracy} of this paper]{van04c}.
For example, two matrices 
commute when their symmetry axes are collinear (parallel or anti-parallel).
\begin{proof}
Given $$\JM{A} = a\,\pauli{0} + \Pv{a} \cdot \Pv{\sigma} \quad\text{and}\quad
\JM{B} = b\,\pauli{0} + \Pv{b} \cdot \Pv{\sigma},$$
 \eqn{multiplication_rule} shows that $\JM{A}\JM{B} = \JM{B}\JM{A}$ if and only if $\Pv{a} \times \Pv{b} = 0$.
\end{proof}
%
Furthermore, if \JM{A} and \JM{B} commute, then they have common eigenvectors.
\begin{proof}
\label{proof:common_eigenvalue}
If $\JM{A}\JM{B} = \JM{B}\JM{A}$ and $\JM{A}\Jcol{v} = \lambda \Jcol{v}$,
then
\begin{align*}
\JM{A}\JM{B}\Jcol{v} &= \JM{B}\JM{A}\Jcol{v} = \lambda \JM{B}\Jcol{v}.
\end{align*}
That is, $\JM{B}\Jcol{v}$ is an eigenvector of \JM{A} with the same eigenvalue $\lambda$;
therefore, $\JM{B}\Jcol{v}=\mu\Jcol{v}$.
\end{proof}

The coherency matrix has an axis of symmetry defined by its associated Stokes polarization vector, which facilitates the identification of the eigenmatrices of a congruence transformation.
A matrix $\cohM{\rho}$ is a congruence eigenmatrix of an invertible matrix $\JM{J}$ with real-valued congruence eigenvalue $\kappa$ if it satisfies the relation
\begin{equation}
\JM{J}\cohM{\rho}\JM{J}^\dagger = \kappa \cohM{\rho}.
\end{equation}
The eigenvectors of a Jones matrix define the singular congruence eigenmatrices of that Jones matrix.
\begin{proof}
\label{proof:eigenvector_to_eigenmatrix}
If $\JM{J} \Jcol{e} = \lambda \Jcol{e}$, where $\lambda \in \mathbb{C}$, then
\begin{align*}
\JM{J}\tilde{\cohM{\rho}}\,\JM{J}^\dagger
&= \JM{J} \Jcol{e} \otimes \Jcol{e}^\dagger \JM{J}^\dagger
    & \peqn{instantaneous_coherency} \\
&= \JM{J} \Jcol{e} \otimes  (\JM{J}\Jcol{e})^\dagger 
    & (\JM{A}\JM{B})^\dagger = \JM{B}^\dagger \JM{A}^\dagger \\
&= \lambda \, \Jcol{e} \otimes \lambda^* \Jcol{e}^\dagger \\
&= |\lambda|^2 \tilde{\cohM{\rho}} & \tilde{\cohM{\rho}} \equiv \Jcol{e} \otimes \Jcol{e}^\dagger \\
&= \kappa \tilde{\cohM{\rho}} & \kappa \equiv |\lambda|^2 \in \mathbb{R}
\end{align*}
\end{proof}
\noindent
Accordingly, a Jones matrix has two congruence eigenmatrices.\footnote{
Although the Mueller matrix associated with \JM{J}
has four eigenvectors, only two of these
describe physical polarization states.
For example, two of the eigenvectors of a rotation are complex-valued.  
Two of the eigenvectors of a Lorentz boost have polarization vectors that lie 
in the plane that is normal to the boost axis; however, these have $S_0 = 0$, which is not physical.   }
%
%
%
%
Conversely, if a singular coherency matrix
$\hat{\cohM{\rho}}=\Jcol{e} \otimes \Jcol{e}^\dagger$ 
is a congruence eigenmatrix of \JM{J}, then $\Jcol{e}$
is an eigenvector of \JM{J}.
\begin{proof}
\label{proof:eigenmatrix_to_eigenvector}
If $\hat{\cohM{\rho}}=\Jcol{e} \otimes \Jcol{e}^\dagger$ and $\JM{J}\hat{\cohM{\rho}}\JM{J}^\dagger = \kappa \hat{\cohM{\rho}}$, then
$$ \JM{J} \left( \Jcol{e} \otimes \Jcol{e}^\dagger \right) \JM{J}^\dagger
=
\JM{J} \Jcol{e} \otimes  (\JM{J}\Jcol{e})^\dagger 
=
\kappa \Jcol{e} \otimes \Jcol{e}^\dagger, $$
and $\JM{J} \Jcol{e}$ and $\Jcol{e}$ must be collinear; i.e.,
$\JM{J} \Jcol{e} = \lambda \Jcol{e}$, where $|\lambda|^2 = \kappa$.
\end{proof}
\noindent
Furthermore, if
$\hat{\cohM{\rho}}=\Jcol{e} \otimes \Jcol{e}^\dagger$ 
commutes with \JM{J}, then $\hat{\cohM{\rho}}$
is a congruence eigenmatrix of $\JM{J}$.
\begin{proof}
\label{proof:if_commutes_then_eigen}
If $\JM{J} \tilde{\cohM{\rho}} = \tilde{\cohM{\rho}} \JM{J}$, then 
$\tilde{\cohM{\rho}}$ and $\JM{J}$ have a common eigenvector (\Proof{common_eigenvalue}).
Therefore,
$$\JM{J} \Jcol{e} = \lambda \, \Jcol{e}$$
where $\Jcol{e}$ is the only eigenvector of $\tilde{\cohM{\rho}}=\Jcol{e} \otimes \Jcol{e}^\dagger$, and
(via \Proof{eigenvector_to_eigenmatrix})
$$\JM{J} \tilde{\cohM{\rho}}\, \JM{J}^\dagger = \kappa\tilde{\cohM{\rho}},$$
where $\kappa=|\lambda|^2$.
\end{proof}

\noindent
Finally, if 
$\JM{J}$ is a normal matrix
and 
$\hat{\cohM{\rho}}$ is a congruence eigenmatrix of $\JM{J}$,
then 
$\hat{\cohM{\rho}}$ commutes with $\JM{J}$.

\begin{proof}
\label{proof:if_eigen_then_commutes}
If $\JM{J}$ is normal,
$\hat{\cohM{\rho}}=\Jcol{e} \otimes \Jcol{e}^\dagger$ 
and
$\JM{J} \tilde{\cohM{\rho}}\, \JM{J}^\dagger = \kappa\tilde{\cohM{\rho}},$
then 
$\JM{J} \Jcol{e} = \lambda \Jcol{e}$ (\Proof{eigenmatrix_to_eigenvector})
and
$$\JM{J} \tilde{\cohM{\rho}} 
  = \JM{J} \Jcol{e} \otimes \Jcol{e}^\dagger 
  = \lambda \, \Jcol{e} \otimes \Jcol{e}^\dagger 
  = \lambda \tilde{\cohM{\rho}}. $$

Also
\begin{align*}
 \tilde{\cohM{\rho}} \JM{J}
  &=  \Jcol{e} \otimes \Jcol{e}^\dagger \JM{J} \\
  &= \Jcol{e} \otimes \left( \JM{J}^\dagger \Jcol{e} \right)^\dagger  \\
  &= \Jcol{e} \otimes \left( \lambda^* \Jcol{e} \right)^\dagger & \text{\Proof{normal_eigenvalue}} \\
  &= \lambda \, \Jcol{e} \otimes  \Jcol{e}^\dagger \\
  &= \lambda \tilde{\cohM{\rho}}
\end{align*}
Therefore, $\JM{J} \tilde{\cohM{\rho}} = \tilde{\cohM{\rho}} \JM{J}.$
\end{proof}

\noindent
In conclusion, if $\JM{J}$ is normal,  then $\hat{\cohM{\rho}}$ is a congruence eigenmatrix of $\JM{J}$ if and only if $\hat{\cohM{\rho}}$ commutes with $\JM{J}$.
Both unitary and Hermitian matrices are normal and their congruence eigenmatrices are explored in the following sections.

\subsubsection{ Unitary Matrices}
\label{app:eigen_unitary}

If $\cohM{\rho}$ commutes with \rotat, then
\begin{equation}
\rotat \, \cohM{\rho} \, \vRotation^\dagger(\phi)
= \cohM{\rho}  \rotat \vRotation^\dagger(\phi) 
= \cohM{\rho};
\end{equation}
that is, $\cohM{\rho}$ is a congruence eigenmatrix of \rotat\ with associated congruence eigenvalue equal to unity.
Conversely, if $\cohM{\rho}$ is a congruence eigenmatrix of \rotat, then
\begin{equation}
\begin{split}
\rotat \, \cohM{\rho} \, \vRotation^\dagger(\phi) & = \kappa \cohM{\rho}  \\
\rotat \, \cohM{\rho} \, \vRotation^\dagger(\phi)\, \rotat & = \kappa \cohM{\rho} \, \rotat \\
\rotat \, \cohM{\rho} & = \kappa \cohM{\rho} \, \rotat
\end{split}
\end{equation}
and $\cohM{\rho}$ commutes with \rotat\ if $\kappa = 1$.
\noindent
The congruence eigenmatrices of a unitary matrix share a single 
degenerate congruence eigenvalue, $\kappa=1$; therefore, any linear combination of the congruence eigenmatrices
is also a congruence eigenmatrix of the transformation.
Considering the equivalent three-dimensional rotation of the Stokes polarization vector $\Pv{S}$;
any vector that lies along the axis of rotation remains unchanged by that rotation, including $\Pv{S}=0$ (unpolarized radiation).
%

\subsubsection{ Hermitian Matrices }
\label{app:eigen_self_adjoint}

Define $b=\cosh\beta$ and $\Pv{b}=\hat{\Pv{m}}\sinh\beta$ such that 
\begin{equation}
    \boost = b\,\pauli{0} + \Pv{b}\cdot\Pv{\sigma}.
\end{equation}
Furthermore, define 
$\tilde{\cohM{\rho}} = \left(\tilde{s}_0\,\pauli{0} + \tilde{\Pv{s}}\cdot \Pv{\sigma}\right)/2$,
and note that $|\tilde{\Pv{s}}| = \tilde{s}_0$ because $\tilde{\cohM{\rho}}$ is singular / purely polarized. 
If $\tilde{\cohM{\rho}}$ and \boost\ commute, then their axes of
symmetry are collinear; i.e.,
\begin{equation}
\begin{split}
\hat{\Pv{m}} \times \tilde{\Pv{s}} &= 0, \\
\hat{\Pv{m}} \cdot \tilde{\Pv{s}} &= \pm|\tilde{\Pv{s}}| = \pm\tilde{s}_0, \\
\hat{\Pv{m}} &= \pm \, \tilde{\Pv{s}} / \tilde{s}_0.
\end{split}
\end{equation}
Using \eqn{multiplication_rule},
\begin{equation} \begin{split}
\boost \tilde{\cohM{\rho}} 
&= (b \tilde{s}_0 + \Pv{b} \cdot \tilde{\Pv{s}})\pauli{0}/2 + (b\tilde{\Pv{s}} + \tilde{s}_0\Pv{b})\cdot\Pv{\sigma}/2 \\
&= (b \tilde{s}_0 \pm |\Pv{b}| \tilde{s}_0)\pauli{0}/2 + (b\tilde{\Pv{s}} \pm |\Pv{b}| \tilde{\Pv{s}})\cdot\Pv{\sigma}/2 \\
&= (b \pm |\Pv{b}|) \tilde{s}_0  \pauli{0}/2 + (b \pm |\Pv{b}|) \tilde{\Pv{s}}\cdot\Pv{\sigma}/2 \\
&= (b \pm |\Pv{b}|) \tilde{\cohM{\rho}} \\
&= (\cosh\beta \pm \sinh\beta) \tilde{\cohM{\rho}} \\
&= e^{\pm\beta} \tilde{\cohM{\rho}}.
\end{split} \end{equation}
Therefore,
\begin{equation} \begin{split}
\boost \, \tilde{\cohM{\rho}} \, \vBoost^\dagger(\beta)
&= \boost \, \tilde{\cohM{\rho}} \, \boost \\
&= \vBoost^2(\beta) \tilde{\cohM{\rho}} \\
&= e^{\pm2\beta} \tilde{\cohM{\rho}}.
\end{split} \end{equation}
Hermitian matrices have two distinct congruence eigenvalues; 
$\kappa=e^{2\beta}$ when the polarization vector $\tilde{\Pv{s}}$
is parallel to the boost axis $\hat{\Pv{m}}$,
and
$\kappa=e^{-2\beta}$ when $\tilde{\Pv{s}}$
is anti-parallel to $\hat{\Pv{m}}$.
Therefore, only these two purely polarized states are congruence eigenmatrices of the transformation.


\subsection{ Pure Mueller Matrices }
\label{app:pure_Mueller_matrices}
Various authors \cite[e.g.,][]{barakat81,sim82,clo86} have considered the necessary and sufficient conditions that must be satisfied by the elements of a Mueller matrix for it to have an equivalent Jones matrix representation.
\cite{sim82} and \cite{clo86} present the most intuitive geometric constraints on 
a pure Mueller matrix based on its equivalent target coherency matrix \citep{clo86}.

For a Jones matrix \JM{J}, the equivalent $4 \times 4$ Hermitian target coherency matrix $\MM{N}$ is defined via the rank 4 tensor $\JT{N}=\outerBilinear{\bf J}{\bf J}^\dagger$, such that
\eqn{tensor_to_Mueller} yields
\begin{equation}
\begin{split}
\label{eqn:target_coherency_matrix}
N_\irow^\icol
& = \frac{1}{2} \dc{\pauli{\irow}}{\dc{\left( \outerBilinear{\bf J}{\bf J}^\dagger \right)}{\pauli{\icol}}} \\
& = \frac{1}{2} \dc{\pauli{\irow}}{{\bf J}} \left( \dc{\pauli{\icol}}{{\bf J}^\dagger} \right) \\
& = \frac{1}{2} k_\irow k^*_\icol.
\end{split}
\end{equation}
Here, $k_\irow = \dc{\pauli{\irow}}{{\bf J}}$ are the components of a complex-valued target four-vector $\Sv{k}$ \citep{clo86}.
\Eqn{target_coherency_matrix} is equivalent to $\MM{N} = \Sv{k} \otimes \Sv{k}^\dagger / 2$ and, as
noted by \cite{sim82}, \MM{N} is a scalar multiple of a projection matrix, such that
\begin{equation}
\MM{N}^2 = \tr{\MM{N}} \MM{N}.
\label{eqn:pure_target_coherency}
\end{equation}

\begin{proof}
Define $\hat{\Sv{k}}=\Sv{k}/|\Sv{k}|$ and the projection $\MM{P} = \hat{\Sv{k}} \otimes \hat{\Sv{k}}^\dagger$ \eqnp{projection}, 
such that
\begin{equation*}
\MM{N}
 = \frac{1}{2}\Sv{k} \otimes \Sv{k}^\dagger \\
 = \frac{1}{2}|\Sv{k}|^2 \hat{\Sv{k}} \otimes \hat{\Sv{k}}^\dagger
 = \frac{1}{2}|\Sv{k}|^2 \MM{P}.
\end{equation*}
Note that 
\begin{equation*}
|\Sv{k}|^2 = k_\mu k_\mu^* = 2 \tr{\MM{N}}.
\end{equation*}
Therefore, $\MM{N} = \tr{\MM{N}} \MM{P}$ and \eqnp[via ]{idempotent_orthogonal}
\begin{equation*}
\MM{N}^2 = \tr{\MM{N}}^2 \MM{P}^2 = \tr{\MM{N}}^2 \MM{P}= \tr{\MM{N}} \MM{N}.
\end{equation*}
\end{proof}

The mapping between any Mueller matrix $\MM{M}$ and its associated target coherency matrix $\MM{N}$ exploits the symmetry of the transpose used to define the $\tilde\otimes$ operator \eqnp{indexSpinorBilinear}.
Let $\mathcal{T}$ represent the transpose of covariant tensor indeces, such that
$\spinorBilinear{\bf A}{\bf B} = \mathcal{T}(\outerBilinear{\bf A}{\bf B})$ and, by symmetry,
$\outerBilinear{\bf A}{\bf B} = \mathcal{T}(\spinorBilinear{\bf A}{\bf B})$.  Using this operator, the rank 4 tensors associated with \MM{M} and \MM{N} are related by $\JT{U} = \mathcal{T}(\JT{N})$, where $\JT{U}$ is defined by \eqn{Mueller_to_tensor}
and 
\begin{equation}
\begin{split}
\JT{N} & = \mathcal{T}(\JT{U}) \\
 & = \mathcal{T}\left( \frac{1}{2} M_\irow^\icol \outerBilinear{\pauli{\irow}}{\pauli{\icol}} \right) \\
 & = \frac{1}{2} M_\irow^\icol \spinorBilinear{\pauli{\irow}}{\pauli{\icol}},
\end{split}
\end{equation}
such that
\begin{equation}
\begin{split}
\label{eqn:Mueller_to_target_coherency}
N_\jrow^\jcol & = \frac{1}{2} \dc{\pauli{\jrow}}{\dc{ \JT{N} }{\pauli{\jcol}}} \\
& = \frac{1}{4} M_\irow^\icol \, \dc{\pauli{\jrow}}{\dc{\left( \spinorBilinear{\pauli{\irow}}{\pauli{\icol}} \right)}{\pauli{\jcol}}} \\
& = \frac{1}{4} M_\irow^\icol \, \dc{\pauli{\jrow}}{ \pauli{\irow}}{\pauli{\jcol}} {\pauli{\icol}} \\
& = \frac{1}{4} M_\irow^\icol \, \tr{ \pauli{\jrow} \pauli{\irow} \pauli{\jcol} \pauli{\icol} }. \\
\end{split}
\end{equation}
\Eqn{Mueller_to_target_coherency} can be written
as $\MM{N} = \MM{\JT{T}} \dcbo \MM{M}$, where $\MM{\JT{T}}$ is the
four-dimensional rank 4 tensor defined by
\begin{equation}
T_{\jrow\icol}^{\jcol\irow} = \frac{1}{4}  \tr{ \pauli{\jrow} \pauli{\irow} \pauli{\jcol} \pauli{\icol} }.
\end{equation}
By symmetry, $\MM{M} = \MM{\JT{T}} \dcbo \MM{N}$; that is, $\MM{\JT{T}}$ is an involution.
As for the target coherency matrix derived from a Jones matrix,
$\MM{N}$ is Hermitian.

\begin{proof}
\begin{align*}
&(N_\jrow^{\jcol})^* = (T_{\jrow\icol}^{\jcol\irow})^* (M_\irow^\icol)^* \\
&= \frac{1}{4} M_\irow^\icol \, \tr{ \pauli{\jrow} \pauli{\irow} \pauli{\jcol} \pauli{\icol} }^* & M_\irow^\icol\in\mathbb{R} \\
&= \frac{1}{4} M_\irow^\icol \, \tr{ \left( \pauli{\jrow} \pauli{\irow} \pauli{\jcol} \pauli{\icol} \right)^\dagger } & \tr{\bf A}^* = \tr{{\bf A}^\dagger} \\
&= \frac{1}{4} M_\irow^\icol \, \tr{ \pauli{\icol}  \pauli{\jcol} \pauli{\irow} \pauli{\jrow}  } & {\bf AB}^\dagger = {\bf B}^\dagger {\bf A}^\dagger \\
&= \frac{1}{4} M_\irow^\icol \, \tr{ \pauli{\jcol} \pauli{\irow} \pauli{\jrow} \pauli{\icol} } & \tr{\bf AB} = \tr{\bf BA} \\
&= N_\jcol^\jrow  & \peqn{Mueller_to_target_coherency}
\end{align*}
\end{proof}

Just as the coherency matrix formed by the outer product of a single Jones vector
$\tilde{\cohM{\rho}} = \hat{\Jcol{e}} \otimes \hat{\Jcol{e}}^\dagger$ corresponds to a purely polarized state
\eqnp[see]{spectral_decomposition_of_coherency_matrix}, the target coherency matrix formed by the outer product of a single target vector $\MM{N} = \Sv{k} \otimes \Sv{k}^\dagger / 2$ corresponds to a pure Mueller matrix.
Therefore, a Mueller matrix is pure if and only if 
its target coherency matrix, $\MM{N} = \MM{\JT{T}} \dcbo \MM{M}$ 
satisfies \eqn{pure_target_coherency}.
\cite{clo86} expressed this definition with the equivalent constraint that a pure Mueller matrix is associated with a target coherency matrix that has only one non-zero eigenvalue, $\lambda$.
Combined with its associated eigenvector, $\hat{\Sv{k}}$,
the target vector $\Sv{k} = \lambda^{1/2} \hat{\Sv{k}}$ can be used to compute (up to arbitrary phase) the Jones matrix associated with a pure Mueller matrix,
\begin{equation}
\JM{J} = \frac{1}{2} k_\mu \pauli{\mu}.
\end{equation}

\subsection{Impure Mueller Matrices}
\label{app:impure}


Some polarimetric transformations are described by impure Mueller matrices that cannot be represented by a linear transformation of the electric field, as described by a Jones matrix.
For example, temporal or spectral depolarization occurs when the response of a system varies as a function of time or frequency on characteristic scales that are smaller than the duration or bandwidth over which the Stokes parameters are integrated.  In this case, the measured Stokes parameters,
\begin{equation}
\Sv{S}'=\langle \MM{M}(t,\nu) \rangle \Sv{S},
\end{equation}
where the angle brackets represent integration over time and/or frequency.  

For example, consider stochastic Faraday rotation that causes fluctuations in the position angle $\Delta\Psi$ that are normally distributed with zero mean and standard deviation $\sigma_\Psi$.
The average Mueller matrix that describes the integrated Stokes parameters at a single radio frequency,

\begin{equation} \begin{split}
\langle \MM{R}_{\hat{\Pvs v}}(2\Delta\Psi) \rangle 
&= \left( \begin{array}{cccc}
1 & 0 & 0 & 0 \\
0 & \langle\cos2\Delta\Psi\rangle & -\langle\sin2\Delta\Psi\rangle & 0\\
0 & \langle\sin2\Delta\Psi\rangle & \langle\cos2\Delta\Psi\rangle & 0 \\
0 & 0 & 0 & 1 
\end{array} \right) \\
&= \left( \begin{array}{cccc}
1 & 0 & 0 & 0 \\
0 & d & 0 & 0\\
0 & 0 & d & 0 \\
0 & 0 & 0 & 1 
\end{array} \right)
\end{split} \end{equation}
where the angle brackets represent integration over time and $d=\exp(-2\sigma_\Psi^2) < 1$ describes the depolarization of Stokes Q and U. Note that this Mueller matrix is able to depolarize a purely polarized state, something that cannot be done by a linear transformation of the electric field (\Proof{only_valid}).

\setcounter{equation}{0}
\section{Example Transformations}
\label{app:example_transformation}

This section demonstrates some basic transformations,
starting with a source described by Stokes parameters
\begin{equation}
    \Sv{S}=[S_0,S_1,0,0]^T
\end{equation}
and coherency matrix
\begin{equation}
    \cohM{\rho}=(S_0 \, \pauli{0} + S_1 \, \pauli{1})/2.
\end{equation}

\subsection{Example Boost}
\label{app:example_boost}

\noindent
Consider a boost along $\hat{\Pv{1}}=(1,0,0)^T$, 
$$\Boost = \vBoost[1](\beta) =\pauli{0}\cosh\beta+\pauli{1}\sinh\beta,$$ 
such that
\begin{equation} \begin{split}
\cohM{\rho}' &= \Boost \, \cohM{\rho} \, \Boost^\dagger \\
&= \Boost \left( S_0 \, \pauli{0}+S_1 \, \pauli{1} \right) \Boost / 2 \\
&= S_0 \Boost \pauli{0} \Boost/2 +S_1 \Boost \pauli{1} \Boost/2.
\end{split} \end{equation}
Using \eqn{boost_power}, the first term of this equation
includes 
\begin{equation} \begin{split}
\vBoost[1](\beta) \; \pauli{0} \; \vBoost[1](\beta)
&=\vBoost[1]^2(\beta) = \vBoost[1](2\beta)
\end{split} \end{equation}
The second term includes
%
%
\begin{equation} \begin{split}
\vBoost[1]&(\beta) \; \pauli{1} \; \vBoost[1](\beta)  \\
&= \left( \pauli{0}\cosh\beta + \pauli{1}\sinh\beta \right) \pauli{1} \left( \pauli{0}\cosh\beta + \pauli{1}\sinh\beta \right) \\
&= \left( \pauli{0}\cosh\beta + \pauli{1}\sinh\beta \right) \left( \pauli{1}\cosh\beta + \pauli{0}\sinh\beta \right) \\
&= \pauli{1}\cosh^2\beta + \pauli{0}\cosh\beta\sinh\beta \\
& \omit\hfill\ensuremath{ + \pauli{0}\sinh\beta\cosh\beta + \pauli{1}\sin^2\beta } \\
&= \pauli{1} \left(\cosh^2\beta + \sinh^2\beta\right) + \pauli{0} \left( 2 \sinh\beta\cosh\beta \right) \\
&= \pauli{1} \cosh 2\beta + \pauli{0} \sinh 2\beta.
\end{split} \end{equation}
Therefore,
\begin{equation} \begin{split}
\cohM{\rho}' &= S_0 \vBoost[1](2\beta)/2 + S_1 \Boost \pauli{1} \Boost/2 \\
&= S_0 \left(  \pauli{0} \cosh 2\beta + \pauli{1} \sinh 2\beta \right)/2 \\
& \omit \quad \quad \quad \quad \ensuremath{+ S_1 \left( \pauli{1} \cosh 2\beta + \pauli{0} \sinh 2\beta \right)/2} \\
&= \pauli{0} \left( S_0 \cosh 2\beta + S_1 \sinh 2\beta \right)/2\\
& \omit \quad \quad \quad \quad \ensuremath{+ \pauli{1} \left( S_0 \sinh 2\beta + S_1 \cosh 2\beta \right)/2}
\end{split} \end{equation}
and
\begin{equation}
\begin{split}
S_0' &= \pauli{0} \dcbo \cohM{\rho}' = S_0 \cosh 2\beta + S_1 \sinh 2\beta \\
S_1' &= \pauli{1} \dcbo \cohM{\rho}' = S_0 \sinh 2\beta + S_1 \cosh 2\beta. \label{eqn:Stokes_boost}
\end{split}
\end{equation}
That is, the Stokes parameters are boosted along the $\hat{\Pv{1}}$
axis with rapidity $-2\beta$ and Lorentz factor $\gamma = \cosh 2\beta.$

\subsection{Example Rotation}
\label{app:example_rotation}

\noindent
Consider a rotation about $\hat{\Pv{2}}=(0,1,0)^T$, 
$$\Rotation = \vRotation[2](\chi) =\pauli{0}\cos\chi+\Ci\pauli{2}\sin\chi,$$ 
such that
\begin{equation} \begin{split}
\cohM{\rho}' &= \Rotation \, \cohM{\rho} \, \Rotation^\dagger \\
&= \Rotation \left( S_0 \, \pauli{0}+S_1 \, \pauli{1} \right) \Rotation^\dagger / 2 \\
&= S_0 \Rotation \pauli{0} \Rotation^\dagger/2 +S_1 \Rotation \pauli{1} \Rotation^\dagger/2.
\end{split} \end{equation}

\noindent
For the first term of this equation, note that rotations have no effect on \pauli{0}; e.g.
\begin{equation} \begin{split}
\vRotation(\phi) \; \pauli{0} \;\vRotation^\dagger(\phi) 
& = \vRotation(\phi) \;\vRotation^\dagger(\phi) \\
& = \vRotation(\phi) \;\vRotation^{-1}(\phi) = \pauli{0}
\end{split} \end{equation}
Noting that \eqnp[see]{rotation_power},
\begin{equation}
\vRotation^{-1}(\phi) = \vRotation(-\phi),
\end{equation}
the second term includes
%
%
\begin{equation}
\begin{split}
\vRotation[2]&(\chi) \; \pauli{1} \; \vRotation[2](-\chi) \\
&= \left( \pauli{0}\cos\chi + \Ci\pauli{2}\sin\chi \right) \pauli{1} \left( \pauli{0}\cos\chi - \Ci\pauli{2}\sin\chi \right) \\
&= \left( \pauli{0}\cos\chi + \Ci\pauli{2}\sin\chi \right) \left( \pauli{1}\cos\chi + \pauli{3}\sin\chi \right) \\
&= \pauli{1}\cos^2\chi + \pauli{3}\cos\chi\sin\chi \\
& \omit\hfill\ensuremath{+ \Ci(-\Ci \pauli{3})\sin\chi\cos\chi + \Ci(\Ci \pauli{1})\sin^2\chi} \\
&= \pauli{1} \left(\cos^2\chi - \sin^2\chi\right) + 2 \pauli{3} \cos\chi\sin\chi \\ 
&= \pauli{1} \cos2\chi  + \pauli{3} \sin2\chi \label{eqn:rotate_q_about_u}
\end{split}
\end{equation}
Therefore,
\begin{equation} \begin{split}
\cohM{\rho}' &= S_0 \, \pauli{0}/2 + S_1 \left( \pauli{1} \cos2\chi  + \pauli{3} \sin2\chi \right)/2
\end{split} \end{equation}
and
\begin{equation} \begin{split}
S_0' &= \pauli{0} \dcbo \cohM{\rho}' = S_0 \\
S_1' &= \pauli{1} \dcbo \cohM{\rho}' = S_1 \cos2\chi \\
S_3' &= \pauli{3} \dcbo \cohM{\rho}' = S_1 \sin2\chi.
\end{split} \end{equation}
That is, the total intensity is unchanged and the Stokes polarization vector is rotated about the $\hat{\Pv{2}}$
axis by $-2\chi$.


\subsection{Complex-valued Polarization Ellipse}
\label{app:analytic_polarization_ellipse}

This section demonstrates the relationship between the 
geometry of the polarization ellipse \eqnsp{psi_rotation}{through}{stokes_ellipse}
and the
spherical coordinates of the Stokes parameters \eqnp{stokes_geometric}
by deriving the analytic representation of a monochromatic electromagnetic wave \eqnp{analytic_field}.
First, consider 
\begin{equation}
\Jcol{e}(t) = \hat{\Jcol{e}}_0 r e^{\Ci (2\pi \nu t + \phi')}
\end{equation}
%
with polarization defined by the unit Jones vector $\hat{\Jcol{e}}_0$,
and associated Stokes polarization vector \eqnp[see]{stokes_geometric},
\begin{equation}
\Pv{S} = r^2 \left( \cos2\chi  \cos2\psi, \; \cos2\chi  \sin2\psi, \; \sin2\chi \right).
\label{eqn:equivalent_stokes}
\end{equation}
As in \Sec{basis_rotation}, consider the transformation from the original basis to one in which the second component of the electric field vector is zero; i.e.,
\begin{equation}
	\Jcol{e}''(t) = \JM{R} \Jcol{e}(t) =
	\left( 
	\begin{array}{c}
         1 \\
		0
	\end{array}
	\right) r e^{\Ci (2\pi \nu t + \phi')}.
\end{equation}
Invert this using \eqn{orthonormal_spherical}, yielding
\begin{equation}
\Jcol{e}(t) = \JM{R}^{-1} \Jcol{e}''(t) = \vRotation[3](-\psi) \vRotation[2](\chi) \Jcol{e}''(t),
\end{equation}
and compute the components of $\hat{\Jcol{e}}_0$, starting with
the transformation of $\Jcol{e}''(t)$ by $\vRotation[2](\chi)$,
\begin{equation} \begin{split}
\Jcol{e}'(t)
&= \vRotation[2](\chi) \Jcol{e}''(t)
 = \left( \pauli{0}\cos\chi + \Ci\pauli{2}\sin\chi \right) \Jcol{e}''(t) \\
&= \left[ \pauliI\cos\chi + \Ci\pauliU\sin\chi \right] \Jcol{e}''(t) \\
&= \left( 
      \begin{array}{ccc}
	\cos\chi & \;\; & \Ci\sin\chi \\
	\Ci\sin\chi & \;\; & \cos\chi
      \end{array}
   \right)
   \left(
      \begin{array}{c}
	1 \\
	0
      \end{array}
   \right) r e^{\Ci (2\pi \nu t + \phi')} \\
&= \left(
      \begin{array}{c}
	\cos\chi \\
	\Ci\sin\chi
      \end{array}
   \right) r e^{\Ci (2\pi \nu t + \phi')}.
\end{split} \end{equation}
The real part of this analytic signal,
\begin{equation}
	\Jcol{\epsilon}'(t)=
	\left( 
	\begin{array}{c}
		r \cos\chi \cos (2\pi \nu t + \phi') \\
		-r \sin\chi \sin (2\pi \nu t + \phi')
	\end{array}
	\right).
\end{equation}
The polarization state of the wave is independent of the absolute phase $\phi'$; therefore,
choose  $\phi'=\phi-\pi/2$ to arrive at \eqn{stokes_ellipse}.
Then transform $\Jcol{e}'$ by $\vRotation[3](-\psi)$ to arrive at \eqn{analytic_field},
\begin{eqnarray}
\Jcol{e}(t)
&=& \vRotation[3](-\psi) \Jcol{e}'(t) \nonumber \\
&=& \left( \pauli{0}\cos\psi - \Ci\pauli{3}\sin\psi \right) \Jcol{e}'(t) \nonumber \\
&=& \left[ \pauliI\cos\psi - \Ci\pauliV\sin\psi \right] \Jcol{e}'(t) \nonumber \\
&=& \left( 
      \begin{array}{ccc}
	\cos\psi & \;\; & -\sin\psi \\
	\sin\psi & \;\; & \cos\psi
      \end{array}
   \right)
   \left(
      \begin{array}{c}
	\cos\chi \\
	\Ci\sin\chi
      \end{array}
   \right) r e^{\Ci (2\pi \nu t + \phi')} \nonumber \\
&=& \left(
      \begin{array}{c}
	\cos\psi\cos\chi -\Ci \sin\psi\sin\chi \\
	\sin\psi\cos\chi +\Ci \cos\psi\sin\chi
   \end{array}
   \right) r e^{\Ci (2\pi \nu t + \phi')} \nonumber \\
&=& \Jcol{e}_0  e^{\Ci (2\pi \nu t + \phi')},
\label{eqn:monochromatic_analytic}
\end{eqnarray}
where $\Jcol{e}_0 = r \hat{\Jcol{e}}_0$ as in \eqn{Jones_vector}.

This result can be verified by demonstrating that the equivalent congruence transformation of the coherency matrix
\begin{equation}
\cohM{\rho}=\vRotation[3](-\psi) \vRotation[2](\chi) \cohM{\rho}'' \vRotation[2](-\chi) \vRotation[3](\psi),
\label{eqn:double_check}
\end{equation}
yields the original polarization vector \Pv{S}.
Starting with 
\begin{equation}
	\cohM{\rho}'' = \left( \pauli{0} + \pauli{1} \right) r^2 / 2,
\end{equation}
expand the rotations from the inside outward.  
Note that rotation of a $S_1$ polarized state about the $S_2$ axis
has already been derived in \eqn{rotate_q_about_u}, where it is shown that
\begin{equation}
\vRotation[2](\chi) \; \pauli{1} \; \vRotation[2](-\chi) = \pauli{1} \cos2\chi  + \pauli{3} \sin2\chi.
\end{equation}
%

As detailed in \App{eigen_unitary}, the $\pauli{3}$ basis matrix is a congruence eigenmatrix of any rotation about the $S_3$ axis; therefore, it is 
necessary to consider only the transformation of the $\pauli{1}$ basis matrix,
\begin{equation} \begin{split}
\vRotation[3]&(-\psi)  \; \pauli{1} \;	\vRotation[3](\psi) \\
&= \left( \pauli{0}\cos\psi - \Ci\pauli{3}\sin\psi \right) \pauli{1} \left( \pauli{0}\cos\psi + \Ci\pauli{3}\sin\psi \right) \\
&= \left( \pauli{0}\cos\psi - \Ci\pauli{3}\sin\psi \right) \left( \pauli{1}\cos\psi + \Ci(-\Ci\pauli{2})\sin\psi \right) \\
&=  \pauli{1}\cos^2\psi +  \pauli{2}\cos\psi\sin\psi \\
& \omit\hfill\ensuremath{- \Ci(\Ci \pauli{2})\sin\psi\cos\psi  - \Ci(-\Ci \pauli{1})\sin^2\psi} \\
&=  \pauli{1} \left(\cos^2\psi - \sin^2\psi\right) + 2 \pauli{2} \cos\psi\sin\psi \\
&=  \pauli{1} \cos2\psi  + \pauli{2} \sin2\psi.
\end{split} \end{equation}
Finally, expand \eqn{double_check}
\begin{equation} \begin{split}
\cohM{\rho}
&=\vRotation[3](-\psi) \vRotation[2](\chi) \cohM{\rho}'' \vRotation[2](-\chi) \vRotation[3](\psi) \\
&=\vRotation[3](-\psi) \vRotation[2](\chi) \left(\pauli{0}+\pauli{1} \right) \vRotation[2](-\chi) \vRotation[3](\psi) \; r^2/2\\
&=\vRotation[3](-\psi)  \left(\pauli{0}+ \pauli{1} \cos2\chi  + \pauli{3} \sin2\chi \right)  \vRotation[3](\psi) \; r^2/2 \\
&=\frac{r^2}{2} \left[ \pauli{0}+ \left( \pauli{1} \cos2\psi  + \pauli{2} \sin2\psi \right) \cos2\chi  + \pauli{3} \sin2\chi \right]  \\
&=\frac{r^2}{2} ( \pauli{0}+ \pauli{1} \cos2\chi \cos2\psi \\
& \omit\hfill\ensuremath{+ \pauli{2} \cos2\chi \sin2\psi  + \pauli{3} \sin2\chi )},
\end{split} \end{equation}
which is consistent with the polarization vector defined in \eqn{equivalent_stokes}.

\subsection{Transformation to a Circular Basis}
\label{app:circular_basis}

\Eqn{circular} is not the only way to represent a pair of orthonormal
circularly-polarized receptors; however, it is the only unimodular unitary matrix that effects the desired cyclic permutation of $\Pv{S}=(Q,U,V)^T$ into $\Pv{S}'=(V,Q,U)^T$ for all values of $Q$, $U$ and $V$ \citep[assuming the Stokes V sign convention of ][]{ieee145}.
To see this, consider the following three derivations of \eqn{circular}, each of which provides a slightly different perspective.

\subsubsection{Direct Derivation}

As discussed in \Sec{polarization_ellipse}, for a left-hand circularly polarized (LCP) wave, the phase of $e_y$ leads that of $e_x$ by $90\deg$.
If the LCP wave is also 100\% polarized, then 
\begin{equation}e_y(t)=\exp(\Ci \pi/2) e_x(t) = \Ci e_x(t),\end{equation} 
which can be expressed as in \eqn{pure_poln_e},
\begin{equation}
\label{eqn:lucky_left}
\Jcol{e}_L(t) = 
\left( \begin{array}{c}1 \\
i \end{array}\right) z(t).
\end{equation}
For RCP, the phase of $e_y$ lags that of $e_x$ by $90\deg$ and
\begin{equation}
\label{eqn:lucky_right}
\Jcol{e}_R(t) = 
\left( \begin{array}{c}
1 \\
-\Ci \end{array}\right) z'(t).
\end{equation}
Unit receptors that have maximal response to LCP and RCP are 
given by the normalized Hermitian transposes of the Jones vectors 
in \eqns{lucky_left}{and}{lucky_right}; i.e.,
\begin{equation}
\label{eqn:lucky_receptor}
\hat{\Jcol{r}}_L = \frac{1}{\sqrt{2}} \left( 1, \, -\Ci \right)
\quad\text{and}\quad
\hat{\Jcol{r}}_R = \frac{1}{\sqrt{2}} \left( 1, \, \Ci \right).
\end{equation}
The unitary matrix comprised of these row vectors,
\begin{equation}
\JM{J}_C
= \left( \begin{array}{c}
    \hat{\Jrow{r}}_L \\
    \hat{\Jrow{r}}_R
\end{array} \right)
= \frac{1}{\sqrt{2}}
  \left( \begin{array}{cc}
    1 & -\Ci \\
    1 & \Ci
  \end{array} \right),
\end{equation}
has a determinant of $\Ci$, and normalizing by $\sqrt{i}$ yields
the unimodular unitary matrix shown in \eqn{circular}.
This transformation effects the desired
cyclic permutation of the Stokes parameters.

\subsubsection{Indirect Derivation}

Either row vector in \Eqn{lucky_receptor} can be multiplied by an arbitrary phase and the pair would continue to be orthonormal.
For example, it would be equally reasonable to start with a more symmetric pair of equations,
\begin{equation}\hat{\Jcol{r}}'_L = \hat{\Jcol{r}}_L =  \frac{1}{\sqrt{2}} \left( 1, \, -\Ci \right)\end{equation}
and
\begin{equation}\hat{\Jcol{r}}'_R = -\Ci \hat{\Jcol{r}}_R = \frac{1}{\sqrt{2}} \left( -\Ci, \, 1 \right).\end{equation}
The unitary matrix comprised of these orthonormal row vectors,
\begin{equation}
\begin{split}
\JM{R}_C
&= \left( \begin{array}{c}
    \hat{\Jrow{r}}'_L \\
    \hat{\Jrow{r}}'_R
\end{array} \right)
= \frac{1}{\sqrt{2}}
  \left( \begin{array}{cc}
    1 & -\Ci \\
    -\Ci & 1
  \end{array} \right) \\
&= \frac{1}{\sqrt{2}} (\pauli{0} - \Ci \pauli{2}) = \vRotation[2](-\pi/4),
\end{split}
\end{equation}
rotates the polarization vector $\Pv{S}=(Q,U,V)^T$ around the $S_2$ axis by $90\deg$, yielding $\Pv{S}'=(V,U,-Q)^T$.
%

%
To arrive at the desired cyclic permutation, note that a $90\deg$ rotation about the $S_1$ axis transforms 
$(V,U,-Q)^T$ 
into 
$(V,Q,U)^T.$
Therefore,
\begin{align}
\JM{R}_b 
  &= \vRotation[1](-\pi/4) \vRotation[2](-\pi/4) \\
  &= 
  \left( \begin{array}{cc}
    e^{-\Ci\pi/4} & 0 \\
    0 & e^{\Ci\pi/4}
  \end{array} \right)
  \frac{1}{\sqrt{2}}
  \left( \begin{array}{cc}
    1 & -\Ci \\
    -\Ci & 1
  \end{array} \right) \nonumber \\
  &=   \frac{\sqrt{-\Ci}}{\sqrt{2}}
  \left( \begin{array}{cc}
    1 & 0 \\
    0 & \Ci
  \end{array} \right)
  \left( \begin{array}{cc}
    1 & -\Ci \\
    -\Ci & 1
  \end{array} \right) 
  =   \frac{1}{\sqrt{2i}}
  \left( \begin{array}{cc}
    1 & -\Ci \\
    1 & \Ci
  \end{array} \right). \nonumber
\end{align}

\subsubsection{Geometric Derivation}

Rather than starting with the electric field, consider the three-dimensional rotation of the Stokes polarization vector that effects the desired cyclic permutation.
%
The eigenvectors of this rotation lie on the axis defined by $Q=U=V$; therefore,
choose $\hat{\Pv{n}}=(1,1,1)^T/\sqrt{3}$ and note that a $120\deg$ rotation about this axis rotates Stokes $V$ into $S'_1$, Stokes $Q$ into $S'_2$, and Stokes $U$ into $S'_3$.
The equivalent unitary transformation of the electric field,
\begin{equation}
\begin{split}
\JM{R}_b 
  &= \vRotation[n](-\pi/3) = \frac{1}{2} \left( \pauli{0} - \Ci \sum_{k=1}^3 \pauli{k} \right) \\
  &= \frac{1}{2}
  \left( \begin{array}{cc}
    1-\Ci & -(1+\Ci) \\
    1-\Ci & 1+\Ci
  \end{array} \right)\\
  &= \frac{1}{\sqrt{2}}
  \left( \begin{array}{cc}
    e^{-i\pi/4} & -e^{i\pi/4} \\
    e^{-i\pi/4} & e^{i\pi/4}
  \end{array} \right)\\
  &= \frac{1}{\sqrt{2i}}
  \left( \begin{array}{cc}
    1 & -i \\
    1 & i
  \end{array} \right).
\end{split}
\end{equation}
%


\setcounter{equation}{0}
\section{Stokes Parameters as Intensity Differences}
\label{app:Stokes_via_intensity_differences}

Currently, at frequencies above the microwave region of the electromagnetic spectrum, it is not possible to directly sample the electric field and therefore not possible to simultaneously compute all four Stokes parameters as in \eqns{StokesI}{through}{StokesV}.
Rather, only $S_0$ and $S_1$ can be directly measured because they are the sums and differences of the flux densities (or intensities) of the radiation after passing through oppositely polarized filters.

In the basis defined by the Cartesian coordinates introduced in \Sec{geometry}, the Stokes polarization vector $\Pv{S}_q=(Q,U,V)^T$, and \eqn{StokesQ} yields
\begin{equation}
\label{eqn:Q_intensity}
	S_1 = \left| e_x \right|^2 -  \left| e_y \right|^2 = I_x - I_y = Q,
\end{equation}
where $I_x$ and $I_y$ are the intensities of the electromagnetic radiation after passing through a linearly polarized filter with its transmission axis oriented along the $x$ and $y$ axes, respectively.

In a basis that is formed by rotating the original basis by $45\deg$ around the line of sight
(e.g.,  the $x'$ and $y'$ axes of \Sec{geometry} when $\psi=\pi/4$),
the Stokes polarization vector $\Pv{S}_u=(U,-Q,V)^T$, and \eqn{StokesQ} yields
\begin{equation}
\label{eqn:U_intensity}
	S_1' = \left| e_x' \right|^2 -  \left| e_y' \right|^2 = I_x' - I_y' = U,
\end{equation}
where $I_x'$ and $I_y'$ are the intensities measured after passing through a linearly polarized filter with its transmission axis oriented along the $x'$ and $y'$ axes, respectively.


In a basis defined by receptors with orthogonal senses of circular
polarization, as discussed in \Sec{feed_basis}, the Stokes polarization vector $\Pv{S}_v=(V,Q,U)^T$, 
and \eqn{StokesQ} yields
\begin{equation}
\label{eqn:V_intensity}
	S_1^c = \left| e_l \right|^2 -  \left| e_r \right|^2 = I_l - I_r = V,
\end{equation}
where $I_l$ and $I_r$ are the intensities after passing through filters that pass left and right circularly polarized radiation, respectively.

In all three cases, the total intensity $S_0$ is given by the sum of the intensities of the orthogonal polarizations, which is independent of the basis in which it is measured.  By measuring the six intensities that appear in \eqns{Q_intensity}{through}{V_intensity}, which correspond to the six special cases plotted in \Fig{polarization_cases}, all four Stokes parameters can be measured.


\setcounter{equation}{0}
\section{Refinements to Existing Definitions}
\label{app:refinements}


This article introduces some minor refinements and corrections to existing definitions that enhance their clarity and accuracy.
First, in \cite{sto52}, the ellipticity angle $\chi$ varies between $-90\deg$ and $90\deg$, as inferred from ``\emph{the numerical value of [$\chi$] being supposed not to lie beyond the limits 0 and 90\deg}'' and ``\emph{polarization is right-handed or left-handed according to the sign of [$\chi$].}''  (The ``numerical value'' is understood to mean the ``absolute value.'')  However, as noted in \Sec{injectivity}, allowing $|\chi| > 45\deg$ leads to model degeneracy; therefore $|\chi| \le 45\deg$ in this work.

The \cite{ieee145} standard defines the axial ratio $r'$ as the ``\emph{ratio
of the major to minor axes of a polarization ellipse}'' that ``\emph{carries a sign that is
taken as plus if the sense of polarization is right-handed and minus if it is left-handed.}''
The caption to \Fig{polarization_ellipse} of this article defines a ratio between semi-minor and semi-major axis $r \equiv \tan\chi = -1/r'$ that is positive for left-handed polarization.
The sign of $r'$ is negated and, relative to $r$, it is effectively inverted in the \cite{ieee145} definition of the Poincar\'{e} sphere, where ``\emph{the latitude is twice the angle whose cotangent is the negative of the axial ratio of the polarization ellipse.}''  That is, the latitude, $2\chi' = -2\cot^{-1} r' = 2\tan^{-1} r = 2\chi$; therefore, this article and the IEEE standard arrive at mutually consistent definitions of latitude in the Poincar\'{e} sphere.
However, though only briefly mentioned in this article, the ratio $r$ is preferred because it does not approach infinity for any polarized states and it requires no negation when it is related to latitude in the Poincar\'{e} sphere.

When defining the axis-angle representations of unitary and Hermitian matrices, \cite{bri00} wrote that unitary transformations rotate the Stokes polarization vector about $\hat{\Pv{n}}$ by $2\phi$, and Hermitian transformations boost the Stokes four-vector along $\hat{\Pv{m}}$ by $2\beta$.  These definitions describe the inverses of the transformations defined in this article and verified in \App{example_transformation}.


\setcounter{equation}{0}
\section{Down-conversion}
\label{app:downconversion}


By the Nyquist Theorem, a signal must be discretely sampled at a rate equal to twice its bandwidth in order to completely represent its information content.
Therefore, subject to the finite recording rate of
digital observatory equipment, a radio astronomical signal must be constrained to a limited portion of the radio spectrum.
The intermediate process by which the signal from the receiver is band-limited and made ready for discrete sampling is known as down-conversion.

Consider the incoming radio signal, $x(t)$, and its Fourier transform, $X(\nu)$.  The band-limited signal of interest, $x_b(t)$, is parameterized by its centre frequency, $\nu_0$, and bandwidth, \bw.  
Baseband down-conversion is the process by which the spectral information originally contained in the range [$\nu_0-\bw/2$, $\nu_0+\bw/2$] is shifted down to [0, \bw]. 


The spectral information is shifted to baseband by demodulating or mixing the radio frequencies (RF) with a local oscillator (LO).  This is equivalent to multiplying the signal, $x(t)$, with a pure tone, $l(t)=a\cos(2\pi\nu_l t + \phi)$.  
By application of the convolution theorem, and with reference to \Tab{fourier}, mixing may also be understood as a convolution with a pair of complex-valued delta
functions in the frequency domain.  This understanding proves useful in the following sections.

In addition to mixing to lower frequencies, the signal must also be band-limited before analog-to-digital conversion.  Otherwise, power from frequencies higher than the Nyquist frequency will be reflected back into the band of interest, a pollution known as aliasing.  
The ideal low-pass filter is represented by the rectangle function (see \Tab{fourier}) so that a bandpass filter with centre frequency, $\nu_0$, and bandwidth, \bw, is given by
\begin{equation}
\Pi \left( \frac{|\nu|-\nu_0}{\bw} \right).
\end{equation}
Note that the absolute value of $\nu$ in the first term of this equation creates a bandpass window at both positive and negative frequency values.

Down-conversion therefore refers to the combined operation of mixing and band-limiting.  The following two subsections describe in detail two commonly used methods of down-conversion: single-sideband (SSB) and
dual-sideband (DSB).  These are also represented graphically in \Figs{usb}{through}{dsb}.  The process of
down-conversion is performed separately and (ideally) identically on each of the two orthogonal senses of polarization from the receiver feed.

\subsection{Single-sideband Down-conversion}
\label{app:ssb}

\begin{figure}
\centerline{\includegraphics[width=0.5\linewidth]{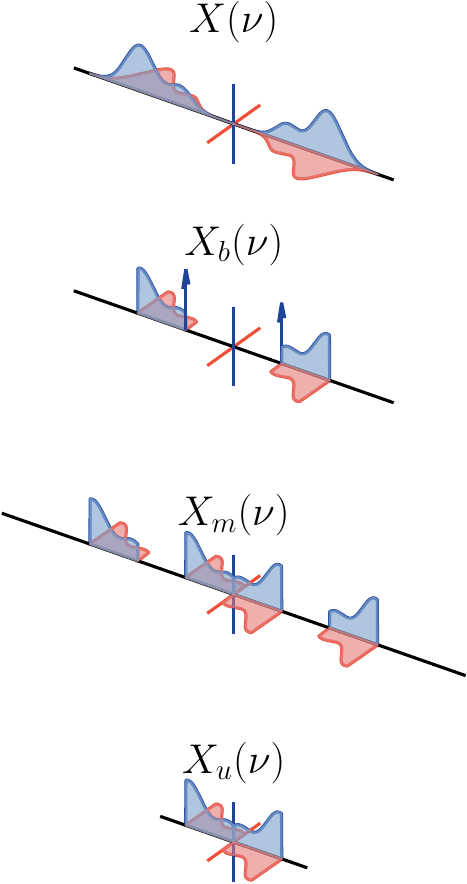}}
\caption{Upper-sideband down-conversion. From the top, the real-valued signal, $x(t)$
has a complex-valued spectrum with conjugate symmetry, such that $X(-\nu)=X^*(\nu)$,
as depicted by $\real[X(\nu)]$ in blue and $\imag[X(\nu)]$ in red.
The signal is bandpass filtered and the band-limited signal, $X_b(\nu)$, 
is mixed with a local oscillator with frequency
$\nu_0-\bw/2$, as depicted by a pair of blue delta functions.
The resulting signal, $X_m(\nu)$, is low-pass filtered, producing $X_u(\nu)$.
For each spectrum, the frequency axis is drawn in black, and the real and
imaginary parts of each complex value are projected along the blue and red
axes (respectively) depicted at the origin (where $\nu = 0$). }
\label{fig:usb}
\end{figure}

\begin{figure}
\centerline{\includegraphics[width=0.5\linewidth]{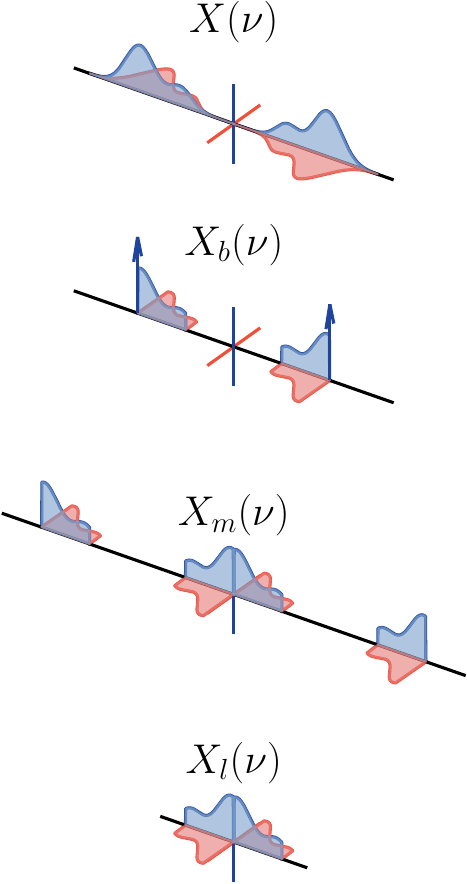}}
\caption{Lower-sideband down-conversion.  As for upper-sideband
except the band-limited signal, $X_b(\nu)$, is mixed with a local
oscillator with frequency $\nu_0+\bw/2$. The resulting
signal, $X_m(\nu)$, is low-pass filtered, producing $X_l(\nu)$.}
\label{fig:lsb}
\end{figure}

During single-sideband down-conversion (SSB), the
signal of interest, $x(t)$, is first bandpass filtered, producing
$x_b(t)$ where \mbox{$X_b(\nu)=X(\nu)\Pi((|\nu|-\nu_0)/\bw)$}.  The
band-limited signal is then mixed with a LO with frequency, $\nu_1$,
producing \mbox{$x_m(t)=x_b(t)\cos(2\pi\nu_1t+\theta)$}, where $\nu_1$ is
set to either 
$\nu_0-\bw/2$ (upper-sideband, shown in \Fig{usb}), producing $x_u(t)$;
or
$\nu_0+\bw/2$ (lower-sideband, shown in \Fig{lsb}), producing $x_l(t)$.
After another stage of low-pass filtering, either $x_u(t)$ or $x_l(t)$
is digitally sampled at the Nyquist rate, $2\bw$.
Compared to $x_u(t)$, the two halves of the spectrum in $x_l(t)$
are swapped, which is equivalent to negating the direction on the
frequency axis and taking the complex conjugate of the spectrum.

During playback, the analytic signal associated with $x(t)$ may be
formed in practice by taking the real-to-complex Fast Fourier
Transform (FFT), followed by the complex-to-complex inverse
FFT.  Most real-to-complex FFT implementations automatically omit the
redundant negative frequencies ($X(-\nu)=X^*(\nu)$) from their output,
implicitly producing the analytic signal.
Since many signal processing operations
(such as phase-coherent dispersion removal)
are performed in the Fourier domain, the cost of calculating the
analytic signal is transparent.

\subsection{Dual-sideband Down-conversion}
\label{app:dsb}

\begin{figure}
\centerline{\includegraphics[width=\linewidth]{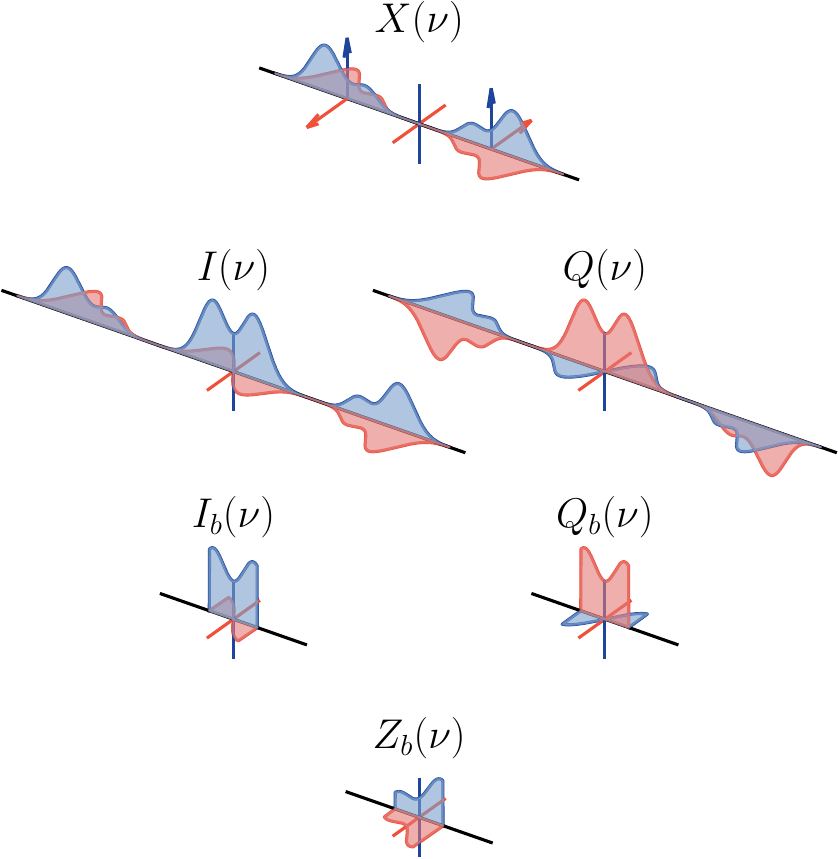}}
\caption{Dual-sideband down-conversion (DSB).  The real-valued
signal, $X(\nu)$, is split before mixing with a local oscillator with frequency $\nu_0$ (blue delta functions) and a $90\deg$ phase-shifted local oscillator (red delta functions), 
yielding the in-phase and quadrature components, I$(\nu)$ and Q$(\nu)$, respectively.
Each of the signals are low-pass filtered, yielding I$_b(\nu)$ and Q$_b(\nu)$ 
with bandwidth $\bw/2$.  The complex signal,
Z$_b(\nu)$=I$_b(\nu)$+\Ci Q$_b(\nu)$, is the analytic
signal associated with $X_u(\nu)$, which is depicted in \Fig{usb}.}
\label{fig:dsb}
\end{figure}

During dual-sideband down-conversion (DSB, see \Fig{dsb}), also known
as quadrature mixing, the voltages from the receiver are split equally
into two signal paths.  One signal is mixed with a local oscillator,
producing
\begin{equation}
	i(t)=x(t)\cos(2\pi\nu_0t).
\end{equation}
The other signal is mixed with the same local oscillator phase-shifted
by 90\deg,
\begin{equation}
q(t)=x(t)\sin(2\pi\nu_0t).
\label{eqn:quadrature}
\end{equation}
Both $i(t)$ and $q(t)$ are low-pass filtered with a cutoff frequency
of $\nu_c=\bw/2$, producing $i_b(t)=i(t)*\sinc(\pi\bw
t)$ and $q_b(t)=q(t)*\sinc(\pi\bw t)$.  The low-pass
filtered signals are then digitally sampled at the Nyquist rate of
$2\nu_c=\bw$.  The signals, $i_b(t)$ and $q_b(t)$ are known as
the in-phase and quadrature components, respectively, of $x(t)$ with
respect to $\nu_0$.

During playback, the analytic signal associated with $x(t)$ is given by
\begin{align}
  z_b(t) & =  i_b(t)+\Ci q_b(t) \nonumber \\
       & =  [x(t)\cos(2\pi\nu_0 t)+\Ci x(t)\sin(2\pi\nu_0 t)] 
		* \sinc(\pi\bw t) \nonumber \\
       & =  [x(t)e^{2\pi\Ci\nu_0t}] * \sinc(\pi\bw t).
\end{align}
In the Fourier domain,
\begin{equation}
	Z_b(\nu)=X(\nu+\nu_0) \Pi(\nu/\bw).
\end{equation}
That is, the spectrum of $z_b(t)$ is equivalent to the band-limited
portion of $x(t)$ centred at $\nu_0$.  The negative frequency
components, centred at $-\nu_0$, have been suppressed by low-pass
filtering, forming the analytic signal associated with $x(t)$.

Note that changing the sign of the $90\deg$ phase shift in \eqn{quadrature}
changes the sign of $q(t)$, which results in complex conjugation of $z_b(t)$.
This is equivalent to modulating $x(t)$ with a local oscillator that has
frequency $-\nu_0$, which shifts the negative half of the spectrum of $X(\nu)$
to baseband.  As in the case of lower-sideband down-conversion, 
the resulting spectrum is complex conjugated and the frequency axis is negated.

During SSB down-conversion, bandpass filtering is performed before mixing, necessitating a filter that is tunable over the range of 
frequencies of interest.  In contrast, the low-pass filter used for DSB down-conversion remains constant regardless of the desired observing frequency.  
For this reason, DSB is often the more economical means of down-conversion.

\subsection{Nyquist Zones}
\label{app:Nyquist}

\begin{figure}
\centerline{\includegraphics[width=0.8\linewidth]{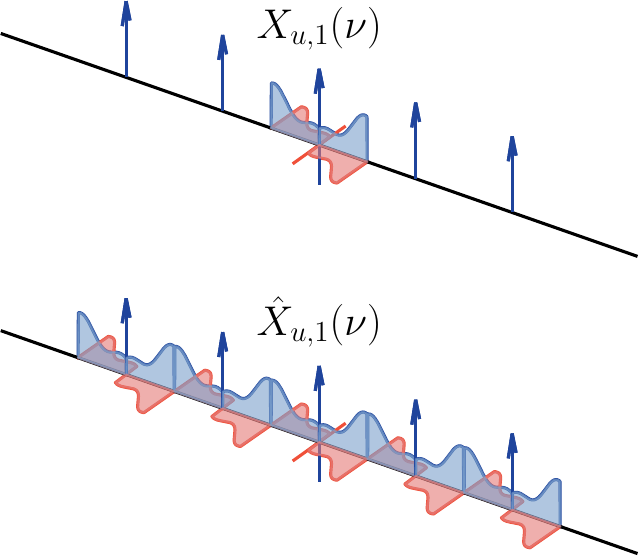}}
\caption{Down-conversion to the first Nyquist zone before sampling.
The band-limited analog signal, $X_{u,1}(\nu)$ is depicted following
upper-sideband down-conversion, along with five of the infinite
series of delta functions separated by the sampling frequency.
The digitized signal $\hat{X}_{u,1}(\nu)$ is the convolution of
the analog signal and the sampling function.}
\label{fig:first_Nyquist}
\end{figure}

In the examples so far, 
the down-converted signal was placed in the first Nyquist zone, which spans
from $-\bw$ to $\bw$ in the case of single-sideband down-conversion,
and from $-\bw/2$ to $\bw/2$ for dual-sideband down-conversion.
The process of discretely sampling the analog signal $x(t)$ at uniformly-spaced instances
in time is equivalent to convolving the band-limited spectrum with an infinite
series of delta functions separated by the sampling frequency.
An example of sampling an analog signal after upper-sideband down-conversion
to the first Nyquist zone is depicted in \Fig{first_Nyquist}.

\begin{figure}
\centerline{\includegraphics[width=0.8\linewidth]{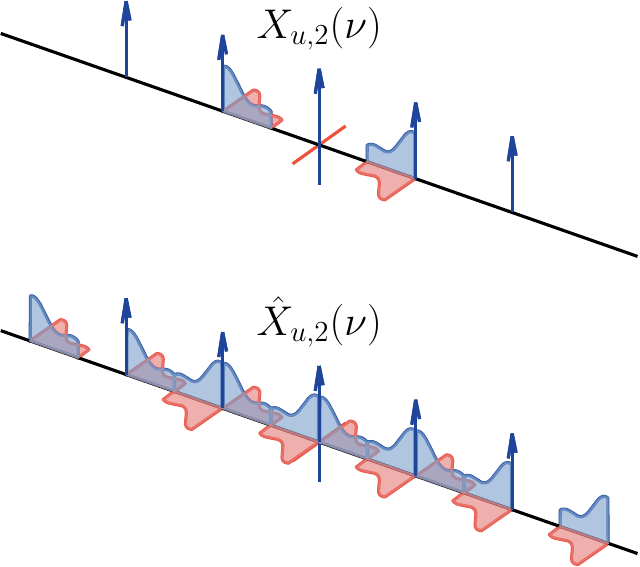}}
\caption{Down-conversion to the second Nyquist zone before sampling.
As for \Fig{first_Nyquist}, except that the analog signal has been
down-converted to occupy the second Nyquist zone before analog-to-digital conversion. }
\label{fig:second_Nyquist}
\end{figure}

It is also possible to place the down-converted and band-limited analog signal in
a region of spectrum offset from the first Nyquist zone.
If the spectrum is offset by $N\bw$,
where $N$ is an integer, then the spectrum is said to be placed in the $(N+1)^\mathrm{th}$
Nyquist zone, and the digitized signal will have frequencies that either increase or decrease
linearly across the spectrum.
An example of sampling an analog signal after upper-sideband down-conversion
to the second ($N$ = 1) Nyquist zone is depicted in \Fig{second_Nyquist}.
With respect to the digitized signal depicted in \Fig{first_Nyquist}, the digitized signal
in \Fig{second_Nyquist} is shifted by $\bw$, which is equivalent to what would
be obtained by digitizing an analog signal after lower-sideband down-conversion
to the first Nyquist zone.
That is, the spectrum is complex conjugated and the frequency axis is negated.

\setcounter{equation}{0}
\section{Reflection as Turning Over the Receiver}
\label{app:yaw_rotation}

Reflection can also be modeled as turning over the receiver, which is equivalent to rotating the reference frame depicted in \Fig{polarization_ellipse}
by $180\deg$ about a transverse axis in the $x$-$y$ plane.
This negates the $z$ axis and results in a two-dimensional reflection in the $x$-$y$ plane 
through the rotation axis.
Let $\hat{\Pv{a}}=(\cos\alpha,\sin\alpha,0)^T$ define the three-dimensional 
rotation axis in the physical space of \Fig{polarization_ellipse}\  
and use Rodrigues' Rotation Formula to express the $3\times3$ matrix 
for a 180$\deg$ rotation about $\hat{\Pv{a}}$.
\begin{equation}
\left( \begin{array}{ccc}
2a_x^2 - 1 & 2a_x a_y    & 0 \\
2a_x a_y   & 2a_x^2 - 1  & 0 \\
0          & 0           & -1 \\
  \end{array} \right)
  \end{equation}
In the $x$-$y$ plane, this transformation reduces to 
\begin{equation}
\JM{W} = \left( \begin{array}{cc}
\cos2\alpha  & \sin2\alpha  \\
\sin2\alpha  & -\cos2\alpha 
  \end{array} \right)
  = \cos2\alpha\, \pauli{1} + \sin2\alpha\, \pauli{2},
\end{equation}
which is a linear combination of the two-dimensional reflection basis matrices 
defined by the two real-valued Pauli matrices.\footnote{
Both \pauli{1}\ and \pauli{2}\ 
satisfy the criteria of a $2\times2$ reflection matrix, which must be orthogonal 
($\JM{W}\JM{W}^T = \JM{W}^T\JM{W} = \pauli{0}$) and
have eigenvalues, $\lambda = \pm 1$, and determinant, $|\JM{W}|=-1$.
}
A reflection in the $x$-$y$ plane negates the \emph{handedness}\footnote{
Strictly, handedness is a property of only three-dimensional
vector spaces; this property is formally known as the orientation of the ordered basis. }
of the two-dimensional basis, which can also be reversed by swapping the cables 
used to propagate the orthogonally polarized signals.
In \citet{vmjr10}, the \emph{feed hand} defines the handedness of the receptor basis
and the reflection is performed through the axis defined by the \emph{symmetry angle}.

The polar decomposition of a two-dimensional reflection through the
axis defined by $(\cos\alpha, \sin\alpha)$ yields a unitary matrix parameterized by 
$\hat{\Pv{n}} = (\cos2\alpha\,\sin2\alpha,0)^T$,
\begin{equation}
\JM{W}(\alpha) = -\Ci\, \vRotation(\pm\pi/2).
\end{equation}

\begin{proof}
Note that $|\JM{W}|^{1/2}=\pm\Ci$ and
\begin{align*}
\JM{W}(\alpha)
    &= -\Ci \, \vRotation(\pi/2) \\
    &= -\Ci \left( \pauli{0}\cos\frac{\pi}{2} 
        + \Ci \, \Pv{n} \cdot \Pv{\sigma} \sin\frac{\pi}{2} \right) \\
    &= \Pv{n} \cdot \Pv{\sigma} \\
    &= \cos2\alpha\, \pauli{1} + \sin2\alpha\, \pauli{2}
\end{align*}
\end{proof}
\noindent
Congruence transformation by $\JM{W}(\alpha)$ rotates $\Pv{S}$ by $180\deg$ about the $\hat{\Pv{n}}$ axis defined by $\alpha$, which negates both the ellipticity angle $\chi$ and the position angle $\psi$.

In a basis defined by linearly-polarized receptors, the nominal axis of symmetry
has a position angle of $45\deg$.
The reflection defined by $\JM{W}(\pi/4)=\pauli{2}$ swaps $e_x$ and $e_y$.
With reference to \eqns{StokesQ}{through}{StokesV}, swapping $e_x$ and $e_y$
negates Stokes Q and V.
This negation is equivalent to a $\pm 180\deg$ rotation of the Stokes polarization vector about $\hat{\Pv{n}} = (0,1,0)^T$,
which is also the result of congruence transformation by $\JM{W}(\pi/4)$.

In the circular basis, the nominal axis of symmetry
has a position angle of $0\deg$, and $\JM{W}(0)=\pauli{1}$ negates $e_y$.
Negating either component 
of the electric field vector reverses the signs of both Stokes U and V, 
which is equivalent to the $\pm 180\deg$ rotation of the Stokes polarization vector about 
$\hat{\Pv{n}} = (1,0,0)^T$ under congruence transformation by $\JM{W}(0)$.



\setcounter{equation}{0}
\section{Fundamental Numerical Instability}
\label{app:instability}

This section considers two forms of instability that arise when fitting a mathematical model of the instrumental response to experimental data.  
%
%
First, model degeneracy arises when there is no unique solution to the measurement equation; in this case, the Hessian matrix is singular and matrix inversion fails.
Second, when one or more model parameters are highly collinear, the Hessian matrix is ill-conditioned, matrix inversion is numerically unstable, and the uncertainties of the collinear model parameters are inflated.

The model under consideration relates the observed Stokes parameters $\Sv{S}'$ to the unknown Stokes parameters $\Sv{S}$ intrinsic to the source by a Mueller matrix $\MM{M}$ that describes the unknown instrumental response via
\begin{equation}
\Sv{S}'(\MM{X}; \mbf{\alpha}) = \MM{M} \, \MM{X} \, \Sv{S}.
\label{eqn:generic_measurement_equation}
\end{equation}
In this measurement equation, $\MM{X}$ is the known constraining transformation, or matrix argument, and $\mbf{\alpha}$ is the set of model parameters that describe both $\MM{M}$ and $\Sv{S}$.
The matrix argument $\MM{X}$ can be the projection between the receptors and the celestial sphere (\Sec{projection}), which can vary with time owing to the rotation of the Earth or mechanical rotation of the feed horn.
Faraday rotation (\Sec{Faraday}) may also serve as the matrix argument under the assumption that the intrinsic Stokes parameters do not vary with radio frequency $\nu$, or are a known function of $\nu$ \citep[e.g.,][]{es04}.

Both forms of numerical instability arise when solving \eqn{generic_measurement_equation} through variation of $\mbf{\alpha}$. Model degeneracy occurs when unknown components of $\MM{M}$ commute with $\MM{X}$, and parameter collinearity arises when unknown components of $\Sv{S}$ are eigenvectors of $\MM{X}$.

\subsection{Commuting with the Matrix Argument}
\label{app:degeneracy}

Appendix B of \cite{van04c} proves that no unique solution to the polarization measurement equation can be derived when only unknown sources are observed at multiple parallactic angles.  Here, the proof is repeated and extended to include impure Mueller matrices (\App{impure}).
Consider  
\begin{equation}
\Sv{S}' = \MM{M} \, \MM{R}(\Phi) \Sv{S}
\label{eqn:parallactic_measurement_equation}
\end{equation}
where $\MM{R}(\Phi)$ represents a rotation about the line of sight by the parallactic angle $\Phi$.

Given $\MM{M}$ and $\Sv{S}$ that satisfy this equation for all $\Phi$, 
it is possible to define a family of solutions, $\MM{M}_u = \MM{M} \, \MM{U}^{-1}$ and $\Sv{S}_u = \MM{U} \, \Sv{S}$,
where $\MM{U}$ is any matrix that commutes freely with $\MM{R}(\Phi)$ for all values of $\Phi$,
%
such that
\begin{equation} \begin{split}
\Sv{S}' & = \MM{M}_u \, \MM{R}(\Phi) \Sv{S}_u \\
 & = \MM{M} \, \MM{U}^{-1} \MM{R}(\Phi) \MM{U} \Sv{S}  \\
 & = \MM{M} \, \MM{U}^{-1}  \MM{U} \, \MM{R}(\Phi) \Sv{S}  \\
 & = \MM{M} \, \MM{R}(\Phi) \Sv{S}.
\end{split} \end{equation}

Pure Mueller matrices that commute with $\MM{R}(\Phi)$ include
rotations about the Stokes~V axis, and 
Lorentz boosts along the Stokes~V axis.
Impure Mueller matrices that commute with $\MM{R}(\Phi)$ can be described using
equation~(46) of \citet[][hereafter LC96]{lc96},
\begin{equation}
\begingroup
\setlength\arraycolsep{5pt}
\MM{M}_\Delta \equiv
\begin{pmatrix}
  1 & \Pv{0}^T \\
  \Pv{P}_\Delta & \PM{M}_\Delta
\end{pmatrix},
\endgroup
\end{equation}
where $\Pv{P}_\Delta = \left( P_{1}, P_{2}, P_{3} \right)^T$
is the polarizance vector that describes 
the conversion of total intensity to polarized flux, 
$|\Pv{P}_\Delta| \le 1$,
$\Pv{0} = (0,0,0)^T$, 
and $\PM{M}_\Delta$ is the $3\times3$
symmetric depolarizer matrix.
This matrix can be diagonalized by a similarity transformation and
written in the form of equation~(43) of LC96,
\begin{equation}
\begingroup
\setlength\arraycolsep{5pt}
\PM{M}_\Delta = {\PM{R}}^{-1}
\begin{pmatrix}
  a & 0 & 0 \\
  0 & b & 0 \\
  0 & 0 & c \\
\end{pmatrix} {\PM{R}}, 
\endgroup
\quad |a|, |b|, |c| \le 1.
\end{equation}
where {\PM{R}} is a $3 \times 3$ rotation matrix.  In this form, it 
can be seen that $\MM{M}_\Delta$ commutes with $\MM{R}(\Phi)$ if
\begin{itemize}
\item only Stokes~V is polarized by $\Pv{P}_\Delta$; and/or
\item ${\PM{R}}$ is a rotation about the Stokes~V axis, and
\begin{itemize}
\item Stokes Q and U are equally depolarized by $\MM{M}_\Delta$, and/or
\item Stokes~V is depolarized by $\MM{M}_\Delta$.
\end{itemize}
\end{itemize}
For linearly-polarized receptors, 
only Stokes~V is polarized if $\Pv{P}_\Delta = \left(0,0,P_{3}\right)^T$.
Furthermore, Stokes Q and U are equally depolarized by $\MM{M}_\Delta$ when $a = b$
(as for the case of stochastic Faraday rotation described in \App{impure}).

Note that depolarization of Stokes~V includes negation ($c = -1$) and,
as described in \Sec{conjugation}, no linear transformation of the electric field 
can negate the sign of only Stokes~V.  For example, if the instrumental response is modeled 
using Jones matrices, then negation of Stokes~V in the unknown $\Sv{S}$
can be compensated only by rotating the reference frame.  However, a $\pm 180\deg$ 
rotation that negates Stokes~V must also negate the position angle and derived
quantities such as the Faraday rotation measure.
Therefore, the Stokes~V sign ambiguity can be eliminated by observing a source for which 
the sign of the position angle and/or rotation measure is known.

Owing to commutation, there is no unique solution to \eqn{parallactic_measurement_equation} 
and other constraints or assumptions must be introduced to constrain
the degenerate dof.
Similar degeneracy will arise whenever the experimental constraints
include only observations of unknown sources of radiation
as a function of a matrix argument with a fixed axis of symmetry.

When modeling variations of the observed Stokes parameters as a function of a matrix argument with a variable axis of symmetry, it is no longer possible for an unknown component of $\MM{M}$ to commute with all values of $\MM{X}$, and the degeneracy is eliminated.
However, as described in the following section, even when the axis of symmetry of $\MM{X}$ is variable, unknown model parameters can be highly collinear and cause numerical instability.

\subsection{Eigenvectors of the Matrix Argument}
\label{app:eigenvectors_of_projection}

In some experiments, the observed Stokes parameters are related to the intrinsic Stokes parameters by a measurement equation that includes a matrix argument with a variable axis of symmetry.
Such a measurement equation may have no fundamental degeneracy; however, collinearity 
between model parameters can arise when one or more unknown components of the polarization states used as constraints are eigenvectors of the matrix argument.

For example, when observations are made over a wide range of hour angles with a fixed dipole array, the geometric projection transformation,
\begin{equation}
\MM{P} \simeq
\MM{B}_L(l,m) \MM{R}(\Phi),
\end{equation}
where $\MM{B}_L(l,m)$ approximates the foreshortening of the projected receptors, 
a direction-dependent Lorentz boost with variable axis of symmetry in the $Q$--$U$ plane,
and $\MM{R}(\Phi)$ models the rotation of the observatory about the line of sight by the parallactic angle $\Phi$,
a rotation with fixed symmetry along the Stokes V axis.
The projection has a symmetry axis that varies with time and therefore there is no fundamental degeneracy.  

However, the Stokes vector, $\Sv{V}=[0,0,0,V]^T$ has a polarization vector that is parallel to the symmetry axis of $\MM{R}$ and perpendicular to the symmetry axis of $\MM{B}_L$. Therefore, it is
an eigenvector of $\MM{P}$ with associated eigenvalue $\lambda = 1$; i.e.,
\begin{equation}
\MM{P} \, \Sv{V}=\Sv{V}.
\end{equation}
Consequently, when the model parameters include unknown circular polarization intrinsic to the sources used as constraints, the Stokes V components
are highly covariant with the unknown instrumental boost along the Stokes V axis, $\MM{B}_V(\beta)$, causing the Hessian matrix to be ill-conditioned.

To demonstrate this, first consider the typical Gauss–Newton approximation of
the Hessian matrix ${\bf H}$.  Here, second-order derivative terms are ignored and
\begin{equation}
    {\bf H} \simeq 2 {\mbf\Upsilon}^T{\mbf\Upsilon},
\end{equation}
where ${\mbf\Upsilon}$ is the Jacobian matrix with elements defined by the partial derivatives
of the predicted Stokes parameters $S'_j$ with respect to the free model parameters $\alpha_k$,
\begin{equation}
    \Upsilon_j^k = \frac{\partial S'_j}{\partial \alpha_k}.
\end{equation}
In principle, the row index $j$ covers all observations of all 4 Stokes parameters of all source states included as constraints.  However, for brevity, the observation index will be ignored and $S$ will loop over the four Stokes parameters, such that $j$ indexes only the source state and $S'_j \in \{ I'_j, Q'_j, U'_j, V'_j \}$.

The Hessian is ill-conditioned if there is a high degree of multicollinearity between the unknown model parameters, such that the gradient vector (or column of the Jacobian) for a model parameter is (nearly) proportional to a linear combination of the gradient vectors for a set of other parameters.

To characterize the multicollinearity between model parameters in the approximation of a fixed dipole array, consider a simplified model of the observed Stokes
parameters in which the unknown instrumental response consists of only a Lorzentz boost along the Stokes V axis.  For the $n^\mathrm{th}$ source used as a constraint,
\begin{equation}
    \Sv{S}'_n = \MM{B}_V(\beta) \MM{P} \, \Sv{S}_n.
\end{equation}
The set of unknown model parameters include the boost rapidity $\beta$ and the Stokes parameters intrinsic to each source.
Let the model of the intrinsic Stokes parameters for each source be decomposed as
\begin{equation}
    \Sv{S}_n = \Sv{S}_{n,L} + \Sv{S}_{n,V},
\end{equation}
where $\Sv{S}_{n,L}=[I_n,Q_n,U_n,0]^T$ and $\Sv{S}_{n,V}=[0,0,0,V_n]^T$,
such that
\begin{equation}
    \Sv{S}'_n = \MM{B}_V(\beta) \left( \MM{P} \,  \Sv{S}_{n,L} +  \Sv{S}_{n,V} \right).
\end{equation}
With reference to \eqn{Stokes_boost},
the partial derivatives of $\Sv{S}'_n$ with respect
to $\beta$ include only
\begin{equation}
\begin{split}
\frac{\partial I'_j}{\partial\beta} &= 2 \left( I''_j \sinh 2\beta + V_j \cosh 2\beta \right) \\
\frac{\partial V'_j}{\partial\beta} &= 2 \left( I''_j \cosh 2\beta + V_j \sinh 2\beta \right), \label{eqn:partial_beta}
\end{split}
\end{equation}
where $I''_j$ is the total intensity of the $j^\mathrm{th}$ source after passing through the projection transformation.
Similarly, the partial derivatives of $\Sv{S}'_j$ with respect
to $V_n$ include only
\begin{equation}
\begin{split}
\frac{\partial I'_j}{\partial V_n} &= \delta_{jn} \sinh 2\beta  \\
\frac{\partial V'_j}{\partial V_n} &= \delta_{jn} \cosh 2\beta.
\end{split}
\end{equation}
To first order, the total intensity changes by only a small fraction as the projection transformation varies.  For example, even when the differential gain is as large as 2, $\gamma = \ln(2)/2 \sim 0.35$ \eqnp{differential_gain} and $\cosh\gamma \sim 1.06$.
Therefore, assume that $I_n''\simeq I_n$ and consider the following linear combination of $V_n$ gradient vectors,
\begin{equation}
\mbf{\Upsilon}_V = \sum_n I_n \mbf{\Upsilon}_{V_n},
\end{equation}
where $\mbf{\Upsilon}_{V_n}$ is the gradient vector defined by the column of the Jacobian matrix corresponding to $V_n$.  The vector $\mbf{\Upsilon}_V$ has components,
\begin{equation}
    \Upsilon_{V,j} = \sum_n I_n \frac{\partial S'_j}{\partial V_n} = I_j \frac{\partial S'_j}{\partial V_j}
\end{equation}
and its inner product with the gradient vector for $\beta$,
\begin{equation} \begin{split}
\mbf{\Upsilon}_V \cdot \mbf{\Upsilon}_\beta 
& = \sum_j I_j \frac{\partial S'_j}{\partial V_j} \frac{\partial S'_j}{\partial \beta}. \\
& \simeq
2 \sum_j \left[ I_j^2 X(\beta) + I_j V_j Y(\beta) \right]
\end{split} \end{equation}
where
\begin{equation} \begin{split}
X(\beta)&=\sinh^2 2\beta + \cosh^2 2\beta  \\
Y(\beta)&=\sinh 2\beta\cosh 2\beta.
\end{split} \end{equation}
Similarly,
\begin{align}
|\mbf{\Upsilon}_V|^2 & \simeq \sum_j I_j^2 X(\beta)
\end{align}
and
\begin{align}
|\mbf{\Upsilon}_\beta|^2 & \simeq 4 \sum_j \left[ (I_j^2+V_j^2) X(\beta) + I_jV_j Y(\beta) \right]
\end{align}
For small $\beta$, $X(\beta)\simeq 1$ and $Y(\beta) \simeq 2\beta$.
If $V_j/I_j$ is also small, then (to first order) terms involving $V_j^2$ and $\beta V_j$ can be ignored, and the cosine similarity between 
$\mbf{\Upsilon}_V$ and $\mbf{\Upsilon}_\beta$,
\begin{equation*}
\frac{\mbf{\Upsilon}_V \cdot \mbf{\Upsilon}_\beta}
     {|\mbf{\Upsilon}_V| |\mbf{\Upsilon}_\beta|} \simeq
     \frac{2 \sum I_j^2}{ \left(\sum I_j^2\right)^\frac{1}{2} \left(4\sum I_j^2\right)^\frac{1}{2} } = 1  
\end{equation*}
Owing to the high multicollinearity between $V_n$ and $\beta$, the Hessian matrix is poorly-conditioned, its inversion is numerically unstable, and the formal uncertainties of $\beta$ and $V_n$ are inflated.

\end{appendix}

\bibliographystyle{aasjournal}
\bibliography{journals,modrefs,psrrefs,local,crossrefs}

\end{document}